%% file: CENSUSII.tex
\documentclass[]{aastex63}
\usepackage{graphicx}
\usepackage{subfigure}
\usepackage{multirow}
\usepackage{mathptmx} 
\usepackage{hyperref}

\hypersetup{linkcolor=blue,citecolor=blue,filecolor=cyan,urlcolor=blue}

\submitjournal{ApJ}

\begin{document}

\title{Surveys of Clumps, Cores, and Condensations in the Cygnus X: 
\\II. Radio Properties of the Massive Dense Cores}

\correspondingauthor{Keping Qiu}
\email{kpqiu@nju.edu.cn}

\author[0000-0001-6630-0944]{Yuwei Wang} 
\affil{School of Astronomy and Space Science, Nanjing University, 163 Xianlin Ave., Nanjing, 210023, P.R. China}
\affil{Key Laboratory of Modern Astronomy and Astrophysics (Nanjing University), Ministry of Education, Nanjing 210023, P.R.China}

\author[0000-0002-5093-5088]{Keping Qiu}
\affil{School of Astronomy and Space Science, Nanjing University, 163 Xianlin Ave., Nanjing, 210023, P.R. China}
\affil{Key Laboratory of Modern Astronomy and Astrophysics (Nanjing University), Ministry of Education, Nanjing 210023, P.R.China}

\author[0000-0002-6368-7570]{Yue Cao}
\affil{School of Astronomy and Space Science, Nanjing University, 163 Xianlin Ave., Nanjing, 210023, P.R. China}
\affil{Key Laboratory of Modern Astronomy and Astrophysics (Nanjing University), Ministry of Education, Nanjing 210023, P.R.China}
\affil{Center for Astrophysics $\vert$ Harvard \& Smithsonian, 60 Garden Street, MS 42, Cambridge, MA 02138, USA}

\author[0000-0002-8691-4588]{Yu Cheng}
\affil{Dept. of Astronomy, University of Virginia, Charlottesville, Virginia 22904, USA}

\author[0000-0002-4774-2998]{Junhao Liu}
\affil{School of Astronomy and Space Science, Nanjing University, 163 Xianlin Ave., Nanjing, 210023, P.R. China}
\affil{Key Laboratory of Modern Astronomy and Astrophysics (Nanjing University), Ministry of Education, Nanjing 210023, P.R.China}

\author[0000-0002-3286-5469]{Bo Hu}
\affil{School of Astronomy and Space Science, Nanjing University, 163 Xianlin Ave., Nanjing, 210023, P.R. China}
\affil{Key Laboratory of Modern Astronomy and Astrophysics (Nanjing University), Ministry of Education, Nanjing 210023, P.R.China}

\begin{abstract}
We have carried out a high-sensitivity and high-resolution radio continuum study towards a sample of 47 massive dense cores (MDCs) in the Cygnus X star-forming complex using the \textit{Karl G. Jansky Very Large Array}, aiming to detect and characterize the radio emission associated with star-forming activities down to $\sim$0.01 pc scales. We have detected 64 radio sources within or closely around the full widths at half-maximum (FWHMs) of the MDCs, of which 37 are reported for the first time. The majority of the detected radio sources are associated with dust condensations embedded within the MDCs, and they are mostly weak and compact. We are able to build spectral energy distributions for 8 sources. Two of them indicate non-thermal emission and the other six indicate thermal free-free emission. We have determined that most of the radio sources are ionized jets or winds originating from massive young stellar objects, whereas only a few sources are likely to be ultra-compact \ion{H}{2} regions. Further quantitative analyses indicate that the radio luminosity of the detected radio sources increases along the evolution path of the MDCs.
\end{abstract}


\section{Introduction} \label{sec:intro}
Massive stars (M$_* \geqslant$ 8 M$_\odot$) play a vital role in shaping galaxies in all stages of their lives (e.g. \citealt{2005IAUS..227....3K}). Yet our understanding of their formation processes remains sketchy. Massive stars are rare. Their birthplaces are always hidden by distant and highly obscured dust. Only at limited wavelengths, e.g. infrared (IR), (sub)millimeter, and radio, and from their violent interactions with the surroundings can we get some hints on their early formation processes (e.g. \citealt{2011ApJ...728....6Q, 2013ApJ...767L..13R}).

In the very early stage of high-mass star formation (HMSF), the (pre-)protostellar cores are cold and are only detectable at (sub-)millimeter and far-infrared wavelengths \citep{2007ARA&A..45..339B, 2010ApJ...715..310R}. With the growth of the embedded massive young stellar objects (MYSOs), the cold dense molecular cores are heated and become observable at mid-infrared wavelengths \citep{1993ApJ...405..706O, 2007PASP..119..855W, 2010A&A...510A..89M, 2013PASA...30...57J}. The so-called ``hot cores'' \citep{2000prpl.conf..299K} formed at this stage have weak free-free emission, which is now detectable by the most sensitive radio telescopes, e.g. \textit{NSF's Karl G. Jansky Very Large Array} (VLA) \citep{1993A&A...276..489O, 2016ApJS..227...25R}. 

In the next evolutionary stage, the abundant UV photons produced by the central high-mass stars begin to ionize the surrounding material, forming ultra-compact (UC) \ion{H}{2} regions \citep{2002ARA&A..40...27C}. UC \ion{H}{2} regions are strong free-free radiation emitters with diameters smaller than 0.1 pc, electron densities ($N_e$) higher than 10$^4$ cm$^{-3}$, and emission measures ($EM$) higher than 10$^7$ pc cm$^{-6}$ \citep{1989ApJS...69..831W}. Morphologically, UC \ion{H}{2} regions are also resolved to have several typical structures, which is a joint effect of the dynamics of the ionized gas, the surrounding material, and the evolutionary stages of UC \ion{H}{2} regions \citep{1989ApJS...69..831W, 2002ARA&A..40...27C, 2007prpl.conf..181H, 2010ApJ...719..831P}. 

After the first discovery of UC \ion{H}{2} regions in the 1960s \citep{1967ApJ...148L..17R, 1979ARA&A..17..345H}, many radio surveys of UC \ion{H}{2} regions have been carried out with single dishes and interferometers. \citet{1982A&A...108..227W} carried out a pioneer single-dish radio continuum survey towards 85 UC \ion{H}{2} regions. \citet{1989ApJS...69..831W} and \citet{1994ApJS...91..659K} selected dozens of UC \ion{H}{2} regions from the IRAS survey and performed high-resolution VLA observations towards these regions. \citet{1998MNRAS.301..640W} carried out a survey towards hundreds of UC \ion{H}{2} regions with the ATCA. \citet{2007A&A...461...11U, 2009A&A...501..539U} and \citet{2013ApJS..208...11L} carried out the Red MSX Source Survey and observed abundant UC \ion{H}{2} regions at the resolution of $\sim$1$''$ to 2$''$ with the ATCA and the VLA, covering both the southern and northern hemispheres. More recently, \citet{2018A&A...615A.103K} has carried out an unbiased high-resolution survey and provided a catalog of more than 200 UC \ion{H}{2} regions in the Galactic plane. 

A denser counterpart of UC \ion{H}{2} region, namely hyper-compact (HC) \ion{H}{2} region, was discovered during the surveys of UC \ion{H}{2} regions (e.g. \citealt{1993ApJ...417..645G, 1995ApJ...438..776G}). HC \ion{H}{2} regions have diameters smaller than 0.05 pc, $N_e$ higher than 10$^6$ cm$^{-3}$, and $EM$ higher than 10$^{10}$ pc cm$^{-6}$ \citep{2007prpl.conf..181H}. They remain unresolved or barely resolved under interferometric observations \citep{2010MNRAS.405.1560M}. Considering the expansion tendency of UC \ion{H}{2} regions, HC \ion{H}{2} regions used to be regarded as a sub-class of UC \ion{H}{2} regions that are formed at the very early stages of ionization \citep{2005IAUS..227..111K, 2019MNRAS.482.2681Y}. Yet now they are considered to be a distinct class because of their broad radio recombination line profiles (about 40--50 km s$^{-1}$) \citep{2004ApJ...605..285S, 2004ApJ...600..286D, 2011ApJS..194...44S} and probably different driving mechanisms \citep{2002ApJ...580..980K, 2004ApJ...610..351I, 2006ApJ...637..850K, 2007prpl.conf..181H}.

Besides heating and ionizing the surroundings, MYSOs drive jets and molecular outflows during their rapid accretion processes. Ionized jets and molecular outflows are ubiquitous in the formation of stars at all mass regimes \citep{1992ApJ...395..494A, 2003ApJ...587..739G} and are keys to understanding the star formation mechanisms \citep{2007prpl.conf..245A, 2014prpl.conf..451F}. Most of the ionized jets have thermal free-free emission and can be best observed in the centimeter wavelengths. For some radio jets with extreme energy, non-thermal emission can be observed \citep{1989ApJ...346L..85R, 2005ApJ...626..953R}. This non-thermal component is commonly interpreted as synchrotron radiation from shocks produced by high-velocity jets shooting into surrounding dense molecular clouds \citep{2010Sci...330.1209C, 2016ApJ...816...64C}.

Previous studies have provided abundant information about high-mass star-forming regions. Whereas the surveys aiming to cover a large area are always subject to incoherent distances and low-sensitivities; the works focusing on obtaining deep insights into a few well-observed cases are limited in statistical significance. To better characterize the radio properties of HMSF and achieve statistically significant results, it requires an unbiased survey with high sensitivity and high resolution towards an adequate number of MDCs at the same distance. We have launched such a survey in the Cygnus X star-forming complex, an ideal laboratory for HMSF studies. The Cygnus X star-forming complex is one of the nearest and most active massive-star-forming regions in our Galaxy \citep{2007A&A...476.1243M, 2012A&A...539A..79R, 2018ARA&A..56...41M}. It exhibits HMSF processes in all evolution stages and has a rich collection of \ion{H}{2} regions, OB associations, and high-mass star-forming sites \citep{1977A&A....58..197H, 1991A&A...241..551W, 2001A&A...371..675U, 2004ApJ...601..952S, 2007A&A...474..873S, 2013ApJ...772...69G}. An average distance of 1.4 kpc \citep{2012A&A...539A..79R} makes it possible to resolve the massive dense cores (MDCs) down to the sub-Jeans scale ($<$ 0.05 pc) with the high-resolution facilities such as the VLA and the Submillimeter Array (SMA). Most of the previous studies focused on the core scale and discussed the relation between the MDCs' properties and HMSF. Yet later high-resolution observations have revealed that the MDCs are more likely to fragment into condensations \citep{2010A&A...524A..18B, 2018A&A...615A..94F, 2019ApJ...878..146S}, which is in agreement with our SMA observations (Qiu et al., in preparation). Here we use ``core'' for the cloud structure of $\sim$0.1 pc, and ``condensation'' for the structure fragmented from the core with a typical size of $\sim$0.01 pc. In the hierarchical scheme of HMSF, condensation is more directly related to the formation of individual stars or binaries. Thus the studies at the condensation scale can provide us a more explicit insight into the immediate environments of HMSF. With the high angular resolution of the VLA, we are now able to study the radio properties of individual condensations.

This is the second paper of our project, the Surveys of {\bf C}lumps, Cor{\bf E}s, and Co{\bf N}den{\bf S}ations in Cygn{\bf US} X (CENSUS, PI: Keping Qiu), which aims to build a hierarchical view of the HMSF in the Cygnus X star-forming complex from 1--10 pc molecular clumps down to 0.01 pc dust condensations. This project uses the data from infrared to radio wavelengths observed by the facilities such as the \textit{Hershel Space Observatory}, SMA, JVLA, JCMT, ALMA, NOEMA, CARMA, and Tianma 65 m radio telescope. The work presented here takes advantage of the VLA data and focuses on the radio properties of the MDCs under high sensitivity and resolution. 

\section{Sample Selection}\label{sec:sample}

We initially built the MDC sample based on the work of \citet{2007A&A...476.1243M} (hereafter Motte07), who identified 40 MDCs with masses higher than 30 M$_\odot$ from an extinction map of the Cygnus X star-forming complex and 1.2 mm dust continuum observations of selected high-extinction regions. The region along the direction of the OB2 association was not chosen for the MDC survey in Motte07 owing to the low extinction. We included a further 8 MDCs in this region, based on the JCMT and Herschel sub-millimeter continuum observations (\citealt{2019ApJS..241....1C}, hereafter Cao19). In a recent work by \citet{2021ApJ...918L...4C} (hereafter Cao21), over eight thousand dense cores were identified from a column density map of the Cygnus X star-forming complex using the \textsl{getsources} software. We then cross-checked and updated the sample with that in Cao21 since the column density map can better characterize the MDCs and the properties of the MDCs are derived with a uniform method. The parameters of the MDCs are also obtained from Cao21 as given in Table \ref{tab:MDC_Cao21}, except for the 24 $\mu$m flux. The 24 $\mu$m fluxes are obtained from the \textsl{Spitzer} Cygnus X Archive Catalog \citep{2010AAS...21541401K, CygnusX_Archive2011} and the MSX6C Infrared Point Source Catalog \citep{2003AAS...203.5708E, MSX6C2015}. The 21.3 $\mu$m fluxes are linear scaled to 24 $\mu$m according to the model of a B3-type embryo \citep{2019ApJS..241....1C}.  The parameters of the MDCs, e.g., FWHMs (full width at half maximum), in Cao21 are slightly different from those in Motte07 and Cao19. Two dense cores, 742 ( NW01 in Motte07) and 839 (NW12 in Motte07) have masses lower than 30 M$_\odot$. We still take them for study because they have been covered by our VLA programs. We do not specifically distinguish these two dense cores from the others in the following content. Finally, we have a sample of 47 MDCs, among which eight are located in the OB2 region. Their detailed physical parameters are listed in Table \ref{tab:MDC_Cao21}.

Cao21 analyzed the 1.3 mm continuum maps of the sampled MDCs using the SMA and identified dust condensations with \textsl{getsources} from each MDC. Dust condensations are $\sim$0.01 pc-scale structures that are fragmented from the MDCs. Their spatial scales are comparable to the resolution of our VLA data, which helps to study the immediate environment of the radio sources.

\begin{longrotatetable}
\begin{deluxetable*}{c|c|c|ccccc|cccccc|l}
\tablecaption{Physical Parameters of the MDCs\label{tab:MDC_Cao21}}
\tablehead{
\colhead{MDC} & \colhead{Cao19} & \colhead{Motte07} & \colhead{RA} & \colhead{Dec} & \colhead{$D_{maj}$} & \colhead{$D_{min}$} & \colhead{PA} & \colhead{Mass} & \colhead{T$_{dust}$} & \colhead{$L_{FIR}$} & \colhead{$F_{\nu, \ 24\ \mu m}$} & \colhead{$N_{H_2}$} & \colhead{$n_{H_2}$} & \colhead{Type} \\
\colhead{} & \colhead{} & \colhead{} & \colhead{(hms)} & \colhead{(dms)} & \colhead{(pc)} & \colhead{(pc)} & \colhead{(deg)} & \colhead{($M_\odot$)} & \colhead{(K)} & \colhead{($L_\odot$)} & \colhead{(Jy)} & \colhead{(cm$^{-2}$)} & \colhead{(cm$^{-3}$)} & \colhead{}
}
\startdata
214 & C03-1 &  \nodata  & 20:30:29.03 & 41:15:57.1 & 0.276 & 0.179 & 133.6 & 143.64 & 21.24 & 1268.2 & 5.85 & 1.7E+23 & 3.6E+05 &  Q \\
220 & N03-1 & N02 N03 & 20:35:34.17 & 42:20:10.8 & 0.287 & 0.247 & 31.0 & 383.1 & 17.75 & 1151.8 & 5.41 & 3.1E+23 & 5.6E+05 &  Q \\
247 &  \nodata  &  \nodata  & 20:30:27.34 & 41:16:13.7 & 0.300 & 0.198 & 84.1 & 43.24 & 21.28 & 385.8 & 50.74\tablenotemark{a} & 4.1E+22 & 8.3E+04 &  B \\
248 & N12-1 & N12 N13 & 20:36:57.55 & 42:11:35.2 & 0.267 & 0.204 & 145.5 & 202.05 & 18.17 & 698.9 & 0.97 & 2.1E+23 & 4.4E+05 &  Q \\
274 & N05-2 & N05 N06 & 20:36:07.41 & 41:39:58.0 & 0.370 & 0.229 & 98.8 & 107.38 & 21.51 & 1022.1 & 122.96\tablenotemark{a} & 7.2E+22 & 1.2E+05 &  B \\
302 & C08-2 &  \nodata  & 20:35:07.72 & 41:14:01.7 & 0.311 & 0.246 & 122.4 & 76.26 & 22.29 & 899.4 &  \nodata  & 5.7E+22 & 9.9E+04 &  C \\
310 & NW12-1 & NW14 & 20:24:31.73 & 42:04:20.4 & 0.240 & 0.195 & 152.2 & 152.78 & 20.48 & 1084.3 & 18.21\tablenotemark{a} & 1.9E+23 & 4.2E+05 &  B \\
327 &  \nodata  & NW02 & 20:19:40.67 & 40:57:08.4 & 0.300 & 0.222 & 95.5 & 120.74 & 20.87 & 958.6 &  \nodata  & 1E+23 & 1.9E+05 &  Q \\
340 & C05-1, 2 &  \nodata  & 20:32:23.46 & 41:07:52.7 & 0.309 & 0.245 & 113.7 & 119.29 & 16.09 & 199.4 & 0.21 & 8.9E+22 & 1.6E+05 &  Q \\
341 & N63-1 & N63 & 20:40:05.40 & 41:32:13.3 & 0.245 & 0.235 & 4.4 & 160.91 & 17.44 & 435.2 & 0.60 & 1.6E+23 & 3.2E+05 &  Q \\
351 & S32-1 & S32 & 20:31:20.72 & 38:57:15.4 & 0.258 & 0.190 & 141.2 & 78.63 & 18.29 & 283.2 & 0.18 & 9.1E+22 & 2E+05 &  Q \\
370 &  \nodata  &  \nodata  & 20:28:09.35 & 40:52:54.0 & 0.245 & 0.184 & 12.8 & 57.1 & 22.73 & 757.5 &  \nodata  & 7.2E+22 & 1.6E+05 &  C \\
507 & S07-1 & S08 S09 & 20:20:38.43 & 39:37:45.4 & 0.322 & 0.231 & 1.2 & 372.46 & 21.4 & 3439.6 & 196.83\tablenotemark{a} & 2.8E+23 & 5.1E+05 &  B \\
509 & S30-1, 2, 3 & S30 S31 & 20:31:13.29 & 40:03:12.6 & 0.341 & 0.287 & 134.6 & 300.64 & 18.18 & 1045.0 & 2.29 & 1.7E+23 & 2.7E+05 &  Q \\
520 & C08-3 &  \nodata  & 20:35:10.60 & 41:13:11.0 & 0.362 & 0.264 & 8.1 & 81.01 & 23.43 & 1289.1 &  \nodata  & 4.8E+22 & 7.6E+04 &  C \\
540 & S43-2, 3 & S42 S43 & 20:32:40.73 & 38:46:26.3 & 0.398 & 0.253 & 6.6 & 240.78 & 15.54 & 326.0 & 2.91 & 1.4E+23 & 2.1E+05 &  Q \\
608 &  \nodata  &  \nodata  & 20:33:59.89 & 41:22:28.9 & 0.247 & 0.185 & 177.8 & 30.6 & 21.12 & 261.1 &  \nodata  & 3.8E+22 & 8.6E+04 &  Q \\
640 &  \nodata  & NW04 & 20:20:30.07 & 41:22:06.8 & 0.256 & 0.190 & 160.5 & 52.79 & 18.61 & 210.9 &  \nodata  & 6.1E+22 & 1.4E+05 &  Q \\
675 & NW04-1 & NW05 NW07 & 20:20:31.39 & 41:21:27.5 & 0.338 & 0.248 & 139.2 & 104.26 & 20.08 & 656.8 & 116.15\tablenotemark{a} & 7.1E+22 & 1.2E+05 &  B \\
684 & N68-1 & N68 & 20:40:33.83 & 41:59:03.5 & 0.257 & 0.203 & 38.8 & 109.49 & 17.62 & 315.5 & 0.20 & 1.2E+23 & 2.5E+05 &  Q \\
698 & DR21-9 & N56 & 20:39:17.42 & 42:16:10.4 & 0.225 & 0.190 & 109.8 & 81.14 & 16.98 & 186.8 & 5.65 & 1.1E+23 & 2.5E+05 &  Q \\
699 & DR21-3, 13 & N38 N48 & 20:39:00.02 & 42:22:16.0 & 0.397 & 0.282 & 128.8 & 1283 & 20.92 & 10332.1 & 38.42\tablenotemark{a} & 6.5E+23 & 9.5E+05 &  B \\
714 & N05-3 & N14 & 20:37:01.07 & 41:35:00.1 & 0.228 & 0.216 & 134.8 & 101.14 & 22.73 & 1340.3 & 26.67\tablenotemark{a} & 1.2E+23 & 2.5E+05 &  B \\
723 & S29-1 & S29 & 20:29:57.75 & 40:15:54.7 & 0.328 & 0.257 & 52.2 & 123.9 & 16.02 & 201.5 & 0.32 & 8.4E+22 & 1.4E+05 &  Q \\
725 & N05-1 & N10 & 20:36:52.18 & 41:36:23.8 & 0.292 & 0.238 & 146.4 & 147.64 & 24.04 & 2737.4 & 134.39\tablenotemark{a} & 1.2E+23 & 2.2E+05 &  B \\
742 & NW01-1 & NW01 & 20:19:38.81 & 40:56:39.2 & 0.292 & 0.189 & 168.3 & 19.44 & 26.97 & 718.8 & 542.07\tablenotemark{a} & 2E+22 & 4.1E+04 &  B \\
753 &  \nodata  & S06 S07 & 20:20:37.57 & 39:38:25.8 & 0.302 & 0.213 & 92.2 & 189.53 & 17.82 & 582.8 &  \nodata  & 1.7E+23 & 3.2E+05 &  Q \\
798 &  \nodata  & S10 & 20:20:44.02 & 39:35:25.1 & 0.341 & 0.223 & 166.6 & 89.27 & 19.19 & 429.1 & 1.55 & 6.7E+22 & 1.2E+05 &  Q \\
801 & N68-2, 5 & N64 N65 & 20:40:28.06 & 41:57:05.7 & 0.331 & 0.236 & 26.9 & 143.46 & 17.84 & 444.4 & 5.64 & 1E+23 & 1.8E+05 &  Q \\
839 &  \nodata  & NW12 & 20:24:13.93 & 42:11:41.4 & 0.189 & 0.181 & 135.9 & 15.83 & 16.19 & 27.4 & 0.03 & 2.6E+22 & 6.9E+04 &  Q \\
892 & DR15-4 & S41 & 20:32:33.85 & 40:16:58.6 & 0.226 & 0.191 & 134.3 & 31.4 & 31.64 & 3031.6 &  \nodata  & 4.1E+22 & 9.7E+04 &  C \\
1018 & DR21-8 & N36 N40 N41 & 20:38:59.33 & 42:23:37.2 & 0.416 & 0.278 & 11.0 & 811.38 & 19.21 & 3922.3 & 0.66 & 4E+23 & 5.7E+05 &  Q \\
1112 & W75N-1 & N30 N31 N32 & 20:38:36.70 & 42:37:48.6 & 0.420 & 0.369 & 109.6 & 499.31 & 28.08 & 23521.8 & \nodata & 1.8E+23 & 2.3E+05 &  B \\
1179 & DR21-10 & N57 & 20:39:19.47 & 42:16:01.8 & 0.280 & 0.211 & 125.9 & 35.9 & 16 & 58.0 &  \nodata  & 3.4E+22 & 6.9E+04 &  N \\
1201 & S106-2 & S18 S20 & 20:27:25.83 & 37:22:53.6 & 0.374 & 0.205 & 90.8 & 52.83 & 33.37 & 7012.6 &  \nodata  & 3.9E+22 & 6.9E+04 &  C \\
1225 & DR15-1 & S34 & 20:31:57.45 & 40:18:29.3 & 0.417 & 0.290 & 74.8 & 206.54 & 14.26 & 166.7 & 0.25 & 9.7E+22 & 1.4E+05 &  Q \\
1243 & DR21-4 & N51 & 20:39:02.41 & 42:25:09.1 & 0.338 & 0.251 & 12.4 & 320.2 & 19.04 & 1465.3 & 50.53\tablenotemark{a} & 2.1E+23 & 3.6E+05 &  B \\
1267 & W75N-2 & N22 N24 & 20:38:05.11 & 42:39:55.6 & 0.339 & 0.288 & 116.2 & 202.54 & 15.33 & 252.7 &  \nodata  & 1.2E+23 & 1.8E+05 &  Q \\
1454 & DR15-2, 6 & S36 S37 & 20:32:21.85 & 40:20:00.7 & 0.508 & 0.414 & 127.8 & 420.81 & 16.17 & 724.4 & 9.21 & 1.1E+23 & 1.2E+05 &  Q \\
1460 & C09-1 &  \nodata  & 20:34:59.19 & 41:34:48.4 & 0.574 & 0.402 & 58.4 & 198.01 & 22.1 & 2216.3 & 25.00\tablenotemark{a} & 4.9E+22 & 4.9E+04 &  B \\
1467 & DR21-2 & N44 & 20:38:59.64 & 42:23:06.9 & 0.341 & 0.233 & 27.6 & 257.92 & 23.4 & 4067.0 & 0.17 & 1.8E+23 & 3.2E+05 &  Q \\
1528 & DR21-1, 6, 20 & N42 N46 N50 & 20:39:00.76 & 42:19:06.4 & 0.535 & 0.336 & 164.2 & 677.32 & 24.81 & 15187.4 &  \nodata  & 2.1E+23 & 2.5E+05 &  C \\
1599 & DR21-7 & N52 N53 & 20:39:03.13 & 42:26:00.0 & 0.232 & 0.186 & 168.5 & 164.34 & 17.19 & 407.9 & 1.27 & 2.2E+23 & 5.1E+05 &  Q \\
2210 & DR15-3 & S38 & 20:32:22.30 & 40:19:19.5 & 0.294 & 0.259 & 114.7 & 70.06 & 15.63 & 98.1 & 0.48 & 5.2E+22 & 9.2E+04 &  Q \\
3188 &  \nodata  & N21 & 20:38:01.59 & 42:39:39.7 & 0.316 & 0.218 & 53.4 & 46.59 & 16.29 & 83.6 & 0.18 & 3.9E+22 & 7.1E+04 &  Q \\
4797 & N68-4 & N69 & 20:40:33.32 & 41:50:46.6 & 0.587 & 0.398 & 44.5 & 378.04 & 17.26 & 960.7 & 2.13 & 9.2E+22 & 9.2E+04 &  Q \\
5417 & DR21-23 & N37 N43 & 20:38:58.30 & 42:24:35.9 & 0.321 & 0.301 & 0.5 & 148.04 & 18.74 & 616.6 &  \nodata  & 8.7E+22 & 1.4E+05 &  B \\
\hline
\enddata
\tablecomments{The coordiantes and physical parameters of the MDC sample. 
 The corresponding MDC names in Cao19 and the dense core names in Motte07 are provided in columns 2 and 3.
 The MDC parameters are taken from Cao21 except for the 24 $\mu$m flux.
 $D_{maj}$ and $D_{min}$ are the major and minor axes of the fitted FWHM of an MDC.
 PA is the position angle of the fitted FWHM.
 Mass and temperature are determined with a graybody thermal dust emission model.
 $L_{FIR}$ is far-infrared luminosity.
 $N_{H_2}$ is column density. 
 $n_{H_2}$ is volume-averaged density. 
 See Cao19 for the derivation of the parameters.
 $F_{\nu,\ 24 \mu m}$ is the \textit{Spitzer} 24 $\mu$m flux. 
 ``Type'' is a classification according to the 24 $\mu$m flux. N, Q, and B represent starless MDC candidates, infrared-quiet MDCs, and infrared-bright MDCs, respectively. C represents MDCs whose 24 $\mu$m fluxes are contaminated by external sources and cannot be obtained. The detailed definitions are given in Section \ref{sec:dis}.
 }
\tablenotetext{a}{Flux scaled from \textit{MSX} 21 $mu$m flux.}
\end{deluxetable*}
\end{longrotatetable}

\section{Observations and Data Reduction} \label{sec:data}
\subsection{VLA X-Band Survey} \label{subsec:3.6}

The VLA X-band radio continuum observations (Project code: 16A--301, PI: Keping Qiu) were made on 29th June 2016 using the VLA. The observations were carried out in the B configuration with all the 27 antennas, which provided a typical angular resolution of $\sim$0$''$.6, corresponding to a linear scale of $\sim$0.004 pc at a distance of 1.4 kpc. The maximum recoverable scale (MRS) was $\sim$17$''$, corresponding to 0.12 pc at a distance of 1.4 kpc. The observations were set in full polarization mode with two 1-GHz basebands centered at 9.5 GHz and 10.5 GHz, respectively. Each baseband was uniformly divided into eight spectral windows, and each spectral window was further divided into 64 2-MHz-wide channels.

The primary calibrator 3C286 was used for bandpass and flux calibration and was observed at the beginning of each schedule block; the secondary calibrator J2007+4029 was used for complex gain calibration and was observed for 80 seconds before and after every seven or eight scans on the targets. There were a total of 22 observing fields covering 29 MDCs. The MDCs not covered were MDC 839, 1112, the eight MDCs in DR21, and the eight MDCs in the OB2 association, whose archival data had similar observational settings. For each observing field, the on-source integration time was 90 seconds, resulting in a sensitivity of $\sim$20 $\mu$Jy beam$^{-1}$.

\subsection{VLA K-Band Survey} \label{subsec:1.3}

The VLA K-band radio continuum and spectral line observations (Project code: 17A--107, PI: Keping Qiu) were made during April and May 2017 using the VLA. Here we only describe the observational settings of the radio continuum observations. The observations were carried out in the D configuration with all the 27 antennas, which provided a typical angular resolution of $\sim$4$''$, corresponding to a linear scale of $\sim$0.02 pc at a distance of 1.4 kpc. The MRS was $\sim$66$''$, corresponding to 0.5 pc at a distance of 1.4 kpc. The observations were set in full polarization mode with two 1-GHz basebands centered at 22.3 GHz and 24.0 GHz, respectively. Each baseband was uniformly divided into eight spectral windows, and each spectral window was further divided into 64 2-MHz-wide channels.

The same set of calibrators as in the K-band survey was adopted. A total of 33 observing fields were observed, in which two were mosaic fields each composed of two single-pointing fields. The entire sample except the MDCs in the DR21 and W75N regions are covered. Each observing field had an on-source integration time of 20 minutes, resulting in a continuum sensitivity of $\sim$15 $\mu$Jy beam$^{-1}$.

\subsection{Archival Data}\label{subsec:archive}

For each target, we have done an extensive search for any existing VLA observations in the archive and from the literature. The data accumulated for decades greatly replenished this work. We selected the archival data according to the following criteria:
\begin{enumerate}
\item The observations should be carried out after 1988 for a stable performance of the VLA.
\item The target MDCs should not be on the edge of the field of views (FOVs).
\item The resolutions should be higher than 14$''$, which corresponds to 0.1 pc at a distance of 1.4 kpc, the typical spatial scale of an MDC.
\item The observations should have enough bandwidth and on-source integration time for good $uv$-coverage and continuum sensitivities.
\end{enumerate}

For the targets observed by more than one project with similar observational settings, we select the ones with higher sensitivities and different resolutions.

The details of the observations are listed in Table \ref{tab:obsInfo}. The information of the data from the literature is obtained from the corresponding papers.

\startlongtable
{\catcode`\&=11
\gdef\Hunter1994{\citet{1994A&A...284..215H}}
\gdef\Molinari1998{\citet{1998A&A...336..339M}}
\gdef\Urquhart2009{\citet{2009A&A...501..539U}}}
\begin{deluxetable*}{c|ccccccp{4cm}}
\tablecaption{Observational Parameters of the Data Used\label{tab:obsInfo}}
\tablehead{
\colhead{Project Code} & \colhead{Date} & \colhead{Config.} & \colhead{Freq.} & \colhead{Flux Calibrator} & \colhead{Gain Calibrator\tablenotemark{a}} & \colhead{$F_\nu$} & \colhead{Covered MDCs}\\
\colhead{} & \colhead{(yyyy mmm)} & \colhead{} & \colhead{(GHz)} & \colhead{} &\colhead{} & \colhead{(Jy)} & \colhead{}
}
\startdata
16A--301 & 16 Jun & B & 10.00 & 3C286 & J2007+4029 & 1.98 & 220 248 274 310 327 341 351 507 509 540 640 675 684 698 714 723 725 742 753 798 801 839 892 1179 1201 1267 1454 2210 3188 4797 \\
17A--107 & 17 Apr, May & D & 23.17 & 3C286 & J2007+4029 & \nodata & The entire sample excluding the regions of DR21 and W75N\\
\hline
12B--140\_0185\tablenotemark{b} & 2012 Nov & A & 5.74 & 3C286 & J2007+4029 & 2.80 & 310 640 675 1112 \\
13A--315\_1898 & 2013 Aug & C & 43.60 & 3C286 & J2007+4029 & 1.22 & 1467\\
13A--373\_1665 & 2013 May & DnC & 23.23 & 3C48 & J2015+3710 & 4.31 & 1243 5417 \\
13A--373\_4594 & 2013 Apr &DnC & 24.37 & 3C48 & J2015+3710 & 4.52 & 1454 2210 \\
13A--373\_7871 & 2013 May & DnC & 24.70 & 3C48 & J2015+3710 & 4.72 & 1243 5417 \\
13A--373\_9583 & 2013 Mar &D & 24.37 & 3C48 & J2015+3710 & 4.65 & 1225 \\
13B--210\_6088 & 2014 Jan & B & 23.20 & 3C286 & J2015+3710 & 3.77 & 274 509\\
14A--092\_7083 & 2014 Mar & A & 44.00 & 3C286 & J2012+4628 & 0.58 & 310 640 675 1112 \\
14A--092\_9305 & 2014 Mar & A & 44.00 & 3C286 & J2012+4628 &0.65 & 640 675 714 1112 \\
14A--241\_0324 & 2014 Jul &D & 23.09 & 3C48 & J2007+4029 & 1.64 & 220 341 699 1018 1112 1467\\
14A--420\_0232 &2015 May & D & 5.80 & 3C48 & J2052+3635 & 2.95 & 4794\\
14A--420\_1115 &2015 May &B & 5.80 & 3C138 & J2007+4029 & 2.60 & 274 1460\\
14A--420\_6898 &2014 Aug & D & 5.80 & 3C48 & J2052+3635 & 2.92 & 798\\
14A--420\_6944 &2014 Aug & D & 5.80 & 3C286 & J2007+4029 & 2.58 & 892\\
14A--420\_7824 &2014 Aug & D & 5.80 & 3C286 & J2007+4029 & 2.59 & 725 \\
14A--420\_8009 &2014 Aug & D & 5.80 & 3C48 & J2052+3635 & 3.16 & 351\\
14A--420\_9375 & 2015 May & B & 5.80 & 3C286 & J2007+4029 & 2.53 & 699 1018 1112 1243 1467 1528 1599 5417 \\
14A--420\_9676 & 2015 May & B & 5.80 & 3C286 & J2007+4029 & 2.59 & 507 753 \\
14A--481\_3009 &2014 May & A & 21.88 & 3C286 & J2015+3710 & 3.81 & 327 742 \\
14B--173\_7197 & 2014 Dec & C & 30.91 & 3C48 & J2038+5119 & 3.42 & 699 1018 1467 \\
15A--059\_5602 & 2015 Feb & B & 30.91 & 3C48 & J2038+5119 & 3.42 & 699 1018 1467\\
\hline
AB515 & 1989 Apr & B & 1.46 & 3C286 & J2052+3635 & 4.94 & 892\\
AB1073 & 2003 Apr &D & 8.46 & 3C48 & J2015+3710 & 4.04 & 220 351 892\\
AC240 & 1989 Mar &B & 8.44 & 3C286 & J2007+4029 & 3.01 & 742 \\
AC240 & 1989 Mar &B & 14.94 & 3C286 & J2007+4029 & 2.87 & 725 \\
AD219 & 1988 Apr & CD & 4.86 & 3C48 & J2007+4029 & 3.40 & 327 507 742 753\\
AF362 & 1999 Jul & A & 8.46 & 3C48 & J2015+3710 & 2.51 & 1201 \\
AF381 & 2001 Apr &B & 4.86 & 3C286 & J2015+3710& 2.45 & 1112 1528 \\
AF381 & 2001 Apr &B & 14.94\tablenotemark{c} & 3C286 & J2015+3710 & 2.71 & 699 1018 1112 1467\\
AG625 & 2002 Aug & B & 8.44 & 3C286 & J2007+4029 & 2.64 & 742 327 \\
AH398 & 1990 Mar & A & 4.86 & 3C286 & J2023+3153 & 2.05 & 725 \\
AH549 & 1995 Aug &A & 4.86 & Manual\tablenotemark{d} & J2025+3343 & 2.74 & 725 \\
AH726 & 2001 Mar & B & 4.86 & 3C286 & J2015+3710 & 2.56 & 327 742 \\
AH869 & 2005 May& B & 4.86 & 3C286 & J2007+4029 & 2.31 & 1112 1528 \\
AJ239 & 1994 Jul & B & 8.44 & 3C48 & J2007+4029 &3.11 & 507 753 \\
AK355 & 1994 Jul & B & 4.86 & 3C48 & J2007+4029 & 4.39 & 725 \\
AK355 & 1994 Jul & B & 8.49 & 3C48 & J2007+4029 & 2.8 & 725 \\
AK355 & 1994 Jul & B & 14.94 & 3C48 & J2007+4029 & 2.25 & 725 \\
AK450 & 1997 Nov & D & 8.44 & 3C286 & J2015+3710 & 2.72 & 327 742 \\
AK477 & 1998 Dec & C & 8.44 & 3C286 & J2015+3710 & 2.72 & 327 742 \\
AM462 & 1995 Jan & CD &14.94 & 3C48 &J2007+4029 & 2.47 & 302 520 \\
AM432 & 1994 Jan & D& 8.44 & 3C48 & J2007+4029 & 3.12 & 274 \\
AM446 & 1994 Apr & A & 1.43 & 3C48 & J2007+4029 & 3.71 & 725 \\
AM446 & 1994 Apr & A &4.86 & 3C48 & J2007+4029 & 4.32 & 725 \\
AM446 & 1994 Apr & A & 8.44 & 3C48 & J2007+4029 & 2.71 & 725 \\
AM446 & 1994 Apr & A & 14.94 & 3C48 & J2007+4029 & 2.17 & 725 \\
AR436 & 2000 Jul & D& 4.86 & 3C48 & J2015+3710 & 1.96 & 220 \\
AR537 & 2004 Mar & C & 43.34 & 3C286 & J2015+3710 & 2.48 & 1201 \\
AS643 & 1998 Jul & AB & 8.46 & 3C286 & J2007+4029 & 2.15 & 1460\\
AS643 & 1998 Jul & AB & 8.46 & 3C286 & J2322+5057 & 1.69 & 1460\\
AS683 & 2000 May & C & 22.46 & 3C48 & J2025+3343 & 2.52 & 725 \\
AS683 & 2000 May & C & 43.34 &3C48 & J2025+3343 & 2.68 & 725 \\
AS831 & 2005 Apr & B & 8.46 & 3C286 & J2007+4029 & 1.00 & 1112 \\
\hline
Araya2009 & 2004 Nov & A & 8.46 & 3C48 & J2007+404 & 2.54 & 1467 \\
Araya2009 & 2005 May & B & 22.4 & 3C48 & J2007+404 & 1.78 & 1467 \\
CG2010 & 2006 Mar & A & 8.46 & 3C286 & J2007+4029 & 2.30 & 1112 \\
CG2015 & 2014 Mar, Apr & A & 6.0 & 3C286 & J2007+4029 & \nodata & 1112 \\
CG2015 & 2014 Mar, Apr & A & 15.0 & 3C286 & J2007+4029 &\nodata & 1112 \\
CG2015 & 2014 Mar, Apr & A & 22.0 & 3C286 & J2007+4029 &\nodata & 1112 \\
CG2015 & 2014 Mar, Apr & A & 44.0 & 3C286 & J2007+4029 &\nodata & 1112 \\
Fontani2012 & 2007 Mar & D & 22.5 & 3C286 & J2015+3710 & 1.39 & 274 \\
Gibb2007 & 1996 Nov & A & 4.86 & 3C48 & J2025+3343 & 2.80 & 1112 \\
Gibb2007 & 1996 Nov & A & 8.46 & 3C48 & J2025+3343 & 2.60 & 1112 \\
Gibb2007 & 1996 Nov & A & 44.49 & 3C48 & J2025+3343 & 2.56 & 1112 \\
Hunter1994 & 92 Nov & A & 8.44 & 3C48 & J2007+4029 & \nodata & 1112 \\
Kurtz1994 & 1989 Mar & B & 8.41 & 3C286 & J2007+4029 & 3.02 & 742 \\
Kurtz1994 & 1989 Mar & B & 14.96 & 3C286 & J2007+4029 & 3.12 & 742 \\
Masque2017 & 2014 May & A & 22.46 & 3C286 & J2015+3710 & 3.86 & 742 \\
Miralles1994 & 1989 Jul & C & 4.9 & 3C286 & J2007+4029 & 2.36 & 274 \\
Molinari1998 & 1994 Aug & CnB & 4.86 & 3C48 & \nodata & \nodata & 507 753 \\ 
Rosero2016 & 2011 Aug & A & 6.15 & 3C286 & J2007+4029 & \nodata & 274 \\
Shepherd2004 & 2001 Mar & B & 4.89 & 3C286 & J2012+4628 & \nodata & 1112 \\
Shepherd2004 & 2001 Mar & B & 14.96 & 3C286 & J2012+4628 & \nodata & 1112 \\
Shepherd2004 & 2000 Apr & C & 43.34 & 3C286 & J2012+4628 & \nodata & 1112 \\
Shepherd2004 & 2001 Mar & B & 43.34 & 3C286 & J2012+4628 & \nodata & 1112 \\
Torrelles1997 & 1996 Dec & A & 22.28 & 3C48 & J2023+3153 & 2.5 & 1112 \\
Urquhart2009 & 2009 Jul & B & 4.86 & 3C286 & \nodata & \nodata & 725 \\
\hline
\enddata
\tablecomments{Basic observational parameters of the data. Column 1 gives the project code or the reference of the data; column 2 gives the date of the observation; column 3 gives the configuration of the VLA antennas; column 4 gives the central frequency of the observation; column 5 and 6 give the flux calibrators and gain calibrators, relatively; column 7 gives the bootstrapped flux density of the gain calibrator, which may not be provided in the reference. Column 8 gives the MDCs covered by the observation. Projects 16A--301 and 17A--107 are the PI observations (PI: Keping Qiu).}
\tablenotetext{a}{ For easy comparison, here we list the J2000 names of the gain calibrators, which may be different from the source names originally adopted in the observational settings. See \url{https://science.nrao.edu/facilities/vla/observing/callist} for reference.}
\tablenotetext{b}{ The code after the underscore is the last four digits of the archive file ID. It helps to fast locate the specific schedule block of a project when searching the NRAO data archive (\url{https://archive.nrao.edu/archive/advquery.jsp}).}
\tablenotetext{c}{ Only the upper--band data were used for imaging.}
\tablenotetext{d}{ The \texttt{setjy} task in CASA failed by using the flux calibrators. We thus specified the flux densities and spectral indexes of the gain calibrator manually. The input values are from observations with similar settings and a close date.}
\tablereferences{
 (1) \citet{2009ApJ...698.1321A}
 (2) \citet{2010AJ....139.2433C}; 
 (3) \citet{2015Sci...348..114C}; 
 (4) \citet{2012MNRAS.423.1691F};
 (5) \citet{2007MNRAS.380..246G}; 
 (6) \Hunter1994;
 (7) \citet{1994ApJS...91..659K};
 (8) \citet{2017ApJ...836...96M};
 (9) \citet{1994ApJS...92..173M};
 (10) \Molinari1998
 (11) \citet{2016ApJS..227...25R}; 
 (12) \citet{2004ApJ...601..952S};
 (13) \citet{2002ApJ...566..931S};
 (14) \citet{1997ApJ...489..744T}; 
 (15) \Urquhart2009;
}
\end{deluxetable*}

\subsection{Data Reduction}\label{subsec:reduc}

Both the historical VLA data and the JVLA data were calibrated and imaged by the Common Astronomy Software Applications (CASA)\footnote{\url{http://casa.nrao.edu}}\citep{2007ASPC..376..127M} v4.6.0 with the standard procedures. Edge channels, radio-frequency interference, and problematic data were flagged before and during data reduction. The Perley-Butler 2013 model \citep{2013ApJS..204...19P} was applied for flux calibration. Imaging was made by the task {\tt CLEAN} using the Briggs weighting with {\tt ROBUST=0.5} for a compromise between sensitivity and angular resolution. We made joint imaging for the visibilities with the same pointing centers, same frequencies, and comparable bandwidths, and performed mosaic imaging for the visibilities with close (but different) pointings and similar sensitivities. The properties of the radio continuum maps are listed in Table \ref{tab:mapInfo}. Data of the X-band survey had issues that severely affected the final product. Only the maps of MDC 274 and MDC 507 in the X-band survey were adopted.

\include{Table3_MapInfo}

\subsection{Source Identification}\label{ssec:rdetect}

Radio sources are identified in each map by searching for structures: (1) emission peak is not lower than five times the image RMS noise ($\sigma$); (2) area of the 3$\sigma$ contour is not smaller than the synthesized beam. Parameters such as position, peak flux density, and total flux density are further determined by two-dimensional Gaussian-fitting and direct measurement.

Gaussian fitting is performed using the task \texttt{IMFIT} of \textsc{MIRIAD}, which can fit one or multiple Gaussian components simultaneously on an image. This method is applied to the sources with an elliptical morphology. The fitting results include the coordinates, convolved and deconvolved sizes, peak flux densities, and total flux densities of the fitted Gaussian components, along with the corresponding fitting errors.

Direct measurement is adopted when the sources are resolved into cometary or irregular structures. This measurement is performed interactively using the task {\tt viewer} of \textsc{CASA}, with which we carefully measure the peaks and total flux densities within certain contour levels, e.g. 3$\sigma$ or 5$\sigma$. Such a direct measurement gives the coordinations of the geometric centers, peaks, and total flux densities without errors. For elliptical sources, we compare the total flux densities derived from the direct measurement with those returned from a Gaussian fitting, and the results have a typical difference of less than 10\%. For very faint sources, the difference reaches 30\% or more. This is due to a cutoff at a certain contour level in the direct measurement.

For a source covered by multiple observations, the obtained coordinates are always slightly different across the maps. In this case, we adopt the coordinates determined from the maps of the highest resolutions.

\section{Results}\label{sec:rst}

\subsection{Radio Flux density}\label{ssec:flux}

\begin{figure}[ht!]
\plotone{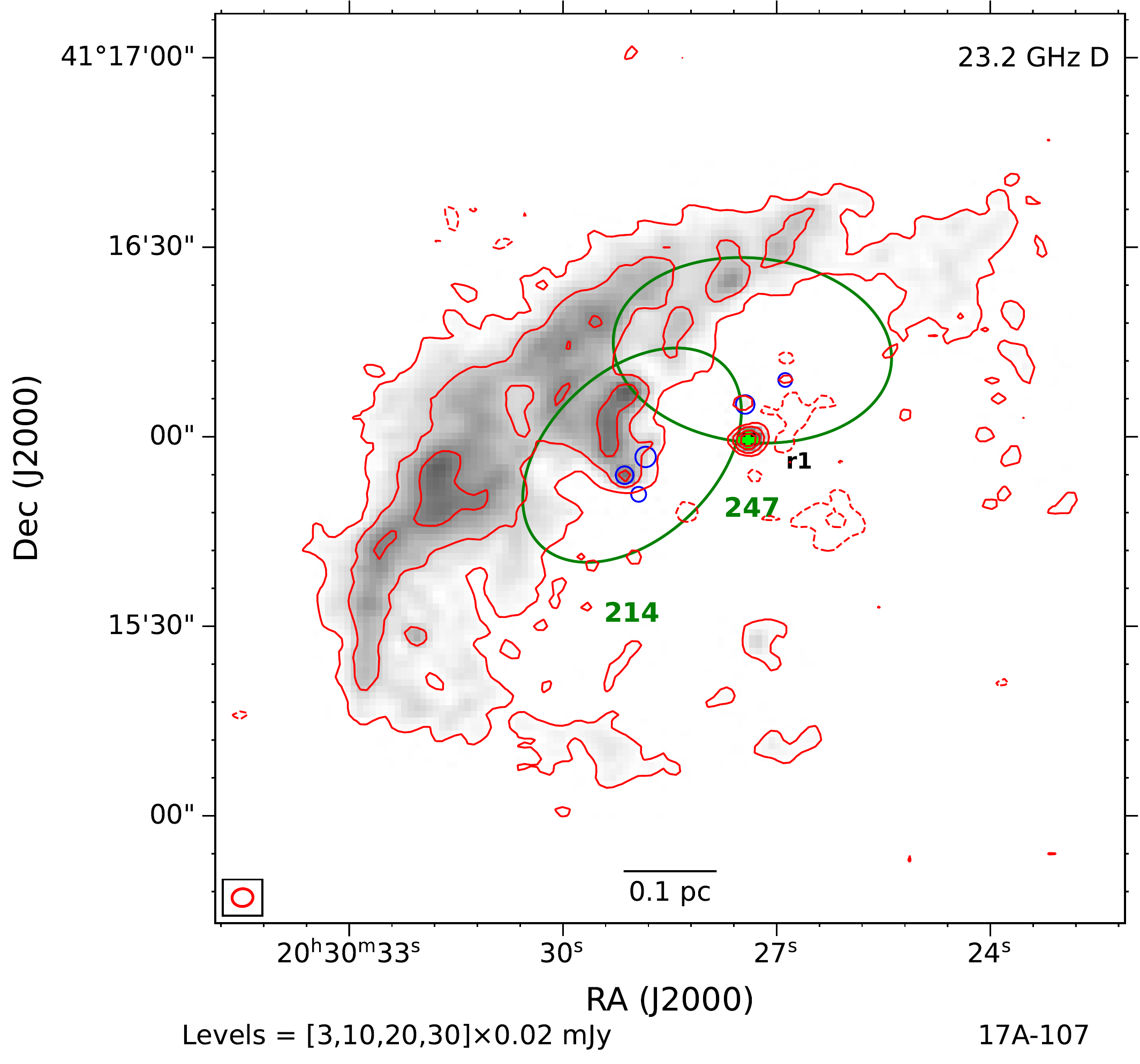}
\caption{Radio continuum map of MDC 214 and 247. The K-band emission is shown in both grayscale and red contours. The grayscale starts at a  3$\sigma$ level. The green ellipses represent the FWHMs of the MDCs in Cao21; the blue solid ellipses represent the dust condensations identified from the SMA 1.3 mm continuum maps. The observing frequency and the VLA configuration are labeled in the upper-right corner. The synthesized beam is shown in the bottom-left corner. The contour levels and the VLA project codes are provided below the map. MDC names are labeled by bold green indices. Radio sources associated with the MDC are marked by light green crosses and indexed by the labels nearby. The complete figure set (96 images) is available in the online journal. \label{fig:radio}}
\end{figure}

We have detected a total of 64 radio sources at $\sim$0.01-pc scale, i.e., the scale of dust condensations, and 17 relatively large-scale (0.1--1 pc) extended sources. Since our primary goal is to study the radio emission at the condensation scale, a detailed analysis of the extended emissions is beyond the scope of this work, and they will not be discussed. The detailed properties of all the detected radio sources are listed in Table \ref{tab:detection}, including the coordinates, observing frequencies, sizes, peak and total flux densities. The radio continuum maps of each MDC are presented in Figure \ref{fig:radio}.

\include{Table4_Detections}

\begin{figure}[ht!]
\epsscale{0.7}\plotone{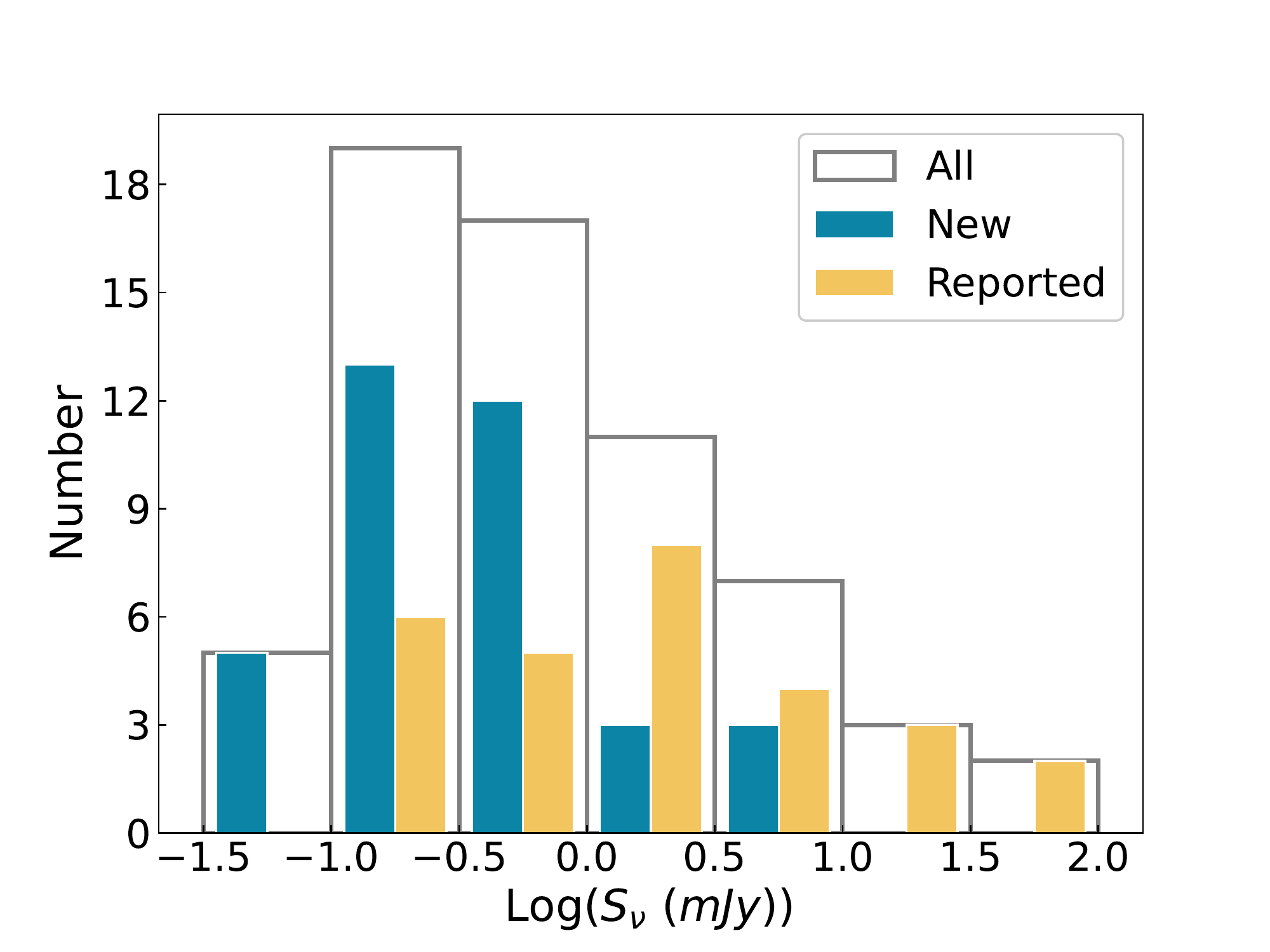}
\caption{Distribution of the K-band flux densities of the radio sources. If a radio source has no K-band detection, the flux density at the nearest frequency is adopted. The histogram in white is the distribution of all the radio sources. The histograms in blue and yellow are the distributions of the newly detected sources and the previously reported ones.\label{dist_NO}}
\end{figure}

Most of the detected radio sources are faint and compact. Their K-band flux density distribution is given in Figure \ref{dist_NO}. For the radio sources with no K-band data, the flux densities at the nearest frequencies are adopted. Most of the radio sources have flux densities ranging from 0.1 mJy to 3 mJy. The distribution peaks at $\sim$0.2 mJy. A few bright sources, e.g. 507-r3 (Figure \ref{fig:radio}.12), 725-r1 (Figure \ref{fig:radio}.22 ), 742-r2 (Figure \ref{fig:radio}.7), 1112-r3 (Figure \ref{fig:radio}.28), and 1467-r1 (Figure \ref{fig:radio}.27), are resolved, and all have flux densities higher than 10 mJy.

By carefully checking the literature, we confirm that 37 radio sources are reported for the first time (see the last column of Table \ref{tab:detection}). The new sources are generally more compact, being point-like or elliptical (Table \ref{tab:new}). They are fainter, with K-band flux densities distributing mainly between 0.1--1.0 mJy and peaking at $\sim$0.2 mJy (Figure \ref{dist_NO}). 

\begin{table}[h!]
\renewcommand{\thetable}{\arabic{table}}
\centering
\caption{Newly detected and the reported sources\label{tab:new}}
\begin{tabular}{c|ccc|c}
\tablewidth{0pt}
\hline
\hline
 & Total & New & Reported & Ratio (N/R) \\
\hline
Compact & 53 & 35 & 18 & 1.9 \\
Extended & 11 & 1 & 10 & 0.1 \\
\hline
Ratio (C/E) & 4.8 & 35 & 1.8 \\
\hline
\hline
\end{tabular}
\tablecomments{`N'' for newly detected sources; ``R'' for reported sources; ``C'' for compact or elliptical sources; ``E'' for extended sources.}
\end{table}

\subsection{Association with MDCs}\label{ssec:rate}

A radio source is considered to be associated with an MDC when it is located within or close to the FWHM boundary of the MDC, and is considered as being associated with a dust condensation when it is abutting or overlapping with the condensation. But in quantitative analyses of the radio luminosity of the MDCs, we only include the radio sources within and right on the FWHM boundaries of the MDCs. The detected 64 radio sources are associated with 34 MDCs, of which 14 MDCs are associated with more than one radio sources. Thirty MDCs have radio sources within the FWHM boundaries, in which 12 have multiple sources. Forty-four (69\%) radio sources are associated with dust condensations and 54 (84\%) are located in the MDCs. The number of radio sources associated with each MDC is given in Table \ref{tab:MDC}, in which we classify the radio sources as being in an MDC and associated with the dust condensations (``$N_A$''), being in the MDCs but associated with no dust condensation (``$N_{NA}$''), and located close but outside of the FWHM boundary of an MDC (``$N_O$''). In this table, we also noted the star-forming indicators such as outflows and masers obtained from the literature.

\startlongtable
{\catcode`\&=11
\gdef\Beuther2002{\citet{2002A&A...383..892B}}
\gdef\Edris2007{\citet{2007A&A...465..865E}}
\gdef\Motte2007{\citet{2007A&A...476.1243M}}
\gdef\Fontani2010{\citet{2010A&A...517A..56F}}
\gdef\Stutzki1982{\citet{1982A&A...111..201S}}
\gdef\Harju1998{\citet{1998A&AS..132..211H}}
\gdef\Larionov1999{\citet{1999A&AS..139..257L}}
\gdef\Palau2007{\citet{2007A&A...465..219P}}
\gdef\Gottschalk2012{\citet{2012A&A...541A..79G}}
\gdef\Duarte2013{\citet{2013A&A...558A.125D}}
\gdef\DuarteC2014{\citet{2014A&A...570A...1D}}
}
\begin{deluxetable*}{c|ccc|c|cccc|l}
\tablecaption{Radio and Infrared Properties of the MDCs\label{tab:MDC}}
\tablehead{
\colhead{MDC} & \colhead{$N_{A}$} & \colhead{$N_{NA}$} & \colhead{$N_O$} & \colhead{Type} & \colhead{Notes}}
\startdata
214 & 0 & 0 & 0 & Q & H$_2$O/CH$_3$OH maser$^{4,10,13,18,26,28}$, outflow$^{10,19}$\\
220 & 0 & 0 & 0 & Q & Outflow$^{21,23,24}$ \\
247 & 0 & 1 & 0 & B & H$_2$O/CH$_3$OH maser$^{3,20}$, outflow$^{27}$ \\
248 & 1 & 0 & 0 & Q & H$_2$O maser$^{26}$, outflow$^{21,23,24}$ \\
274 & 1 & 1 & 0 & B & H$_2$O maser$^4$, outflow$^{6,7,21}$ \\
302 & 0 & 0 & 0 & C & \\
310 & 2 & 0 & 0 & B & H$_2$O maser$^{4,16,18,20}$, outflow$^{12,19,21}$ \\
327 & 0 & 1 & 0 & Q & \\
340 & 0 & 0 & 1 & Q & \\
341 & 1 & 0 & 0 & Q & Outflow$^{21,23,24}$ \\
351 & 1 & 0 & 0 & Q & \\
370 & 0 & 0 & 0 & C & Outflow$^{25}$ \\
507 & 2 & 1 & 0 & B & H$_2$O/CH$_3$OH/OH maser$^{4,10,13,15,16,18,20}$, outflow$^{10,19}$ \\
509 & 4 & 1 & 2 & Q & H$_2$O/CH$_3$OH maser$^{8,20,26,29}$, outflow$^{7,8,14,19,29}$\\
520 & 0 & 0 & 0 & C & H$_2$O maser$^4$ \\
540 & 0 & 0 & 3 & Q & \\
608 & 0 & 0 & 0 & Q & \\
640 & 0 & 0 & 0 & Q & \\
675 & 1 & 0 & 0 & B & H$_2$O maser$^{17,21}$, outflow$^{25}$ \\
684 & 1 & 0 & 0 & Q & Outflow$^{21}$ \\
698 & 1 & 0 & 0 & Q & H$_2$O maser$^{20}$, outflow$^{21,25}$ \\
699 & 0 & 1 & 0 & B & H$_2$O/CH$_3$OH maser$^{2,5,16,20,23,24}$, outflow$^{27}$ \\
714 & 3 & 0 & 0 & B & H$_2$O maser$^{20,26}$, outflow$^{21,27}$ \\
723 & 1 & 1 & 0 & Q & \\
725 & 2 & 0 & 0 & B & H$_2$O/CH$_3$OH maser$^{13,16,18}$, outflow$^{21}$ \\
742 & 2 & 1 & 0 & B & Outflow$^{25}$ \\
753 & 0 & 0 & 1 & Q & H$_2$O/CH$_3$OH/OH maser$^{4,28}$, outflow$^{30}$ \\
798 & 1 & 0 & 0 & Q & \\
801 & 1 & 0 & 0 & Q & Outflow$^{21}$ \\
839 & 1 & 0 & 0 & Q & \\
892 & 0 & 0 & 0 & C & \\
1018 & 0 & 0 & 0 & Q & H$_2$O maser$^{13}$, outflow$^{23,24}$ \\
1112 & 3 & 3 & 0 & B & H$_2$O/CH$_3$OH/OH maser$^{4,16,20,26}$, outflow$^{5,21,25,27}$ \\
1179 & 0 & 0 & 0 & N & \\
1201 & 0 & 1 & 0 & C & H$_2$O maser$^{1,4,13,16,20,26}$, outflow$^1$ \\
1225 & 2 & 0 & 0 & Q & \\
1243 & 1 & 0 & 0 & B & H$_2$O/CH$_3$OH maser$^{20,26}$, outflow$^{25,27}$ \\
1267 & 1 & 0 & 1 & Q & \\
1454 & 3 & 2 & 0 & Q & H$_2$O maser$^{26}$ \\
1460 & 0 & 0 & 0 & B & H$_2$O maser$^4$ \\
1467 & 0 & 0 & 2 & Q & H$_2$O/CH$_3$OH/OH maser$^{2,4,17,26}$, outflow$^{22,25}$ \\
1528 & 0 & 0 & 0 & C & H$_2$O/CH$_3$OH maser$^{2,4,5,9,16}$, outflow$^{5,21}$ \\
1599 & 0 & 0 & 0 & Q & H$_2$O/CH$_3$OH maser$^{4,16}$, outflow$^{23,24}$ \\
2210 & 1 & 0 & 0 & Q & \\
3188 & 0 & 1 & 0 & Q & Outflow$^{21}$ \\
4797 & 0 & 1 & 0 & Q & Outflow$^{21}$ \\
5417 & 1 & 0 & 0 & B & H$_2$O/CH$_3$OH/OH maser$^{11,16,17}$ \\
\hline
\enddata
\tablecomments{A summary of the radio detections, infrared classifications, and star-formation indicators of the MDCs.
 $N_{A}$ is the number of radio sources in an MDC and associated with dust condensations.
 $N_{NA}$ is the number of radio sources in an MDC that are associated with no dust condensation;
 $N_{O}$ is the number of radio sources located close but outside of the FWHM boundary of an MDC.
 Column ``Type'' is the classification based on the 24 $\mu$m fluxes and is the same as that in Table \ref{tab:MDC_Cao21}.
 In the ``Notes'' column, we summarized the star--formation indicators obtained from the literature. }
\tablerefs{(1) \Stutzki1982; (2) \citet{1990ApJ...364..555P}; (3) \citet{1994ApJS...92..173M}; (4) \Harju1998;(5) \Larionov1999; (6) \citet{2002ApJ...576..313K}; (7) \Beuther2002; (8) \citet{2004ApJ...608..330B}; (9) \citet{2005AJ....130..711K};(10) \citet{2005ApJ...625..864Z}; (11) \citet{2005MNRAS.356..637H}; (12) \citet{2006ApJ...643..978K}; (13) \citet{2007PASJ...59.1185S}; (14) \Palau2007;(15) \Edris2007; (16) \Motte2007; (17) \citet{2008MNRAS.384..719H}; (18) \Fontani2010; (19) \citet{2010MNRAS.404..661V};(20) \citet{2011MNRAS.418.1689U}; (21) \Gottschalk2012; (22) \citet{2012ApJ...744...86Z}; (23) \Duarte2013; (24) \DuarteC2014;(25) \citet{2015MNRAS.450.4364N}; (26) \citet{2015MNRAS.453.4203X}; (27) \citet{2015MNRAS.453..645M}; (28) \citet{2016ApJS..222...18G}; (29) \citet{2017ApJS..233....4R};(30) Yang et al., in preparation.}
\end{deluxetable*}

\subsection{Spectral Indices} \label{ssec: idx}

Radio continuum emission produced by star-forming activities can be either thermal free-free emission or non-thermal synchrotron radiation. In the centimeter wavelengths, spectral energy distributions (SEDs) of the two mechanisms both can be modeled by a power-law relation of $S _ { \nu } \propto \nu ^ { \alpha }$ \citep{1975MNRAS.170...41W}. The spectral indices of thermal free-free emission arising from UC \ion{H}{2} regions can vary from $-0.1$ to 2 depending on the optical depth \citep{1993RMxAA..25...23R, 2004ApJ...612L..69S}. The measured spectral indices of thermal ionized jets/winds range from 0.1 to $\sim$1 \citep{2018A&ARv..26....3A}. Non-thermal radiation is characterized by a negative spectral index ($\alpha < -0.1$) \citep{2005ApJ...626..953R, 2010Sci...330.1209C}.

SEDs derived from interferometric data can be affected by the spatial filtering effect, especially for the resolved sources. We use the size (in arc sec$^2$) of the source in a map as the direct indicator of the probed spatial scale. The SEDs are fitted only when there are data covering at least two VLA observing bands and the sizes of the sources vary within a factor of two. The error of flux densities is a quadrature summation of the calibration error and the measurement error \citep{2001AJ....121.1556B}: the calibration error is assumed to be 5\% at frequencies lower than 15 GHz, 10\% between 15 GHz and 35 GHz, and 15\% at higher frequencies; the measurement error is estimated as the image RMS times the source sizes in units of beam size.

Among the 64 radio sources, we are able to characterize the SEDs of 8 sources. Two of the SEDs are flat ($-0.1 \leqslant \alpha \leqslant 0.2$); four are positive ($\alpha > 0.2$); two are negative ($\alpha < -0.1$). The flat and positive spectral indices indicate thermal emission from \ion{H}{2} regions, or ionized jets and winds  \citep{1986ApJ...304..713R, 2018A&ARv..26....3A}, whereas the negative spectral indices suggest the presence of non-thermal emission presumably arising from a synchrotron jet (e.g., \citealt{2017A&A...597A..43G}).

During SED fitting, we notice that for the same source, the spectral indices obtained by different works can sometimes be inconsistent, e.g. the radio sources associated with MDC 1112 (see the Appendix for a detailed discussion). Some of the differences cannot be simply explained by uncertainties but are caused by various reasons. A most common issue in interferometric observations is the spatial filtering effect, which can lead to, e.g., $\sim$60\% flux density missing \citep{2004ApJ...601..952S} and can severely affect the fitting results \citep{1999ApJ...514..232K}. Thus it is essential to use data sensitive to the same spatial scales. An additional point worth consideration is time variability. Previous observations and simulations have revealed that some of the UC \ion{H}{2} regions can show time variation in their flux densities and morphologies within a typical period of ten years (\citealt{2004ApJ...604L.105F, 2005A&A...431..993V, 2008ApJ...674L..33G, 2010RMxAA..46..253N, 2011MNRAS.416.1033G, 2012ApJ...758..137K, 2018A&ARv..26....3A} and the references therein). The time variability of radio jets has also been confirmed (e.g. \citet{2018A&ARv..26....3A, 2010AJ....139.2433C}). The VLA data we adopted have a long time span of thirty years, within which the VLA had a major upgrade from the HVLA to the JVLA. The time span is even longer considering the data taken from the literature. Besides, for the data obtained from the literature, the measurement and uncertainty estimations are carried out by different methods. We thus carefully check the probed spatial scales and refrain from using data with long time spans for SED fitting considering possible time variation. 

\section{Discussion} \label{sec:dis}
\subsection{Overall Performance of the Dataset} \label{ssec:data}
\begin{figure*}[ht!]
\epsscale{0.8}\plotone{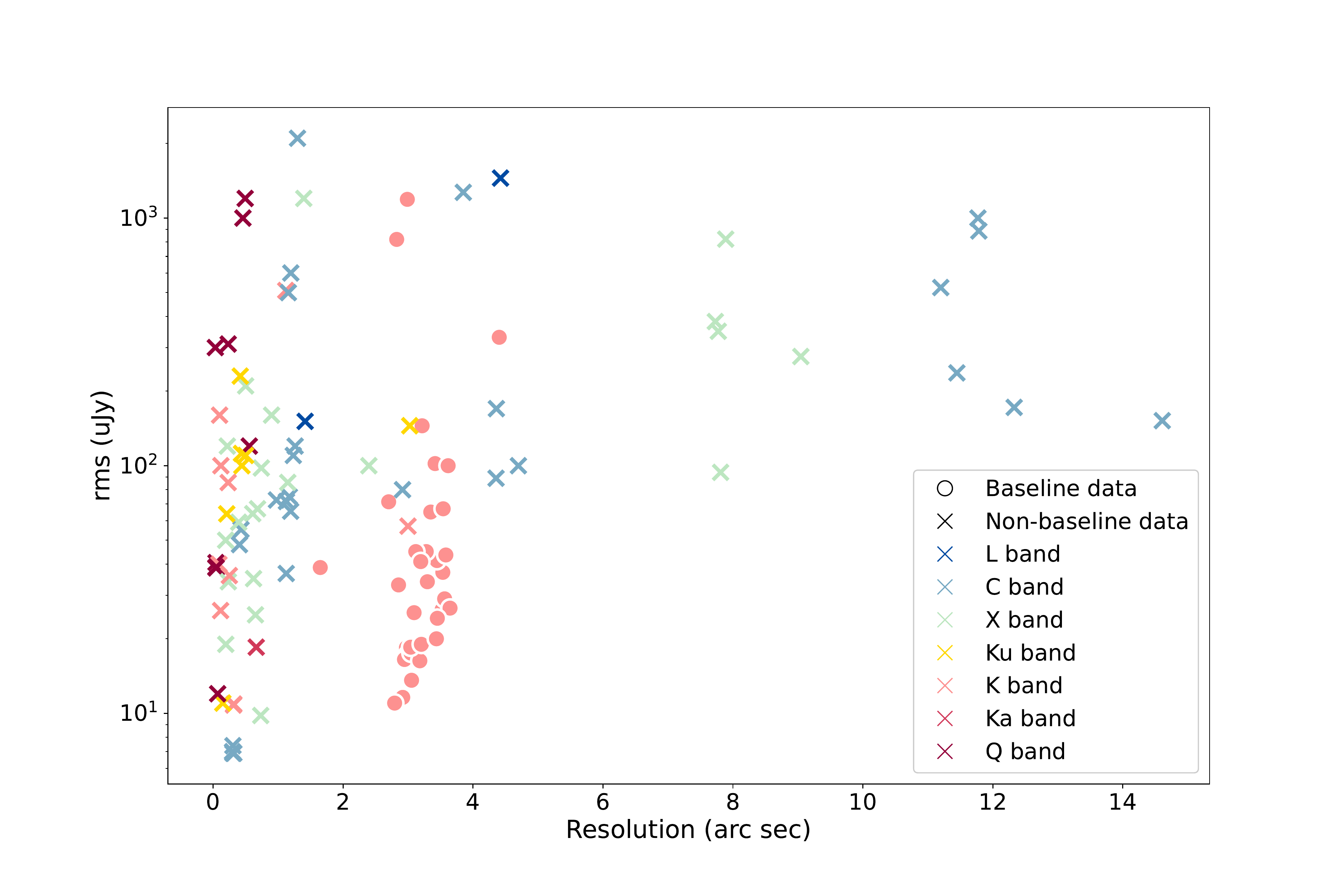}
\caption{Comparison of the RMS and the resolution between the baseline data and the other archival data. The dots show the baseline data, and the crosses indicate the archival data. The observing bands are represented by different colors. \label{cmp}}
\end{figure*}

\begin{figure}[ht!]
\epsscale{0.8}\plotone{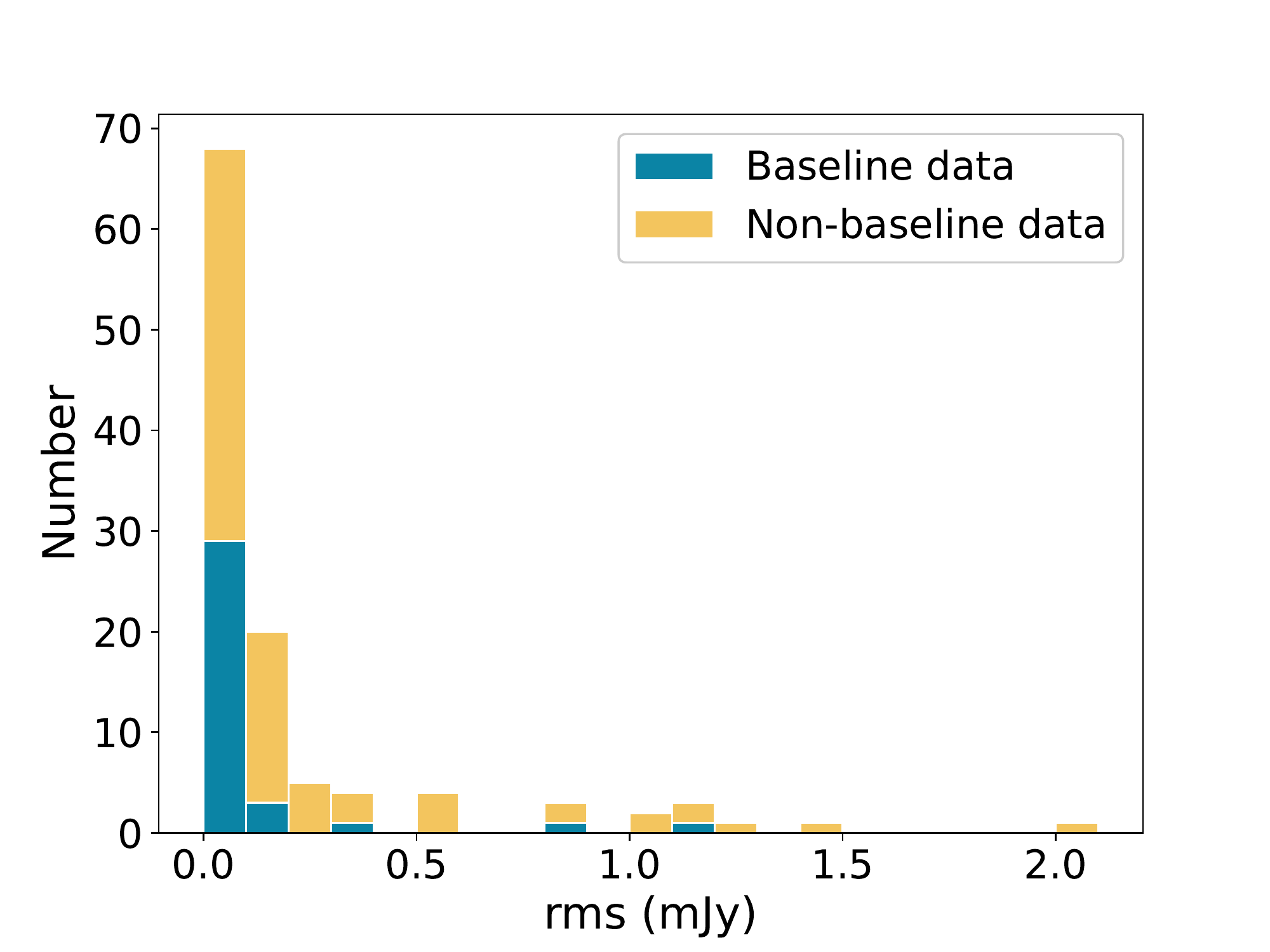}
\caption{Stacked distribution of the RMS noises. The vertical axis is the number of radio continuum maps. The histograms in blue and yellow are of the baseline data and the archival data, respectively. \label{dist_rms_cmp}}
\end{figure}

\begin{figure}[ht!]
\epsscale{0.8}\plotone{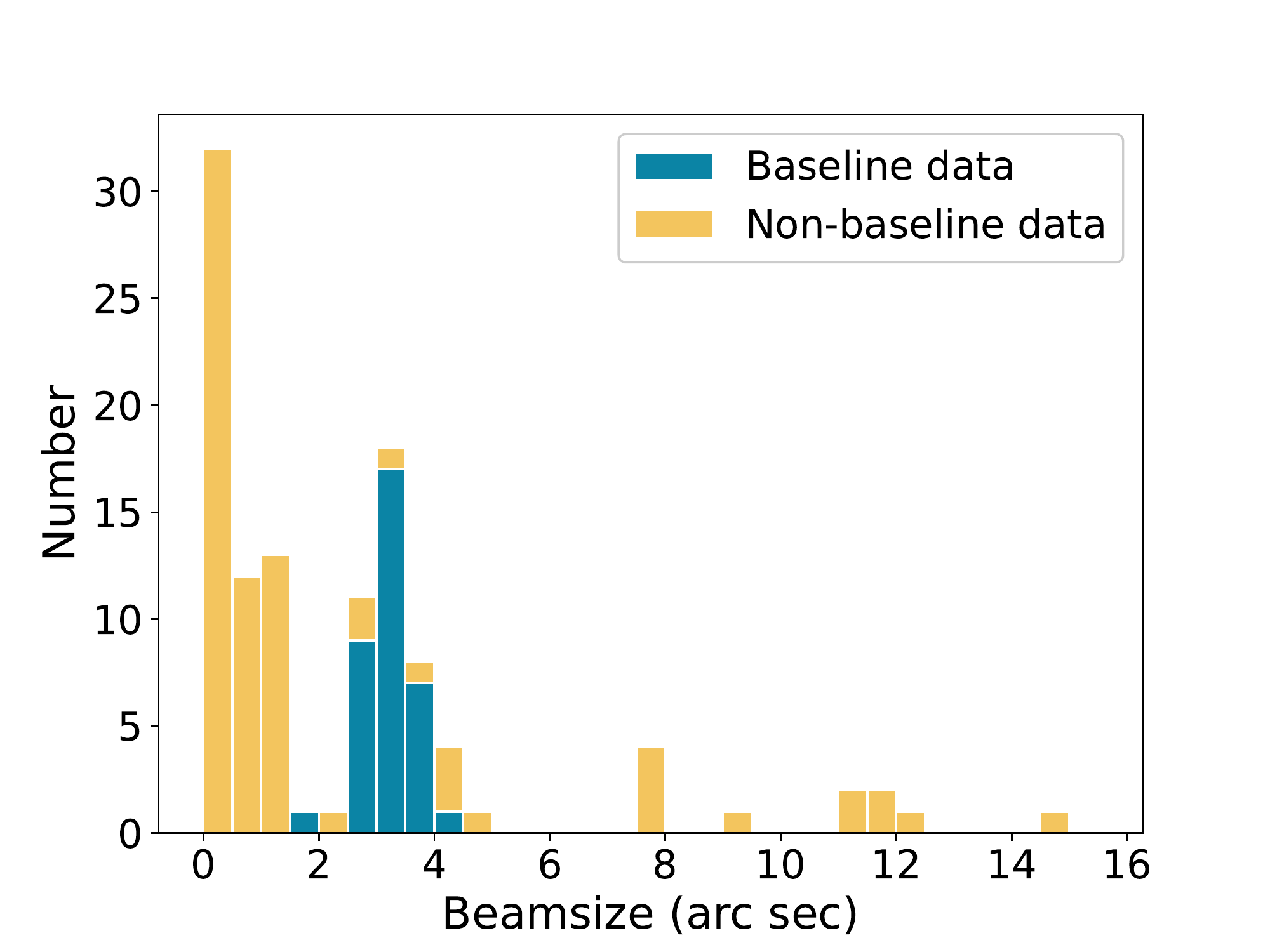}
\caption{Stacked distribution of the beam sizes. The vertical axis is the number of radio continuum maps. The colors are the same as in Figure \ref{dist_rms_cmp}.\label{dist_beam_cmp}}
\end{figure}

We detect a large number of faint radio sources, of which more than half are new detections. We quantitatively characterize the detection capability of the entire work by looking into the observations that cover the majority of the MDCs. Projects 17A--107, 14A--241, and 13A--373 are all 23.2 GHz observations with similar sensitivities and resolutions (Figure \ref{cmp}). Together they cover the entire sample. We take them as the ``baseline data'' to assess the overall detection capability. The other data may have finer resolutions and even higher sensitivities but only cover a small number of MDCs and may have quite different observational settings. Thus those data do not significantly affect the detection rate of the whole sample. The baseline data have concentrated RMS noises of several tens of micro Janskies (also see Figure \ref{dist_rms_cmp}). A few maps have RMS noises higher than 100 $\mu$Jy because of containing bright extended sources and strong sidelobes. After excluding these maps, the average RMS noise of the baseline data is 30 $\mu$Jy/beam, corresponding to a 5$\sigma$ sensitivity of 0.15 mJy/beam. They have similar angular resolutions of about 3$''$, corresponding to 0.02 pc at a distance of 1.4 kpc (also see Figure \ref{dist_beam_cmp}). Although this is not the highest resolution the dataset has achieved, it is sufficient for detecting source multiplicity within a 0.1-pc scale and resolving positional offsets between the radio sources and the dust condensations.

\begin{figure}[ht!]
\epsscale{0.8}\plotone{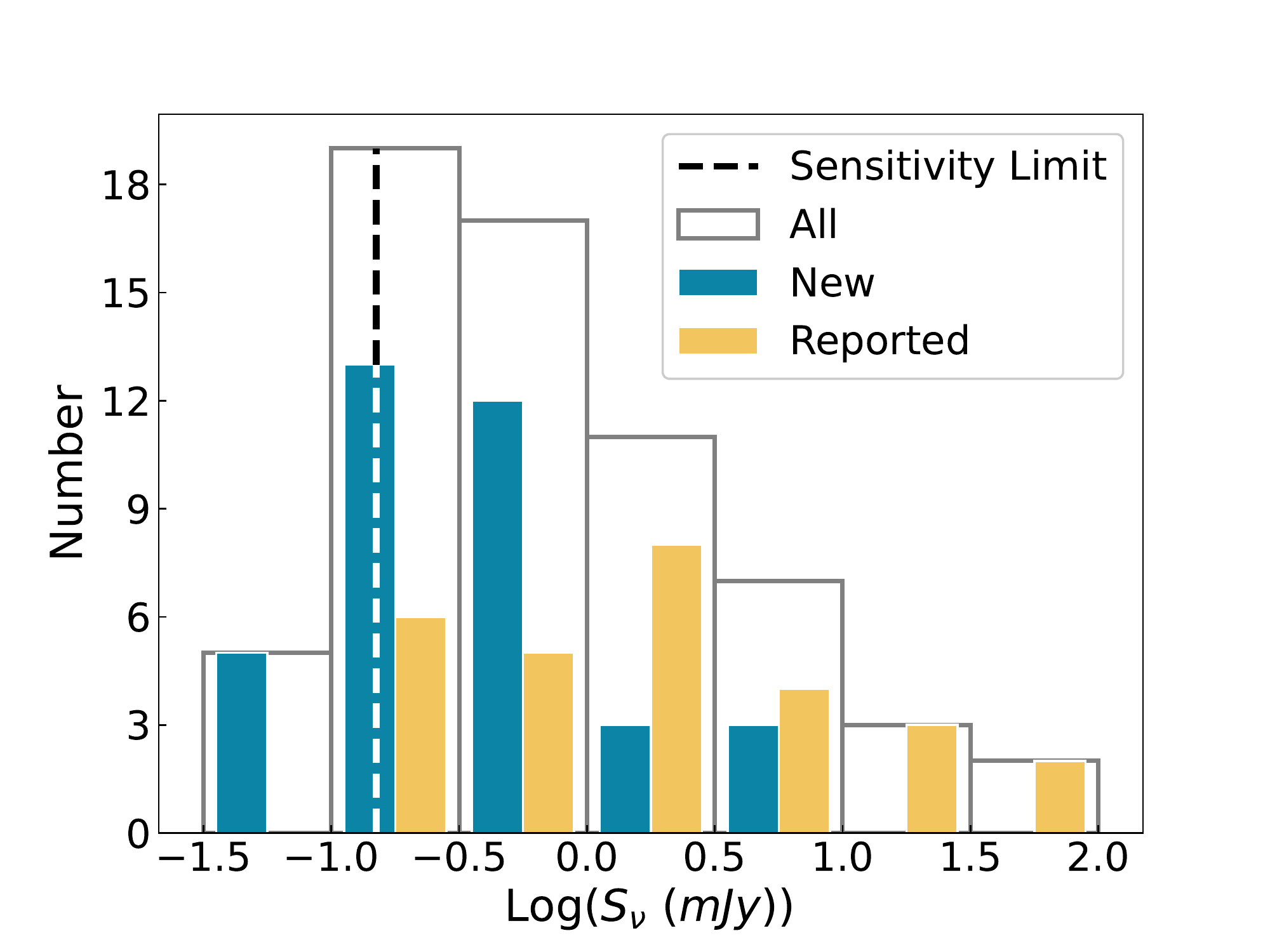}
\caption{Colors are the same as Figure \ref{dist_NO}. The vertical axis is the number of radio detections. The dashed line represents the overall sensitivity limit.\label{dist_NO_withLimit}}
\end{figure}

Among the 64 radio sources, 57 are detected by the baseline data, in which 34 are new detections. For the other seven sources, two are so weak that are only detected with the data of higher sensitivities; five are contaminated by the nearby extended sources and are only distinguishable in the maps that have filtered out the extended emission. Moreover, the flux density distribution peaks near the overall sensitivity limit determined by the baseline data (Figure \ref{dist_NO_withLimit}). We conclude that our work is capable of detecting radio continuum sources with 23.2 GHz peak flux densities no less than 0.15 mJy/beam and resolving multiple sources with separations larger than 0.02 pc.

\subsection{Nature of the Radio Sources} \label{ssec:nature}
Radio continuum emission can be produced by several processes during HMSF. Thermal free-free emission is observed in \ion{H}{2} regions, ionized jets, disc-winds, and the radio knots produced by the interaction between jets and the surrounding material \citep{1986ApJ...304..713R, 2005IAUS..227..120R}. Non-thermal synchrotron components are found in the jet knots produced by jets with extremely high energy and speed \citep{2014ApJ...792L..18A, 2018MNRAS.474.3808V}. In the low-mass regime, Class I objects can produce the non-thermal gyrosynchrotron radiation \citep{2013ApJ...775...63D}. Objects such as planet nebulae (PNe) and extra-galactic objects also produce radio continuum emissions. Since the gyrosynchrotron radiation emitted by Class I objects is too weak to be detected by our data \citep{2013ApJ...775...63D}, we only inspect the possible contamination from extra-galactic objects and PNe before further discussion.

The number of possible extra-galactic objects is estimated using the models from \citet{2005A&A...431..893D}. The number density of extra-galactic objects at 23.2 GHz is interpolated using the models at 20 GHz and 30 GHz and is obtained to be 1.3$\times$10$^6$ sr$^{-1}$. Our search area is 1.5 times the total area enclosed by the FWHMs of all the surveyed MDCs, and is about $2.4 \times 10^{-6}$ sr. Thus the total number of possible extra-galactic sources is about three, which is negligible for statistical analysis. 

We check the HASH PN database\footnote{http://202.189.117.101:8999/gpne/} \citep{2016JPhCS.728c2008P} for PNe with radio fluxes above our detection threshold. No record is found in our FOV. Moreover, PNe are produced during the late stages of massive star evolution. Since their surrounding dust and gas cocoons should have already dispersed, we are not likely to detect them in the dusty MDCs.

In conclusion, the contamination from extra-galactic objects and PNe is negligible. The detected radio sources in this work should be small \ion{H}{2} regions or ionized jets/winds, and a more detailed discussion in this context is presented below.

\subsubsection{\ion{H}{2} Regions}\label{sssec:hii}
A straightforward way to identify \ion{H}{2} region candidates is by morphology. In high-resolution observations, developed UC \ion{H}{2} regions are found to have several typical morphologies, including cometary, core-halo, shell-like, and irregular \citep{1989ApJS...69..831W}. We detect two cometary (699-r1, 742-r2) and three core-halo (725-r1, 1467-r1, 2210-r1) sources. Moreover, the complex radio continuum emission in MDC 1528 may be composed of at least two cometary UC \ion{H}{2} regions or a result of the interaction between the stellar wind and the surrounding material. High radio flux densities can also suggest the existence of UC \ion{H}{2} regions. In our detected radio sources, 507-r3 (Figure \ref{fig:radio}.12), 725-r1 (Figure \ref{fig:radio}.22), 742-r2 (Figure \ref{fig:radio}.7), 1201-r1 (Figure \ref{fig:radio}.29), and 1112-r3 (Figure \ref{fig:radio}.28) have radio flux densities of several tens of milli Janskies, much higher than the others. However, these diagnostics do not work for compact and faint sources. We thus look further into their physical parameters.

UC \ion{H}{2} regions have been quantitatively characterized as a pack of ionized gas with size $<$0.1 pc, $N_e \geq 10^4$ cm$^{-3}$, and $EM \geq 10^7$ pc cm$^{-6}$ and HC \ion{H}{2} regions with size $<$0.05 pc, $N_e \geq 3 \times 10^5$ cm$^{-3}$, and $EM \geq 10^8$ pc cm$^{-6}$ \citep{2010MNRAS.405.1560M}. We calculate these physical parameters of the radio sources on each map with the UC \ion{H}{2} region model of \citet{1967ApJ...147..471M} and \citet{1969ApJ...156..269S}, which assumes a spherically symmetric, optically thin UC \ion{H}{2} region of uniform density and electron temperature of 10$^4$ K. The required parameters for calculation are distances (1.4 kpc), H$_2$ column densities (see Table \ref{tab:MDC_Cao21}), observing frequencies, radio flux densities, and source sizes. The results are listed in columns 6--8 of Table \ref{tab:parameters_full}, which are $EM$, $N_e$, and the excitation parameter ($U$), respectively. The maps with UC- or HC- \ion{H}{2} region candidates are marked by one or two asterisks, respectively.

Only nine sources meet the criteria of UC- or HC- \ion{H}{2} regions in at least one map. Two of them are the most luminous radio sources in MDC 1112, whose natures have been long under debate \citep{1997ApJ...489..744T, 2001ApJ...546..345S, 2010AJ....139.2433C, 2020MNRAS.496.3128R}. Source 675-r1 reaches the criteria only on the Q-band A-configuration map, which only picks up the most compact components. The result of source 1201-r1 is consistent with the fact that it is a very dense and bright ionizing source \citep{2018A&A...617A..45S}. Moreover, this method will underestimate the parameters for the sources that do not meet the optically-thin criterion as the model requires, which can lead to some young UC/HC \ion{H}{2} regions being misclassified as unqualified. Meanwhile, the uncertainty in deriving the source size by deconvolution, especially for the barely resolved sources (e.g., 310-r2 and 1112-r6), can bring large errors.

\startlongtable
\begin{deluxetable*}{cccccccc}
\centering
\tablecaption{Physical Parameters Derived from UC \ion{H}{2} model\label{tab:parameters_full}}
\tabletypesize{\scriptsize} 
\tablehead{
\colhead{MDC} & \colhead{Radio ID} & \colhead{Spectral Index} & \colhead{Frequency} & \colhead{Project Code} & \colhead{$EM$} & \colhead{$N_e$} & \colhead{$U$} \\
\colhead{} & \colhead{} & \colhead{} & \colhead{(GHz)} & \colhead{} & \colhead{(10$^{5}$ pc cm$^{-6}$)} & \colhead{(10$^{3}$ cm$^{-3}$)} & \colhead{(pc cm$^{-2}$)}
}
\startdata
248 & 1 & \nodata & 23.2 & 17A--107 & 19.57 & 13.05 & 0.98 \\
\hline
274 & 1 & $-$0.11 & 5.8 & 14A--240\_1115 & 15.22 & 17.78 & 0.53 \\
 &  &  & 8.4 & AM432 & 1.16 & 2.18 & 0.65 \\
 &  &  & 10.0 & 16A--301 & 13.84 & 15.11 & 0.59 \\
 &  &  & 22.5 & Fontani2012 & 1.54 & 2.73 & 0.62 \\
 &  &  & 23.2 & 17A--107 & 9.03 & 10.83 & 0.58 \\
 &  &  & 23.2 & 13B--210 & 24.03 & 23.58 & 0.55 \\
\hline
310 & 1 & \nodata & 23.2 & 17A--107 & 5.26 & 6.44 & 0.67 \\
 & 2 & 0.75 & 23.2 & 17A--107 & 1.35 & 3.18 & 0.44 \\
 &  &  & 44.0 & 14A--092** & 21824.93 & 3931.42 & 0.53 \\
\hline
340 & 1 & \nodata & 23.2 & 17A--107 & 0.77 & 2.49 & 0.35 \\
\hline
341 & 1 & \nodata & 23.2 & 14A--241 & 7.42 & 8.71 & 0.64 \\
351 & 1 & \nodata & 23.2 & 17A--107 & 0.28 & 1.05 & 0.41 \\
507 & 2 & \nodata & 5.8 & 14A--420\_9676 & 38.12 & 24.53 & 0.86 \\
 &  &  & 8.4 & AJ239 & 40.22 & 30.01 & 0.69 \\
 &  &  & 10.0 & 16A--301 & 47.61 & 38.45 & 0.58 \\
 & 3 & \nodata & 5.8 & 14A--420\_9676* & 100.76 & 30.96 & 1.67 \\
 &  &  & 8.4 & AJ239* & 102.78 & 31.44 & 1.65 \\
\hline
509 & 1 & \nodata & 23.2 & 17A--107 & 0.9 & 2.34 & 0.45 \\
 & 2 & \nodata & 23.2 & 17A--107 & 0.17 & 0.73 & 0.39 \\
 & 4 & \nodata & 23.2 & 17A--107 & 0.19 & 0.82 & 0.38 \\
 & 5 & \nodata & 23.2 & 17A--107 & 0.11 & 0.52 & 0.39 \\
 & 6 & \nodata & 23.2 & 17A--107 & 1.6 & 2.17 & 0.88 \\
 & 7 & \nodata & 23.2 & 17A--107 & 24.79 & 24.66 & 0.53 \\
\hline
675 & 1 & \nodata & 23.2 & 17A--107 & 27.67 & 17.97 & 0.9 \\
 &  &  & 44.0 & 14A--092** & 7973.45 & 1518.3 & 0.69 \\
\hline
714 & 2 & \nodata & 23.1 & 17A--107 & 1.16 & 3.1 & 0.4 \\
\hline
725 & 1 & $-$0.01 & 1.4 & AM446 & 90.41 & 28.69 & 1.74 \\
 &  &  & 22.5 & AS683* & 1619.91 & 182.75 & 2.4 \\
\hline
742 & 2 & $-$0.03 & 4.9 & AD219 & 16.47 & 6.3 & 2.29 \\
 &  &  & 8.5 & AK450 & 14.57 & 5.61 & 2.32 \\
\hline
753 & 1 & $-$0.65 & 4.9 & AD219 & 13.65 & 11.6 & 0.84 \\
 &  &  & 4.9 & Molinari1998 & 22.49 & 16.17 & 0.89 \\
\hline
839 & 1 & \nodata & 23.2 & 17A--107 & 1.64 & 3.75 & 0.43 \\
\hline
1112 & 1 & \nodata & 4.9 & Gibb2007* & 546.25 & 216.59 & 0.68 \\
 &  &  & 5.8 & 14A--420\_9375* & 114.81 & 57.61 & 0.83 \\
 &  &  & 8.5 & CG2010* & 240.03 & 94.29 & 0.89 \\
 &  &  & 8.5 & Gibb2007* & 538.71 & 263.92 & 0.51 \\
 &  &  & 14.9 & AF381* & 131.56 & 64.05 & 0.8 \\
 &  &  & 22.3 & Torrelles1997* & 768.67 & 184.09 & 1.13 \\
 & 2 & \nodata & 5.8 & 12B--140 & 75.07 & 53.04 & 0.61 \\
 &  &  & 14.9 & AF381 & 85.19 & 52.09 & 0.68 \\
 & 3 & \nodata & 4.9 & Gibb2007** & 1311.0 & 489.61 & 0.55 \\
 &  &  & 5.8 & 12B--140* & 836.87 & 289.51 & 0.7 \\
 &  &  & 8.5 & CG2010** & 1256.46 & 322.16 & 0.9 \\
 &  &  & 8.5 & Gibb2007** & 1183.97 & 366.42 & 0.72 \\
 &  &  & 22.3 & Torrelles1997** & 16808.85 & 1675.03 & 1.3 \\
 &  &  & 43.5 & Gibb2007** & 47115.54 & 4389.09 & 0.99 \\
 & 4,5 & \nodata & 5.8 & AF381 & 36.54 & 23.17 & 0.89 \\
 &  &  & 5.8 & 14A--420\_9375 & 21.73 & 18.33 & 0.72 \\
 &  &  & 5.8 & AH869 & 24.37 & 16.3 & 0.95 \\
 &  &  & 8.5 & CG2010 & 88.33 & 46.15 & 0.85 \\
 & 6 & \nodata & 5.8 & 12B--140* & 135.78 & 97.7 & 0.49 \\
 &  &  & 14.9 & AF381 & 32.36 & 30.59 & 0.53 \\
\hline
1201 & 1 & 0.73 & 8.5 & AF362** & 12123.85 & 1394.39 & 1.24 \\
\hline
1225 & 1 & \nodata & 24.4 & 13A--373\_9583 & 0.61 & 2.21 & 0.33 \\
\hline
1243 & 1 & \nodata & 23.8 & 13A--373 & 3.52 & 5.82 & 0.52 \\
\hline
1267 & 2 & \nodata & 23.2 & 17A--107 & 1.44 & 4.09 & 0.34 \\
\hline
1454 & 1 & \nodata & 23.2 & 17A--107 & 0.54 & 1.7 & 0.41 \\
 &  &  & 24.4 & 13A--373\_4954 & 0.96 & 3.08 & 0.33 \\
 & 2 & \nodata & 23.2 & 17A--107 & 1.73 & 4.56 & 0.35 \\
 & 3 & \nodata & 23.2 & 17A--107 & 0.29 & 1.03 & 0.42 \\
 &  &  & 24.4 & 13A--373\_4954 & 0.98 & 2.96 & 0.35 \\
\hline
1467 & 1 & 0.61 & 7.0 & 14A--420\_9375 & 53.74 & 29.65 & 0.94 \\
 &  &  & 30.9 & 14B--173* & 683.99 & 154.29 & 1.26 \\
 & 2 & \nodata & 8.5 & Araya2009 & 94.87 & 95.18 & 0.35 \\
\hline
2210 & 1 & \nodata & 23.2 & 17A--107 & 0.11 & 0.43 & 0.52 \\
 &  &  & 24.4 & 13A--373\_4954 & 0.18 & 0.85 & 0.34 \\
\hline
3188 & 1 & \nodata & 23.2 & 17A--107 & 1.07 & 3.42 & 0.32 \\
\hline
4797 & 1 & \nodata & 23.3 & 17A--107 & 0.15 & 0.67 & 0.39 \\
\hline
5417 & 1 & \nodata & 7.0 & 14A--420\_9375 & 74.15 & 45.42 & 0.73 \\
 &  &  & 23.8 & 13A--373 & 9.68 & 9.58 & 0.73 \\
 \hline
\enddata
\tablecomments{Physical parameters assuming that the radio detections are UC/HC \ion{H}{2} regions. Columns 6, 7, and 8 are emission measure, electron density, and excitation parameter, respectively. Project code with one or two asterisks means the parameters derived from this map meets the criteria of an ultra- or hyper compact \ion{H}{2} region.}
\end{deluxetable*}

\begin{figure}[ht!]
\epsscale{0.8}\plotone{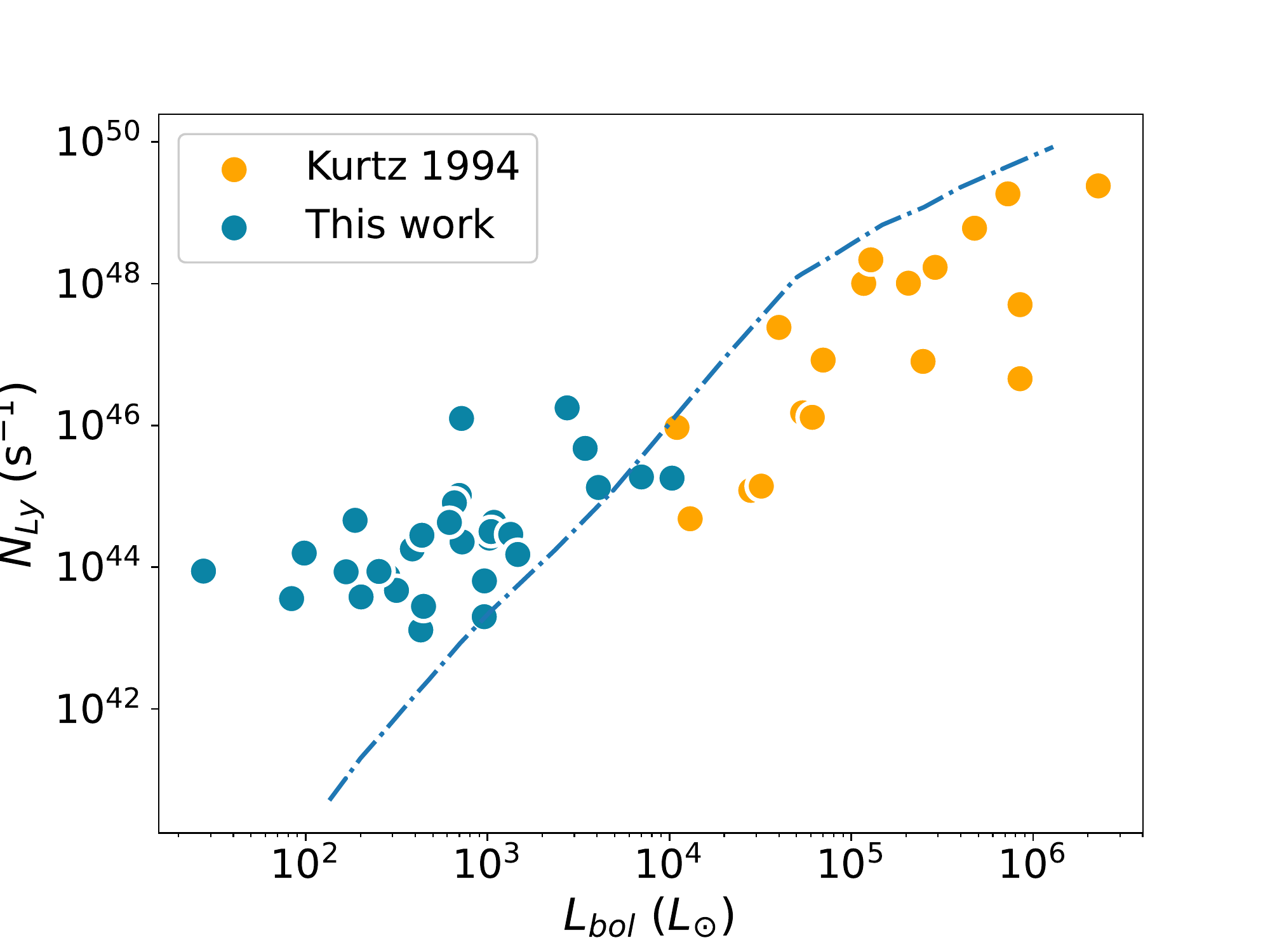}
\caption{Lyman continuum flux as a function of bolometric luminosity. The blue dots represent the data from this work; the orange dots represent the UC \ion{H}{2} regions in \citet{1994ApJS...91..659K}; the blue dash-dotted line represents the expected Lyman continuum flux of a single ZAMS star \citep{1984ApJ...283..165T}. The bolometric luminosities of the data in this work are approximated by the far-infrared luminosities in Cao21. \label{fig:nly}}
\end{figure}

We further inspect the Lyman continuum flux as a function of the bolometric luminosity. Lyman continuum flux is calculated with the K-band flux density, assuming an optically thin, spherical \ion{H}{2} region with an electron temperature of 10$^4$ K \citep{1974A&A....32..269M}. For the MDCs containing multiple compact radio sources, their radio flux densities are summed up or re-measured on the maps of the lowest resolutions, in order to minimize the loss of flux due to spatial filtering. Bolometric luminosity is approximated by the far-infrared (FIR) luminosity from Cao21, considering that FIR luminosity is the dominant contributor of the bolometric luminosity of MDCs at early evolution stages. This approximation have larger uncertainties for MDCs at relatively late evolution stages, e.g., when a bright UC \ion{H}{2} region has formed. Lyman continuum flux as a function of bolometric luminosity is shown in Figure \ref{fig:nly}, where we also plot the data of the UC \ion{H}{2} regions obtained from \citet{1994ApJS...91..659K} and the expected Lyman continuum flux of a single ZAMS star \citep{1984ApJ...283..165T}. The ZAMS line is a theoretical upper limit of the Lyman continuum flux produced by photoionization. Our sample and the UC \ion{H}{2} regions show clearly different distributions. The trend of our data does not follow that of the ZAMS stars, either. Thus most of our detected sources have different ionization mechanisms rather than photoionization. Moreover, this method may underestimate the Lyman continuum flux because part of the extended radio fluxes are filtered out by the interferometer \citep{1999ApJ...514..232K, 2020MNRAS.492..895D} and our sources may not be optically thin \citep{2013MNRAS.435..400U}. This means that the blue dots in Figure \ref{fig:nly} would be pushed even further above the dash-dotted line, which strengthens the conclusion that most of the detected sources are not UC \ion{H}{2} regions.

\subsubsection{Radio Jets and Winds} \label{sssec:jet}
Considering the generally low radio flux densities (no more than a few milli Janskies, see Figure \ref{dist_NO_withLimit}) of the radio sources, we look into the possibility of ionized jets/winds \citep{2009A&A...501..539U}. In high-resolution observations, radio jets can be resolved into elongated or string-like structures. A typical example in our detection is source 753-r1 (Figure \ref{fig:radio}.12). However, most sources are not sufficiently resolved. Ionized jets and winds can be associated with molecular outflows and/or shock-induced water masers \citep{1995A&AS..112..299T}.  A total of 32 MDCs in our sample are found to be associate with either molecular outflows or water masers (see the last column of Table \ref{tab:MDC}). Among the 34 MDCs with radio detections, 24 are associated with outflows or water masers, or both. 

\begin{figure*}[ht!]
\epsscale{0.8}\plotone{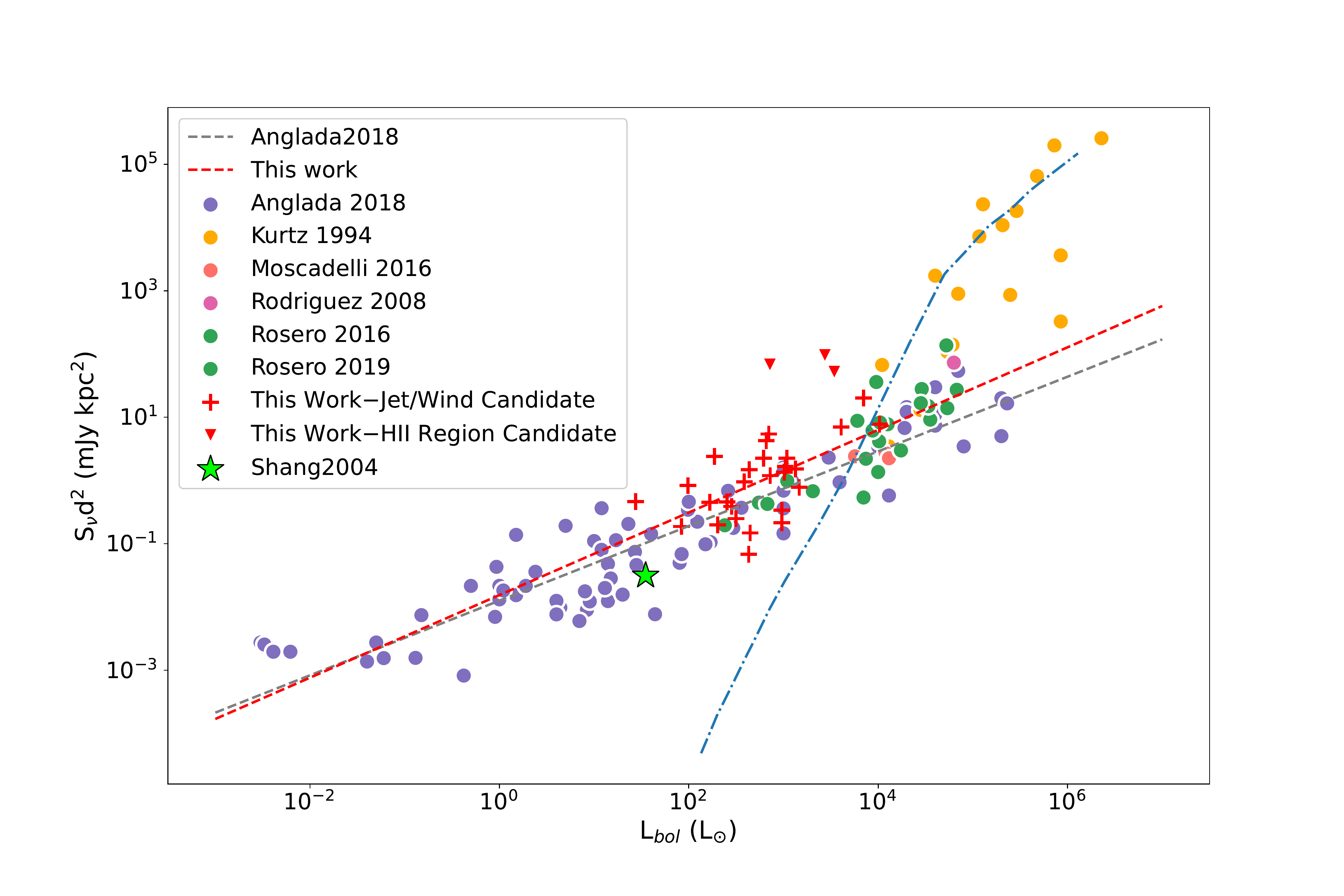}
\caption{\label{fig:radio_bol}Radio luminosity of thermal jets at 8.4 GHz as a function of bolometric luminosity of YSO. The red inverted triangles represent the MDCs with UC \ion{H}{2} region candidates in our sample. The red crosses represent the other MDCs. All of their bolometric luminosities are approximated by the far-infrared luminosities in Cao21. The orange dots represent the UC \ion{H}{2} regions provided by \citet{1994ApJS...91..659K}; the dots in other colors represent radio jets obtained from the literature. The green star marks the highest possible radio luminosity of a jet originating from a low-mass YSO \citep{2004ApJ...612L..69S}. The blue dash-dotted line is the expected radio luminosity from a single ZAMS star, which is a theoretical upper limit of the radio luminosity caused by photoionization \citep{1984ApJ...283..165T}. The grey dashed line is the empirical correlation established by \citet{2018A&ARv..26....3A}; the red dashed line is the fitting result of our data with the UC \ion{H}{2} region candidates excluded.}
\end{figure*}

Statistically, an empirical relation between the radio luminosity of thermal radio jets and the bolometric luminosity of YSOs has been established by \citet{2018A&ARv..26....3A} and is valid across the entire mass regime:
\begin{eqnarray} \label{eq:anglada}
\left(\frac{S_{v} d^{2}}{\mathrm{mJy kpc}^{2}}\right)=10^{-1.90 \pm 0.07}\left(\frac{L_{b o l}}{L_{\odot}}\right)^{0.59 \pm 0.03}\ .
\end{eqnarray}
In Figure \ref{fig:radio_bol}, we plot the radio luminosity at 8.4 GHz as a function of the bolometric luminosity of the corresponding MDCs. 

The data include our sources (red crosses and inverted triangles) as well as the UC \ion{H}{2} regions (orange dots) and radio jets (other colored dots) from the literature \citep{2008AJ....135.2370R, 2016A&A...585A..71M, 2016ApJS..227...25R, 2018A&ARv..26....3A, 2019ApJ...880...99R}. All the radio luminosities are scaled to 8.4 GHz assuming a spectral index of 0.6, which is a theoretical value for a wind \citep{1975MNRAS.170...41W}. The \citet{2018A&ARv..26....3A} relation is represented by the grey dashed line. The bolometric luminosities and radio flux densities used in this figure are the same as those used to calculate the Lyman continuum rate in Figure \ref{fig:nly}. It is clear that except the candidate UC/HC \ion{H}{2} regions (MDC 507, 725, and 742, marked by the red inverted triangles), our data well follow the radio jet trend derived by \citet{2018A&ARv..26....3A}, and a linear fit to the data yield:
\begin{eqnarray} \label{eq:our}
\left(\frac{S_{v} d^{2}}{\mathrm{mJy kpc}^{2}}\right)=10^{-1.81 \pm 0.98}\left(\frac{L_{b o l}}{L_{\odot}}\right)^{0.65 \pm 0.35},
\end{eqnarray}
a relation similar to that of \citet{2018A&ARv..26....3A}. The fitting result supports that our sources are mostly ionized jets/winds. Moreover, by comparing with a theoretical upper limit of the radio luminosity of an ionized jet originating form a low-mass YSO (see the green star in Figure \ref{fig:radio_bol}), these jets/winds are driven from intermediate- to high-mass YSOs.

\subsection{On the evolution of MDCs}\label{ssec:evolve}
Following Cao19, we classify the sample into three categories as starless, IR-quiet, and IR-bright, a supposed evolution sequence determined by the 70 $\mu$m and 24 $\mu$m fluxes. The starless (or presumably pre-stellar given very high masses) cores have no compact emissions at 70 $\mu$m, 24 $\mu$m, or protostars \citep{2014AJ....148...11K}; the IR-quiet cores exhibit 24 $\mu$m and/or 70 $\mu$m sources and their 24 $\mu$m fluxes are lower than 23 Jy, which corresponds to an 8 $M_\odot$ stellar embryo at a distance of 1.4 kpc \citep{2019ApJS..241....1C}; the IR-bright cores are MDCs with at least one 24 $\mu$m source that has fluxes exceeding 23 Jy. With strong radio continuum emission from UC \ion{H}{2} regions characterizing relatively more advanced evolution stages \citep{2002ARA&A..40...27C, 2017A&A...599A.139K}, weak radio emission from ionized jets/winds are strong indicators of ongoing star-forming activities across the pre-UC \ion{H}{2} phases. 

\begin{figure}[ht!]
\epsscale{0.8}\plotone{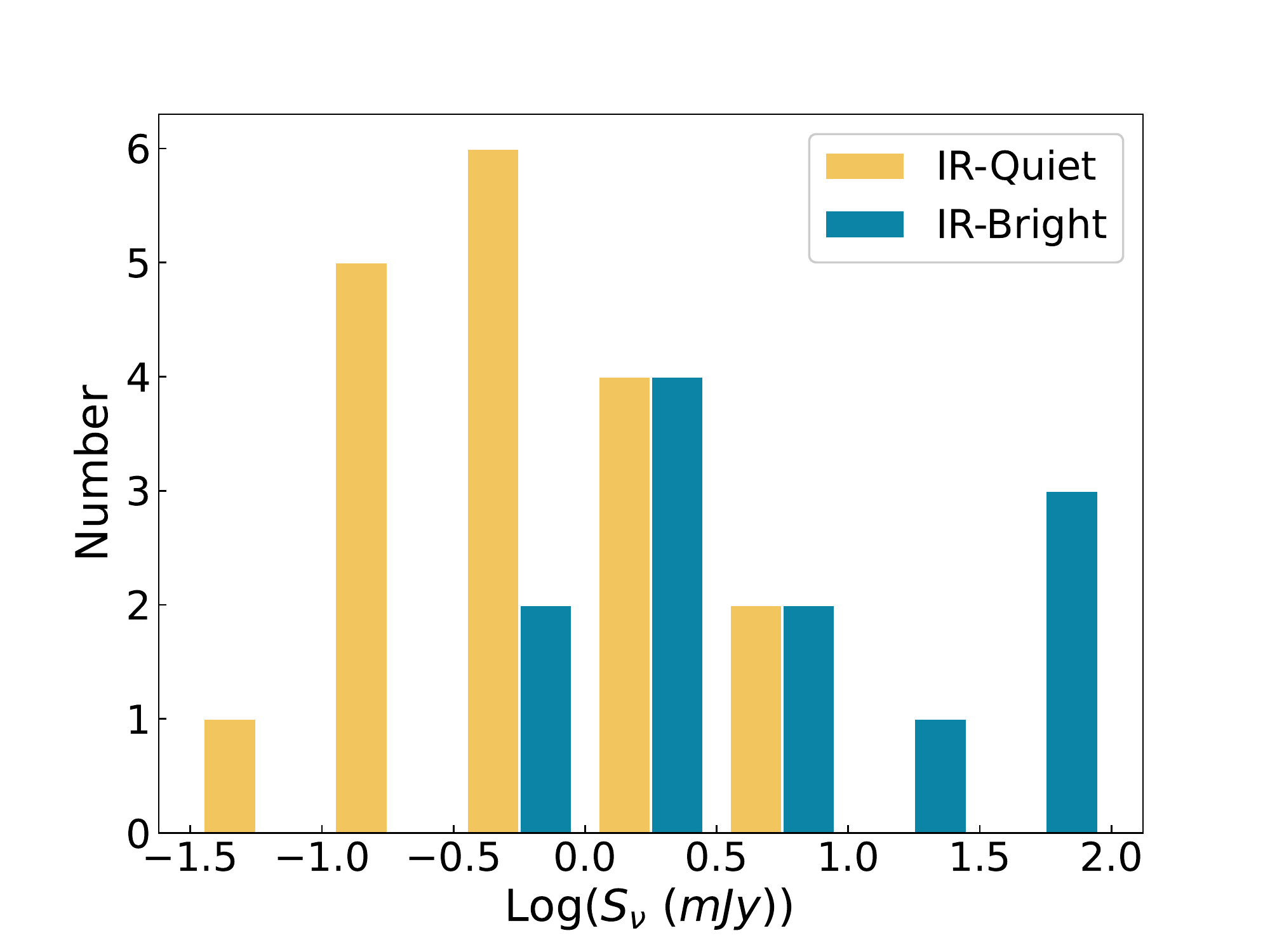}
\caption{Radio flux density distribution of different MDC categories at 23.2 GHz. Green, yellow, and blue histograms represent starless, IR-quiet, and IR-bright MDCs, respectively.\label{fig:sv_cate}}
\end{figure}

The number of MDCs with radio detection in each category is listed in Table \ref{tab:ir}.
\begin{table}[h!]
\renewcommand{\thetable}{\arabic{table}}
\centering
\caption{Radio detections in the three types of MDCs\label{tab:ir}}
\begin{tabular}{c|c|c|c}
\tablewidth{0pt}
\hline
\hline
Type & Total & Radio Detected & Rate \\
\hline
Starless & 1 & 0 & 0\% \\
IR-quiet &  27 & 18 & 67\% \\
IR-bright & 14 & 13 & 93\% \\
\hline
\hline
\end{tabular}
\end{table}
Although there have been previous studies claiming increasing radio detection rates along with the evolutionary stages of the MDCs \citep{1994ApJS...91..659K, 1998A&A...336..339M, 2016ApJS..227...25R}, the apparent trend of the radio detections is not statistically significant in this work according to the result of the Barnard's test. However, the IR-bright MDCs have radio flux densities higher than those of the IR-quiet MDCs (Figure \ref{fig:sv_cate}).

Starless (or pre-stellar) cores should have no radio emission because there are no star-forming activities yet. On the contrary, IR-bright cores are likely to present luminous radio continuum emission since the central young stars or protostars may have attained masses greater than 8 M$_\odot$, and thus ionizing the surroundings to form UC \ion{H}{2} regions. Among the 14 IR-bright MDCs, we have detected radio sources in 13 of them. The most luminous and extended sources are all associated with the IR-bright MDCs. For the only MDC without radio detection at 0.01 pc scale, MDC 1460 has a 0.1-pc UC \ion{H}{2} region at its center, which may prevent us from detecting radio sources at 0.01 pc scale. 

\begin{figure*}[ht!]
\plotone{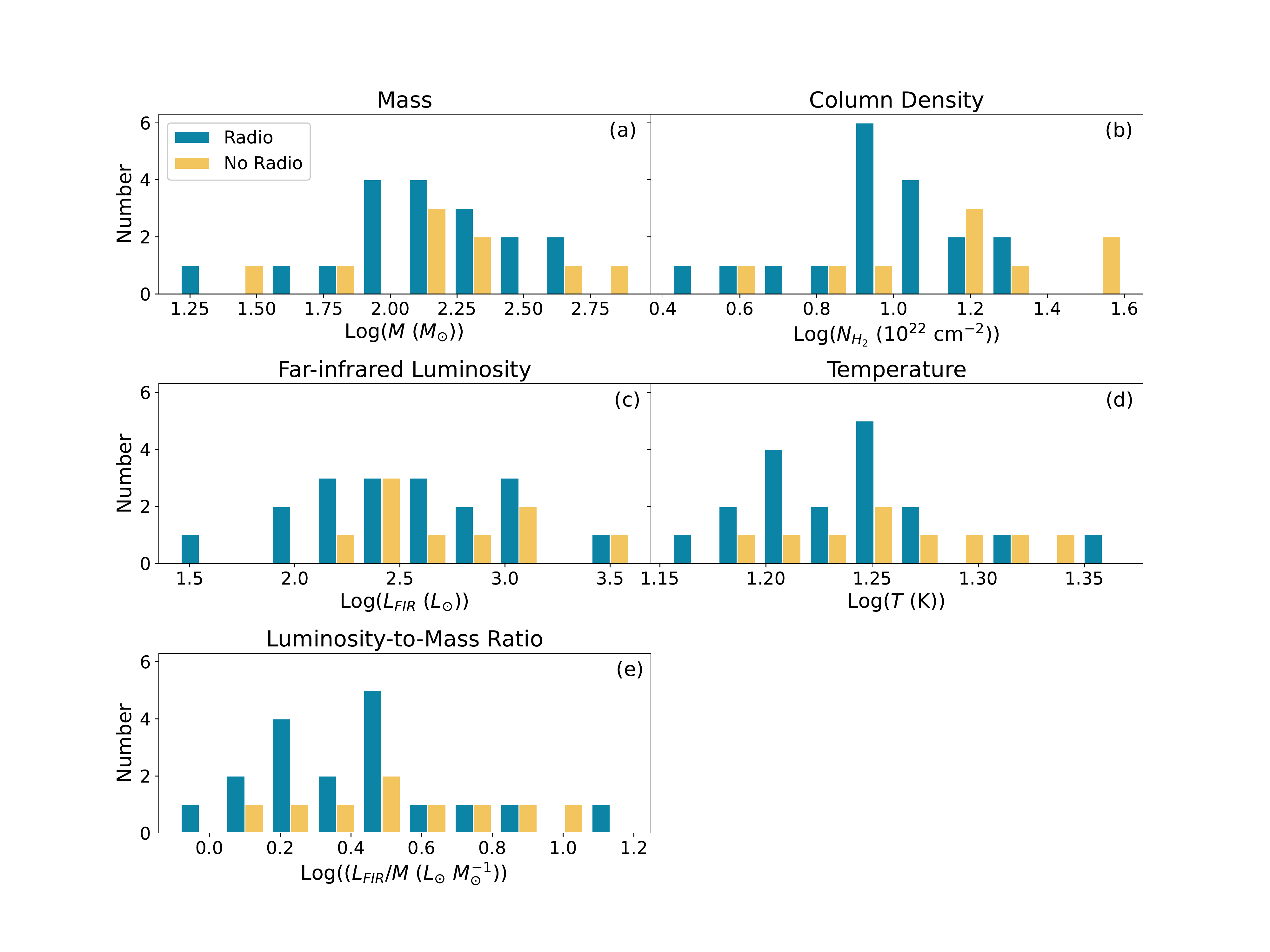}
\caption{Distributions of the physical parameters between the IR-quiet MDCs with and withoud radio detection.\label{fig:dist_RvsQ}}
\end{figure*}

As an intermediate evolution stage, we have detected radio emission in 17 of the 27 IR-quiet MDCs. We first inspect if there are any intrinsic differences between the IR-quiet MDCs with and without radio detection. Figure \ref{fig:dist_RvsQ} shows the distributions of the physical parameters between the two groups of MDCs. The inspected physical parameters are mass, column density, far-infrared luminosity, dust temperature, and luminosity-to-mass ($L_{FIR}/M_{core}$) ratio. We use the Kolmogorov-Smirnov test and Kruskal-Wallis \textsc{H}-test to test if the distributions are statistically different. As a result, the distributions of all the five parameters are statistically the same.

To better understand the results, we inspect the radio and far-infrared environments of each IR-quiet MDC. We notice that five out of the ten IR-quiet MDCs without radio detection are located adjacent to large-scale radio sources, and their far-infrared counterparts are also affected by the large-scale structures (see Cao19). Three MDCs are associated with radio sources adjacent to the FHWM boundaries. We thus propose that, for our sample, the IR-quiet MDCs with and without radio detection are intrinsically the same; the non-detection of radio sources in some IR-quiet MDCs can be a result of the contamination from large-scale radio continuum sources and the real radio detection rate of IR-quiet MDCs are likely to be higher than what we have obtained. Comparison between the IR-quiet MDCs with and without radio sources requires a sample with clean radio backgrounds.

Finally, we briefly discuss whether radio luminosity can refine the evolution stages of the MDCs with radio detection. During the evolution of MDCs, the dominant mechanism of radio continuum emission changes from shock-ionization to photoionization \citep{2002ApJ...580..980K, 2009ApJ...699L..31S, 2016ApJ...818...52T, 2016MNRAS.460.1039P}, and the bolometric luminosity changes from accretion-dominated to radiation-dominated \citep{2018A&ARv..26....3A}. Observationally, jets/winds are direct indicators of accretion processes \citep{1996A&A...311..858B, 2013A&A...558A.125D}. We focus on the MDCs associated with candidate ionized jets/winds.

\begin{figure}
\epsscale{.6}\plotone{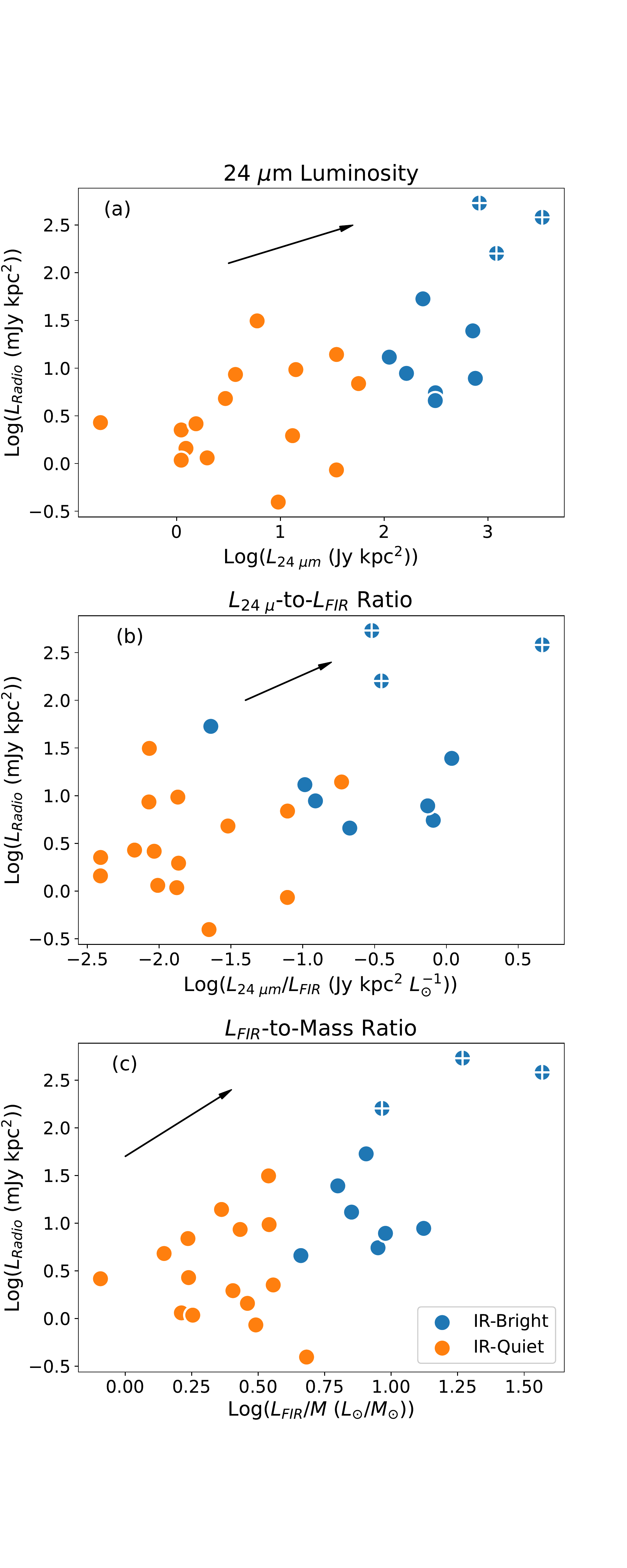}
\caption{Radio luminosity of the radio sources versus the three parameters that may probe the evolutionary stages of the MDCs. Orange and blue dots represent the IR-quiet and IR-bright MDCs, respectively. Blue dots centered with white crosses denote MDCs associated with candidate UC \ion{H}{2} regions. The arrows mark the hypothetical evolution trends. (a) Radio luminosity against 24 $\mu$m luminosity. (b) Radio luminosity against the ratio of 24 $\mu$m luminosity to far-infrared luminosity. (c) Radio luminosity against $L_{FIR}/M_{core}$ ratio.\label{fig:indicator}}
\end{figure}

We then select three parameters to probe the evolution phases of the MDCs: the 24 $\mu$m luminosity, the 24-$\mu$m-luminosity-to-far-infrared-luminosity ($L_{24\ \mu m} / L_{FIR}$) ratio, and the $L_{FIR}/M_{core}$ ratio. The 24 $\mu$m luminosity is a strong indicator of the mass of the central YSO; the $L_{24\ \mu m} / L_{FIR}$ ratio can probe the SED of the MDC; the $L_{FIR}/M_{core}$ ratio is often adopted for a tracer of evolution stages (e.g. \citealt{2018ApJS..235....3Y}). Relations between the radio luminosity and these three parameters are shown in Figure \ref{fig:indicator}. IR-quiet and IR-bright MDCs are represented by orange and blue dots, respectively. For comparison, we also plot the MDCs associated with candidate UC \ion{H}{2} regions (blue dots centered by white crosses). MDC 753 is excluded because its 24 $\mu$m source does not associate with any radio sources in it.

The radio luminosity appears to increase with all the three probes of the evolutionary stages, as indicated by the arrows in Figure \ref{fig:indicator}. However, linear fittings to the data yield low coefficients (all less than 0.6) and large uncertainties. We then apply the Kendall's $\tau$ test to check if the general trends are statistically significant. The results ($p$-value less than 0.05) reveal that radio luminosity is positively correlated with 24 $\mu$m luminosity. All the three correlations are significant if the MDCs associated with UC \ion{H}{2} region candidates are included. This is consistent with the expectation that the MDCs harboring UC \ion{H}{2} regions are more evolved than those only associated with compact ionized jets/winds, and that the radio luminosities of UC \ion{H}{2} regions are higher than ionized jets/winds.

Based on the above analysis, we conclude that the radio detection rate alone is insufficient to discriminate the levels of star-forming activities between the IR-quiet and IR-bright MDCs. Instead, radio luminosity is a better indicator. The relatively lower radio detection rate of IR-quiet MDCs is likely ascribed to background contamination and the limited observational sensitivity. In our sample, the detected radio sources may provide hints on the evolutionary status of the associated MDCs. During the shock-dominated phase, the radio luminosity increases along the evolution trend continuously from IR-quiet to IR-bright. At a later evolutionary phase characterized by observable UC \ion{H}{2} regions, the radio luminosity is dominated by photoionization and is on average higher than that at the shock-dominated phase.

\section{Summary} \label{sec:sum}
As the second part of the CENSUS project, we have characterized the radio continuum emission at 0.01 pc scales towards a sample of 47 MDCs with the VLA. Our results are summarized as follows:
\begin{enumerate}
\item We have detected a total of 64 radio sources, of which 37 are reported for the first time. The radio sources are mostly faint and compact, with flux densities mostly ranging from 0.1 mJy to a few mJy. The newly reported sources are even fainter, with flux densities mostly of a few tenths of milli Janskies. 
\item Forty-four (69\%)  radio sources are associated with dust condensations; fifty-four (84\%) are located within the FWHM of the MDCs. Twelve MDCs are observed to have multiple radio sources.
\item We have obtained the spectral indices of 8 radio sources, in which two are negative (less than $-$0.1), two are flat (between $-$0.1 and 0.2), and four are positive (higher than 0.2). By comparing our results with the literature, we notice that there can be discrepancies across different works on the same source. We suggest that cautions should be taken in fitting the SEDs of the radio sources, considering that both potential time variability of the sources and different observational settings can induce large uncertainties in the derived spectral indices.
\item We have investigated the nature of the radio sources. Only several sources can be identified as UC \ion{H}{2} region candidates. The majority of the detected radio sources are most likely to be ionized jets or winds originating from massive YSOs.
\item In our sample, the radio detection rate increases from starless to IR-quiet, and IR-bright MDCs. However, this trend is statistically insignificant. Background contaminations and limited sensitivities hamper a meaningful interpretation of the data.
\item Focusing on the MDCs all associated with candidate ionized jets/winds, the radio luminosity appears to increase with the advancing evolution phase.
\end{enumerate}

\acknowledgments
{ 
Y.W., K.Q., Y.C., J.L., and B.H. are supported by the National Key R\&D Program of China No. 2017YFA0402600. We acknowledge the support from the National Natural Science Foundation of China (NSFC) through grants U1731237, 11473011, 11590781, and 11629302. 
Y.C. is partially supported by the Scholarship No. 201906190105 of the China Scholarship Council and the Predoctoral Program of the Smithsonian Astrophysical Observatory (SAO).
This research made use of Astropy,\footnote{http://www.astropy.org} a community-developed core Python package for Astronomy \citep{2013A&A...558A..33A, 2018AJ....156..123A}. 
This research made use of APLpy, an open-source plotting package for Python \citep{2012ascl.soft08017R}. 
This research made use of SciPy \citep{JonesE2001, 2020SciPy-NMeth}.
This research made use of Matplotlib \citep{4160265},.
This research made use of Numpy \citep{2011CSE....13b..22V, 2020NumPy-Array}. 
This research made use of Pandas \citep{pandas_mckinney}. 
This research made use of Scikit-learn \citep{scikit-learn}. 
This research made use of MIRIAD \citep{1995ASPC...77..433S}.
This research made use of the Common Astronomy Software Applications package (CASA) \citep{2007ASPC..376..127M}. 
This research made use of SAOImage DS9 \citep{2000ascl.soft03002S, 2003ASPC..295..489J}. 
This research has made use of the SIMBAD database, operated at CDS, Strasbourg, France. 
The National Radio Astronomy Observatory is a facility of the National Science Foundation operated under cooperative agreement by Associated Universities, Inc.

\facilities{NSF's Karl G. Jansky Very Large Array (VLA)}

\software{CASA, MIRIAD, ds9, \texttt{Astropy}, \texttt{APLpy}, \texttt{Pandas}, \texttt{SciPy}, \texttt{Matplotlib}, \texttt{Numpy}, \texttt{Scikit-learn}}
}

\bibliography{CENSUSII}{}
\bibliographystyle{aasjournal}

\appendix
\include{Individual_MDCs}

\end{document}

%% file: Table3_MapInfo.tex
\startlongtable
\begin{deluxetable*}{lrrrrrr}
\tablecaption{Properties of the Radio Continuum Maps\label{tab:mapInfo}}
\tablehead{
\colhead{MDC} & \colhead{Project Code} & \colhead{Frequency} & \colhead{B$_{maj}$} & \colhead{B$_{min}$} & \colhead{BPA} & \colhead{RMS} \\
\colhead{} & \colhead{} & \colhead{(GHz)} & \colhead{($''$)} & \colhead{($''$)} & \colhead{($^\circ$)} & \colhead{(Jy/Beam)} 
}
\startdata
214, 247 & 17A--107 & 23.17 & 3.32 & 2.81 & $-$84.4 & 1.36E$-$05 \\
\hline
220 & AR436 & 4.86 & 15.96 & 13.37 & $-$87.2 & 1.52E$-$04\\
 & AB1073 & 8.46 & 7.79 & 7.76 & $-$11.4 & 3.49E$-$04 \\
 & 14A--241 & 23.09 & 2.83 & 2.58 & $-$46.4 & 7.15E$-$05 \\
\hline
248 & 17A--107 & 23.17 & 3.39 & 3.31 & 52.4 & 6.50E$-$05 \\
\hline
274 & Miralles1994 & 4.9 & 4.8 & 4.6 & 81.0 & 1.0E$-$04 \\
 & 14A--420\_1115 & 5.80 & 1.45 & 0.95 & 79.3 & 7.50E$-$05 \\
 & Rosero2016 & 6.15 & 0.32 & 0.28 & 54.2 & 7.0E$-$06 \\
 & AM432 & 8.44 & 10.71 & 7.64 & 70.9 & 2.76E$-$04 \\
 & Fontani2012 & 22.46 & 3.02 & 2.98 & \nodata & 5.7E$-$05 \\
 & 16A--301 & 10.00 & 0.85 & 0.50 & $-$82.7 & 2.50E$-$05 \\
 & 17A--107 & 23.17 & 4.31 & 2.90 & $-$77.2 & 3.71E$-$05 \\
 & 13B--210 & 23.20 & 0.43 & 0.23 & $-$81.6 & 1.08E$-$05 \\
\hline
302, 520 & 17A--107 & 23.17 & 4.18 & 2.84 & $-$74.7 & 4.14E$-$05\\
 & AM462 & 14.94 & 5.09 & 1.80 & 62.3 & 1.45E$-$04 \\
\hline
310 & 12B--140 & 5.80 & 0.31 & 0.30 & 70.6 & 6.90E$-$06 \\
 & 16A--301 & 10.00 & 0.76 & 0.51 & $-$87.9 & 3.5E$-$05 \\
 & 17A--107 & 23.17 & 3.43 & 2.79 & $-$76.7 & 2.55E$-$05 \\
 & 14A--092\_7083 & 44.00 & 0.04 & 0.04 & 63.1 & 4.60E$-$05 \\
 & 14A--092\_9305 & 44.00 & 0.06 & 0.04 & 84.5 & 6.20E$-$05 \\
 & 14A--092\tablenotemark{a} & 44.00 & 0.05 & 0.04 & 88.2 & 4.06E$-$05 \\
\hline
327, 742 & AD219 & 4.86 & 4.41 & 4.31 & 77.7 & 1.70E$-$04 \\
 & AH726 & 4.86 & 1.34 & 1.19 & 34.4 & 1.20E$-$04 \\
 & AC240 & 8.44 & 0.81 & 0.68 & $-$51.9 & 9.79E$-$05 \\
 & AG625 & 8.46 & 0.79 & 0.68 & $-$77.2 & 9.80E$-$06 \\
 & AK450 & 8.46 & 8.08 & 7.55 & $-$175.9 & 9.40E$-$05 \\
 & AK477 & 8.46 & 2.52 & 2.28 & $-$171.4 & 1.00E$-$04 \\
 & Kurtz1994 & 8.41 & 0.9 & 0.9 & \nodata & 1.60E$-$04 \\
 & Kurtz1994 & 14.96 & 0.5 & 0.5 & \nodata & 1.10E$-$04 \\
 & Masque2017 & 21.96 & 0.134 & 0.100 & $-$78.8 & 2.6E$-$05 \\
 & 14A--481 & 22.00 & 0.1 & 0.08 & 81.9 & 4.00E$-$05 \\
 & 17A--107 & 23.17 & 3.83 & 2.81 & $-$76.8 & 4.50E$-$05 \\
\hline
340 & 17A--107 & 23.17 & 3.57 & 2.81 & 82.22 & 1.75E$-$05 \\
\hline
341 & 14A--241 & 23.23 & 2.78 & 2.93 & $-$31.9 & 3.30E$-$05 \\
\hline
351 & 14A--420\_8009 & 5.86 & 12.48 & 10.05 & 86.4 & 5.24E$-$04 \\
 & AB1073 & 8.46 & 7.82 & 7.64 & 1.7 & 3.82E$-$04 \\
 & 17A--107 & 23.17 & 3.16 & 2.80 & $-$89.5 & 1.84E$-$05 \\
\hline
370 & 17A--107 & 23.17 & 3.71 & 2.79 & $-$70.23 & 1.4495E$-$4 \\
\hline
507, 753 & AD219 & 4.86 & 4.41 & 4.30 & 81.3 & 8.90E$-$05 \\
 & Molinari1998 & 4.86 & $\sim$5 & 1.7 & \nodata & 8E$-$05 \\
 & 14A--420\_9676 & 5.80 & 1.37 & 1.04 & $-$65.5 & 6.55E$-$05 \\
 & AJ239 & 8.44 & 0.72 & 0.65 & $-$10.2 & 6.70E$-$05 \\
 & 16A--301 & 10.00 & 0.73 & 0.51 & $-$89.7 & 6.40E$-$05 \\
 & 17A--107 & 23.17 & 3.94 & 2.76 & $-$82.8 & 3.40E$-$05 \\
\hline
509 & 17A--107 & 23.17 & 3.24 & 2.77 & 88.3 & 1.73E$-$05 \\
 & 13B--210 & 23.20 & 0.44 & 0.23 & $-$81.0 & 1.09E$-$05\\
\hline
540 & 17A--107 & 23.17 & 3.12 & 2.78 & 84.96 & 1.65E$-$05 \\
\hline
608 & 17A--107 & 23.17 & 4.75 & 2.64 & $-$72.02 & 6.7E$-$05 \\
\hline
640, 675 & 12B--140 & 5.80 & 0.32 & 0.32 & 42.0 & 6.90E$-$06 \\
 & 17A--107 & 23.17 & 3.45 & 2.82 & $-$81.6 & 4.50E$-$05 \\
 & 14A--092\_7083 & 44.00 & 0.04 & 0.04 & 63.68 & 7.50E$-$05 \\
 & 14A--092\_9306 & 44.00 & 0.06 & 0.04 & 86.07 & 6.40E$-$05 \\
 & 14A--092\tablenotemark{a} & 44.00 & 0.05 & 0.04 & 86.00 & 3.866E$-$05 \\
\hline
684 & 17A--107 & 23.17 & 4.42 & 2.81 & $-$75.3 & 2.62E$-$05 \\
\hline
698, 1179 & 17A--107 & 23.17 & 3.37 & 3.03 & 49.7 & 4.10E$-$05 \\
\hline
714 & 17A--107 & 23.17 & 4.19 & 3.03 & 71.2 & 2.90E$-$05 \\
\hline
723 & 17A--107 & 23.17 & 3.27 & 2.84 & 86.8 & 1.75E$-$05 \\
\hline
725 & AM446 & 1.43 & 1.76 & 1.14 & $-$12.6 & 1.51E$-$04 \\
 & AH398 & 4.86 & 0.41 & 0.34 & $-$53.1 & 3.90E$-$05 \\
 & AH549 & 4.86 & 0.41 & 0.36 & $-$44.2 & 6.50E$-$05 \\
 & AK355 & 4.86 & 1.27 & 1.15 & $-$2.5 & 1.70E$-$04 \\
 & AM446 & 4.86 & 0.43 & 0.35 & $-$7.5 & 8.90E$-$05 \\
 & HVLA--C\tablenotemark{b} & 4.86 & 0.45 & 0.41 & $-$39.4 & 5.52E$-$05 \\
 & Urquhart2009 & 4.86 & 1.3 & 1.1 & $-$43.9 & 6E$-$04 \\
 & 14A--420\_7824 & 5.80 & 14.01 & 9.89 & $-$85.9 & 1.45E$-$03 \\
 & AC240 & 8.44 & 0.82 & 0.67 & $-$61.8 & 2.30E$-$04 \\
 & AK355 & 8.44 & 0.72 & 0.64 & 5.5 & 1.20E$-$04 \\
 & AM446 & 8.44 & 0.26 & 0.21 & 25.2 & 7.40E$-$05 \\
 & HVLA--X\tablenotemark{c} & 8.44 & 0.42 & 0.37 & 22.5 & 5.91E$-$05 \\
 & AC240 & 14.94 & 0.43 & 0.37 & $-$37.5 & 5.40E$-$04 \\
 & AK355 & 14.94 & 0.54 & 0.35 & $-$3.2 & 2.00E$-$04 \\
 & AM446 & 14.94 & 0.14 & 0.12 & 30.8 & 1.00E$-$04 \\
 & HVLA--U\tablenotemark{d} & 14.94 & 0.22 & 0.20 & 18.0 & 6.39E$-$05 \\
 & AS683 & 22.46 & 1.39 & 0.90 & 87.5 & 5.10E$-$04 \\
 & 17A--107 & 23.17 & 5.27 & 3.68 & $-$70.3 & 3.30E$-$04 \\
 & AS683 & 43.34 & 0.61 & 0.40 & $-$86.0 & 1.20E$-$03 \\
\hline
798 & 17A--107 & 23.17 & 3.85 & 2.63 & 80.0 & 1.63E$-$05 \\
 & 14A--420\_6898 & 5.80 & 15.2 & 10.0 & 80.0 & 1.72E$-$04 \\
\hline
801 & 17A--107 & 23.17 & 4.57 & 2.91 & $-$74.9 & 2.66E$-$05 \\
\hline
839 & 17A--107 & 23.17 & 3.34 & 2.77 & $-$80.7 & 1.85E$-$05 \\
\hline
892 & AB515 & 1.46 & 4.88 & 4.01 & 38.4 & 1.45E$-$03 \\
 & AB544 & 4.86 & 4.20 & 3.53 & $-$44.1 & 1.27E$-$03 \\
 & 14A--420\_6944 & 5.86 & 13.14 & 10.57 & 83.0 & 8.87E$-$04 \\ 
 & AB1073 & 8.46 & 8.00 & 7.78 & $-$4.4 & 8.22E$-$04 \\
 & 17A--107 & 23.17 & 3.93 & 2.97 & $-$88.6 & 1.02E$-$04 \\
\hline
1112 & Gibb2007 & 4.86 & 0.50 & 0.33 & $-$35 & 4.8E$-$05 \\
  & AF381 & 4.89 & 1.15 & 0.92 & $-$78.03 & 1.194E$-$04 \\
 & Shepherd2004 & 4.89 & 1.36 & 1.12 & 29.2 & 1.1E$-$04 \\
 & 12B--140 & 5.80 & 0.31 & 0.30 & 62.2 & 7.40E$-$06 \\
 & 14A--420\_9375 & 5.80 & 1.28 & 0.99 & $-$83.3 & 3.67E$-$05 \\
 & Hunter1994 & 8.44 & 0.57 & 0.44 & \nodata & 2.10E$-$04 \\
 & AS831 & 8.46 & 1.15 & 1.09 & $-$57.2 & 4.28E$-$05 \\
 & CG2010 & 8.46 & 0.22 & 0.17 & 18 & 5E$-$05 \\
 & Gibb2007 & 8.46 & 0.29 & 0.19 & $-$32 & 3.4E$-$05 \\
 & AF381 & 14.94 & 0.49 & 0.40 & $-$78.2 & 1.00E$-$04 \\
 & Shepherd2004 & 14.96 & 0.46 & 0.38 & 34.5 & 2.3E$-$04 \\
 & CG2015\tablenotemark{e} & 15.0 & 0.15 & 0.15 & \nodata & 1.10E$-$05 \\
 & Torrelles1997 & 22.28 & $\sim$0.1 & $\sim$0.1 & \nodata & 1.60E$-$04 \\
 & CG2015\tablenotemark{e} & 22.0 & 0.12 & 0.12 & \nodata & 1.00E$-$04 \\
  & 14A-241 & 23.1 & 2.84 & 2.57 & $-$47.8 & 1.43E$-$04 \\
 & Shepherd2004 & 43.34 & 0.27 & 0.20 & 89.4 & 3.1E$-$04 \\
 & Gibb2007 & 43.49 & 0.04 & 0.03 & $-$63 & 3.00E$-$04 \\
 & 14A--092 & 44.00 & 0.05 & 0.04 & 88.16 & 3.89E$-$05 \\ 
 & CG2015\tablenotemark{e} & 44.00 & 0.07 & 0.07 & \nodata & 1.20E$-$05 \\
\hline
1225 & 17A--107 & 23.17 & 4.38 & 2.93 & $-$82.1 & 4.36E$-$05 \\
 & 13A--373\_9583 & 24.37 & 3.52 & 2.42 & 83.5 & 1.16E$-$05 \\
\hline
1267, 3188 & 17A--107 & 23.17 & 3.39 & 3.03 & 52.9 & 1.90E$-$05 \\
\hline
1460 & 14A--420\_1115 & 5.86 & 1.37 & 0.93 & 81.5 & 7.17E$-$05 \\
 & AS643 & 8.46 & 2.00 & 0.66 & 63.9 & 8.56E$-$05 \\
 & 17A--107 & 23.17 & 4.92 & 2.66 & $-$71.3 & 1.00E$-$04 \\
\hline
1528 & AF381 & 8.46 & 1.51 & 1.29 & $-$88.6 & 1.20E$-$03 \\
 & 14A--420\_9375 & 5.86 & 1.37 & 0.98 & $-$82.2 & 5.00E$-$04 \\
\hline
4797 & 14A--420\_0232 & 5.80 & 13.0 & 10.08 & 85.2 & 2.37E$-$04 \\
 & 17A--107 & 23.17 & 4.3 & 2.77 & $-$74.1 & 2.42E$-$05 \\
\hline
 DR15: & 17A--107 & 23.17 & 4.06 & 2.91 & 83.53 & 2.00E$-$05 \\
 1454, 2210 & 13A--373$\_4954$ & 24.37 & 3.16 & 2.47 & 89.23 & 1.10E$-$05 \\
\hline
DR21: & AF381U & 14.94 & 0.49 & 0.39 & 84.52 & 1.12E$-$04 \\
699, 1018, & 14A--420\_9375 & 6.95 & 1.16 & 0.82 & 80.11 & 7.27E$-$05 \\
1243, 1467,  & Araya2009 & 8.46 & 0.20 & 0.19 & $-$18.2 & 1.9E$-$05 \\
1599, 5417 & Araya2009 & 22.4 & 0.25 & 0.25 & 62.1 & 3.6E$-$05 \\
 & 14A--241 & 23.23 & 2.99 & 2.67 & 16.64 & 8.20E$-$04 \\
 & 13A--373\_1665 & 23.23 & 2.71 & 1.00 & 71.2 & 5.58E$-$05\\
 & 13A--373\_7871 & 23.23 & 2.78 & 1.00 & 66.7 & 3.92E-05 \\
 & 13A--373\tablenotemark{f} & 23.23 & 2.72 & 1.00 & 69.2 & 3.88E$-$05 \\
 & 15A--059 & 25.06 & 0.27 & 0.20 & 22.49 & 8.55E$-$05 \\
 & 14B--173 & 30.90 & 0.76 & 0.58 & 82.80 & 1.85E$-$05 \\
 & 13A--315 & 43.60 & 0.69 & 0.45 & 86.30 & 1.20E$-$04 \\
\hline
S106: & AF362 & 8.46 & 0.24 & 0.20 & 23.1 & 1.20E$-$04 \\
1201 & 17A--107 & 23.17 & 3.18 & 2.81 & 72.73 & 1.19E$-$03 \\
 & AR537 & 43.34 & 0.49 & 0.43 & $-$16.0 & 1.00E$-$03 \\
\hline
\hline
\enddata
\tablecomments{The basic parameters of the cleaned maps, which are grouped by the corresponding MDCs. Column 1 gives the name of the MDC, which is covered by the maps in the same block; column 2 gives the corresponding project code of the map; column 3 gives the central frequency of the map; column 4 to 6 give the major axis, the minor axis, and the position angle of the synthesised beam; column 7 gives the RMS of the map.}
\tablenotetext{a}{Combined image of project 14A--092\_7083 and 14A--092\_9375.}
\tablenotetext{b}{Combined image of the C band data in projects AH398, AH549, AK355, and AM446.}
\tablenotetext{c}{Combined image of the X band data in projects AC240, AK355, and AM446.}
\tablenotetext{d}{Combined image of the Ku band data in projects AC240, AK355, and AM446.}
\tablenotetext{e}{Flux density is interpolated from the original data; the other values are the averages of the original data.}
\tablenotetext{f}{Combined image of project 13A--373\_1665 and 13A--373\_7871.}
\end{deluxetable*}

%% file: Table4_Detections.tex
\begin{longrotatetable}
\begin{deluxetable*}{p{0.9cm}p{0.6cm}p{1.4cm}p{1.4cm}p{0.9cm}p{1.4cm}p{1.4cm}p{0.4cm}p{0.4cm}p{0.9cm}p{0.9cm}p{0.4cm}p{0.4cm}p{1.8cm}p{0.2cm}p{0.2cm}p{0.2cm}p{0.2cm}}
\centerwidetable
\movetabledown=8mm
\tablecaption{Radio Sources\label{tab:detection}}
\tabletypesize{\scriptsize}
\tablehead{
\colhead{MDC} & \colhead{ID} & \colhead{RA} & \colhead{DEC} & \colhead{Freq.} & \colhead{$D_{maj}$} & \colhead{$D_{min}$} & \colhead{$D_{maj,D}$} & \colhead{$D_{min,D}$} & \colhead{$S_\nu$} & \colhead{$S_{\nu,p}$} & \colhead{err$_{S_{\nu,p}}$} & \colhead{RMS} & \colhead{Proj. Code} & \colhead{Method} & \colhead{Pos.} & \colhead{Assoc.} & \colhead{New}\\
\colhead{} & \colhead{} & \colhead{J2000} & \colhead{J2000} & \colhead{(GHz)} & \colhead{($''$)} & \colhead{($''$)} & \colhead{($''$)} & \colhead{($''$)} & \colhead{(mJy)} & \colhead{(mJy} & \colhead{(mJy} & \colhead{(mJy} & \colhead{} & \colhead{} & \colhead{} & \colhead{} & \colhead{}\\
\colhead{} & \colhead{} & \colhead{(h m s)} & \colhead{($^\circ$ $'$ $''$)} & \colhead{} & \colhead{} & \colhead{} & \colhead{} & \colhead{} & \colhead{} & \colhead{/Beam)} & \colhead{/Beam)} & \colhead{/Beam)} & \colhead{} & \colhead{} & \colhead{} & \colhead{} & \colhead{}
}
\startdata
247 & 1 & 20:30:27.40 & +41:15:59.52 & 23.2 & 3.41$\pm$0.23 & 2.73$\pm$0.13 & * & * & 0.896 & 0.895 & 0.04 & 0.02 & 17A--107 & G & In & No & No \\
\hline
248 & 1 & 20:36:57.65 & +42:11:29.97 & 23.2 & 3.63$\pm$0.16 & 3.27$\pm$0.15 & 1.30 & 1.02 & 5.083 & 4.512 & 0.19 & 0.06 & 17A--107 & G & In & Yes & Yes \\
\hline
274 & 1 & 20:36:07.24 & +41:39:52.67 & 4.9 & \nodata & \nodata & \nodata & \nodata & 1.1 & \nodata & \nodata & 0.1 & Miralles1994 & L & In & No & No \\
 &  & \nodata & \nodata & 5.8 & 1.55$\pm$0.12 & 1.14$\pm$0.09 & 0.83 & 0.28 & 0.796 & 0.606 & 0.04 & 0.05 & 14A--240\_1115 & G & & & \\
 &  & \nodata & \nodata & 6.2 & \nodata & \nodata & \nodata & \nodata & 1.054 & \nodata & \nodata & 0.007 & Rosero2016 & L & & & \\
 &  & \nodata & \nodata & 8.4 & 12.31$\pm$1.23 & 7.82$\pm$0.80 & 6.21 & 0.97 & 1.516 & 1.290 & 0.04 & 0.28 & AM432\tablenotemark{a} & G & & & \\
 &  & \nodata & \nodata & 10.0 & 1.02$\pm$0.10 & 0.83$\pm$0.08 & 0.67 & 0.55 & 1.087 & 0.554 & 0.03 & 0.03 & 16A--301 & G & & & \\
 &  & \nodata & \nodata & 22.5 & \nodata & \nodata & 4.3 & 1.0 & 1.3 & 0.8 & \nodata & 0.06 & Fontani2012 & L & & & \\
 &  & \nodata & \nodata & 23.2 & 4.40$\pm$0.13 & 3.04$\pm$0.010 & 1.19 & 0.50 & 1.053 & 0.982 & 0.03 & 0.04 & 17A--107 & G & & & \\
 &  & \nodata & \nodata & 23.2 & 0.64$\pm$0.05 & 0.47$\pm$0.04 & 0.48 & 0.39 & 0.881 & 0.295 & 0.01 & 0.01 & 13B--210\tablenotemark{b} & G & & & \\
 & 2 & 20:36:07.53 & +41:40:09.04 & 23.2 & 4.22$\pm$0.30 & 2.74$\pm$0.16 & ? & ? & 0.216 & 0.234 & 0.01 & 0.04 & 17A--107 & G & In & Yes & No \\
 &  & \nodata & \nodata & 23.2 & 0.44$\pm$0.04 & 0.22$\pm$0.02 & * & * & 0.114 & 0.119 & 0.01 & 0.01 & 13B--210\tablenotemark{b} & G & & & \\
\hline
310 & 1 & 20:24:31.66 & +42:04:22.71 & 23.2 & 3.70$\pm$0.09 & 3.03$\pm$0.08 & 1.48 & 1.09 & 1.662 & 1.416 & 0.03 & 0.03 & 17A--107 & G & In & Yes & No \\
 & 1-1 & 20:24:31.672 & +42:04:22.37 & 44.0 & 0.06$\pm$0.01 & 0.04$\pm$0.00 & 0.03 & 0.02 & 0.446 & 0.345 & 0.01 & 0.05 & 14A--092 & G & & & \\
 & 1-2 & 20:24:31.666 & +42:04:22.32 & 44.0 & 0.05$\pm$0.01 & 0.05$\pm$0.01 & * & * & 0.391 & 0.289 & 0.01 & 0.05 & 14A--092 & G & & & \\
 & 2 & 20:24:31.55 & +42:04:13.49 & 5.8 & 0.31$\pm$0.01 & 0.30$\pm$0.01 & * & * & 0.180 & 0.186 & 0.00 & 0.01 & 12B--140 & G & In & Yes & No \\
 &  & \nodata & \nodata & 10.0 & 0.72$\pm$0.05 & 0.54$\pm$0.03 & ? & ? & 0.271 & 0.269 & 0.01 & 0.04 & 16A--301 & G & & & \\
 &  & \nodata & \nodata & 23.2 & 3.78$\pm$0.24 & 3.00$\pm$0.23 & 1.60 & 1.11 & 0.469 & 0.395 & 0.01 & 0.03 & 17A--107 & G & & & \\
 &  & \nodata & \nodata & 44.0 & 0.05$\pm$0.00 & 0.04$\pm$0.00 & 0.02 & 0.01 & 0.802 & 0.685 & 0.03 & 0.05 & 14A--092 & G & & & \\
\hline
327 & 1 & 20:19:39.32 & +40:57:01.92 & 8.5 & 0.76$\pm$0.12 & 0.68$\pm$0.11 & ? & ? & 0.110 & 0.016 & 0.01 & 0.01 & AG625\tablenotemark{c} & G & In & No & No \\
\hline
340 & 1 & 20:32:21.04 & +41:07:54.45 & 23.2 & 4.05$\pm$0.42 & 3.06$\pm$0.25 & 2.13 & 0.73 & 0.235 & 0.190 & 0.01 & 0.02 & 17A--107 & G & Out & Yes & Yes \\
\hline
341 & 1 & 20:40:05.38 & +41:32:13.05 & 23.2 & 3.32$\pm$0.11 & 2.59$\pm$0.08 & 1.81 & 0.53 & 1.394 & 1.144 & 0.03 & 0.03 & 14A--241 & G & In & Yes & Yes \\
\hline
351 & 1 & 20:31:20.67 & +38:57:23.24 & 23.2 & 4.536$\pm$0.603 & 3.531$\pm$0.606 & 3.38 & 1.95 & 0.365 & 0.201 & 0.01 & 0.02 & 17A--107 & G & In & Yes & Yes \\
\hline
507 & 1 & 20:20:39.31 & +39:37:57.84 & 10.0 & 0.73$\pm$0.07 & 0.560$\pm$0.04 & * & * & 0.442 & 0.406 & 0.04 & 0.06 & 16A--301 & G & In & Yes & No \\
 &  & \nodata & \nodata & 23.2 & 5.6 & 4.4 & \nodata & \nodata & 1.119 & 1.037 & \nodata & 0.04 & 17A--107 & D & & & \\
 & 2 & 20:20:39.33 & +39:37:52.88 & 5.8 & 1.72$\pm$0.26 & 1.12$\pm$0.17 & 1.054 & 0.382 & 3.453 & 2.545 & 0.342 & 0.07 & 14A--420\_9676 & G & In & Yes & No \\
 &  & \nodata & \nodata & 8.4 & 1.04$\pm$0.63 & 0.76$\pm$0.51 & 0.80 & 0.25 & 1.7443 & 1.045 & 0.378 & 0.07 & AJ239 & G & & & \\
 &  & \nodata & \nodata & 10.0 & 0.84$\pm$0.08 & 0.64$\pm$0.05 & 0.52 & 0.20 & 1.055 & 0.734 & 0.06 & 0.06 & 16A--301 & G & & & \\
 & 3 & 20:20:39.27 & +39:37:50.82 & 5.8 & 1.73$\pm$0.04 & 1.49$\pm$0.03 & 1.12 & 0.99 & 25.138 & 13.924 & 0.288 & 0.07 & 14A--420\_9676 & G & In & No & No \\
 &  & \nodata & \nodata & 8.4 & 1.34$\pm$0.04 & 1.17$\pm$0.04 & 1.13 & 0.96 & 24.175 & 7.297 & 0.220 & 0.07 & AJ239 & G & & & \\
 &  & \nodata & \nodata & 10.0 & 2.45 & 4.43 & \nodata & \nodata & 20.793 & 4.988 & \nodata & 0.06 & 16A--301 & D & & & \\
\hline
509 & 1 & 20:31:12.90 & +40:03:22.46 & 23.2 & 3.67$\pm$0.38 & 3.20$\pm$0.22 & 1.75 & 1.57 & 0.487 & 0.372 & 0.02 & 0.02 & 17A--107 & G & In & Yes & No \\
 & 2 & 20:31:12.53 & +40:03:21.05 & 23.2 & 5.81$\pm$2.31 & 3.60$\pm$1.66 & 4.99 & 1.91 & 0.310 & 0.133 & 0.02 & 0.02 & 17A--107 & G & In & Yes & No \\
 & 3 & 20:31:12.04 & +40:03:12.44 & 23.2 & 6.1 & 3.2 & \nodata & \nodata & 0.146 & 0.163 & \nodata & 0.02 & 17A--107 & D & In & Yes & Yes \\
 & 4 & 20:31:13.38 & +40:03:10.91 & 23.2 & 4.46$\pm$1.31 & 3.95$\pm$0.90 & 3.49 & 2.26 & 0.292 & 0.149 & 0.01 & 0.02 & 17A--107 & G & In & Yes & No \\
 & 5 & 20:31:13.85 & +40:03:02.37 & 23.2 & 5.57$\pm$1.97 & 4.52$\pm$1.33 & 4.70 & 3.35 & 0.333 & 0.119 & 0.01 & 0.02 & 17A--107 & G & In & No & Yes \\
 & 6 & 20:31:11.20 & +40:03:09.73 & 23.2 & 4.74$\pm$0.16 & 4.40$\pm$0.15 & 3.76 & 3.09 & 3.647 & 1.568 & 0.04 & 0.02 & 17A--107 & G & Out & No & No \\
 & 7 & 20:31:10.31 & +40:03:16.46 & 23.2 & 3.52$\pm$0.09 & 2.77$\pm$0.07 & 1.39 & 0.12 & 0.810 & 0.744 & 0.02 & 0.02 & 17A--107 & G & Out & No & No \\
\hline
540 & 1 & 20:32:38.24 & +38:46:09.41 & 23.2 & 3.27$\pm$0.22 & 2.75$\pm$0.16 & * & * & 0.267 & 0.232 & 0.01 & 0.01 & 17A--107 & G & Out & No & Yes \\
 & 2 & 20:32:40.98 & +38:46:56.43 & 23.2 & 4.59$\pm$0.51 & 2.75$\pm$0.39 & * & * & 0.201 & 0.128 & 0.01 & 0.01 & 17A--107 & G & Out & No & Yes \\
 & 3 & 20:32:42.09 & +38:46:08.94 & 23.2 & 3.21$\pm$0.22 & 2.48$\pm$0.14 & ? & ? & 0.134 & 0.164 & 0.01 & 0.01 & 17A--107 & G & Out & Yes & Yes \\
\hline
675 & 1 & 20:20:30.60 & +41:21:26.27 & 5.8 & 0.36$\pm$0.02 & 0.31$\pm$0.02 & * & * & 0.139 & 0.123 & 0.01 & 0.01 & 12B--140 & G & In & Yes & Yes \\
 &  & \nodata & \nodata & 23.2 & 3.61$\pm$0.10 & 2.92$\pm$0.08 & 1.10 & 0.67 & 3.995 & 3.698 & 0.08 & 0.05 & 17A--107 & G & & & \\
 &  & \nodata & \nodata & 44.0 & 0.06$\pm$0.01 & 0.05$\pm$0.01 & 0.04 & 0.03 & 1.758 & 1.060 & 0.04 & 0.04 & 14A--092 & G & & & \\
\hline
684 & 1 & 20:40:33.49 & 41:59:01.28 & 23.3 & 4.36$\pm$0.48 & 4.36$\pm$0.48 & * & * & 0.234 & 0.219 & \nodata & 0.03 & 17A--107 & G & In & Yes & Yes \\
\hline
698 & 1 & 20:39:16.72 & +42:16:08.95 & 23.2 & 3.65$\pm$0.16 & 2.98$\pm$0.12 & * & * & 2.259 & 2.119 & 0.08 & 0.04 & 17A--107 & G & In & Yes & Yes \\
\hline
699\tablenotemark{d} & 1 & 20:39:02.24 & +42:21:59.60 & 7.0 & 2.2 & 1.6 & \nodata & \nodata & 4.145 & 3.473 & \nodata & 0.07 & 14A--420\_9375 & D & In & No & Yes \\
 &  & \nodata & \nodata & 14.9 & 1.7 & 0.9 & \nodata & \nodata & 5.675 & 1.114 & \nodata & 0.11 & AF381U & D & & & \\
 &  & \nodata & \nodata & 30.9 & 2.5 & 2.5 & \nodata & \nodata & 8.653 & 7.409 & \nodata & 0.02 & 14B--173 & D & & & \\
\hline
714 & 1 & 20:37:00.96 & +41:34:55.80 & 23.1 & 4.26$\pm$0.27 & 2.90$\pm$0.45 & * & * & 0.937 & 0.965 & 0.11 & 0.03 & 17A--107 & G & In & Yes & Yes \\
 & 2 & 20:37:00.87 & +41:34:59.47 & 23.1 & 4.43$\pm$0.70 & 3.20$\pm$0.89 & 1.47 & 1.00 & 0.335 & 0.300 & 0.03 & 0.03 & 17A--107 & G & In & Yes & Yes \\
 & 3 & 20:37:01.10 & +41:34:53.06 & 23.2 & 6.66$\pm$9.57 & 2.02$\pm$0.85 & * & * & 0.159 & 0.150 & 0.15 & 0.03 & 17A--107 & G & In & Yes & Yes \\
\hline
723 & 1 & 20:29:58.30 & +40:15:57.99 & 23.2 & 1.2 & 0.9  & \nodata & \nodata & 0.098 & 0.094 & \nodata & 0.02 & 17A--107 & D & In & Yes & Yes \\
 & 2 & 20:29:57.57 & +40:15:33.28 & 23.2 & 1.0 & 0.9 & \nodata & \nodata & 0.088 & 0.101 & \nodata & 0.02 & 17A--107 & D & In & No & Yes \\
\hline
725 & 1 & 20:36:52.18 & +41:36:24.33 & 1.4 & 1.98$\pm$0.03 & 1.81$\pm$0.03 & 1.44 & 0.84 & 28.364 & 15.961 & 0.24 & 0.15 & AM446 & G & In & Yes & No \\
 &  & \nodata & \nodata & 4.9 & 1.1 & 0.6 & \nodata & \nodata & 50.8 & 33.6 & \nodata & 0.6 & Urquhart2009 & L & & & \\
 &  & \nodata & \nodata & 4.9 & 1.9 & 1.3 & \nodata & \nodata & 60.914 & 18.826 & \nodata & 0.06 & HVLA--C & D & & & \\
 &  & \nodata & \nodata & 5.8 & 16.25$\pm$0.46 & 9.75$\pm$0.16 & * & * & 54.474 & 55.236 & 1.2 & 1.45 & 14A--420\_7824 & G & & & \\
 &  & \nodata & \nodata & 8.4 & 2.1 & 1.3 & \nodata & \nodata & 61.928 & 24.774 & \nodata & 0.06 & HVLA--X & D & & & \\
 &  & \nodata & \nodata & 14.9 & 1.5 & 1.0 & \nodata & \nodata & 45.842 & 14.218 & \nodata & 0.06 & HVLA--U & D & & & \\
 &  & \nodata & \nodata & 22.5 & 1.513$\pm$0.04 & 0.982$\pm$0.02 & 0.59 & 0.40 & 75.114 & 63.170& 2.5 & 0.46 & AS683 & G & & & \\
 &  & \nodata & \nodata & 23.2 & 5.27$\pm$0.17 & 3.71$\pm$0.13 & * & * & 77.993 & 77.307 & 2.50 & 0.33 & 17A--107 & G & & & \\
 &  & \nodata & \nodata & 43.3 & 1.0 & 0.8 & \nodata & \nodata & 68.417 & 48.406 & \nodata & 0.46 & AS683 & D & & & \\
 & 2 & 20:36:52.92 & +41:36:17.37 & 23.2 & 6.27$\pm$3.19 & 3.80$\pm$1.42 & * & * & 9.379 & 7.638 & 2.56 & 0.33 & 17A--107 & G & In & Yes & Yes \\
\hline
742 & 1 & 20:19:38.85 & +40:56:36.66 & 8.5 & 0.7 & 0.7 & \nodata & \nodata & 0.047 & 0.068 & \nodata & 0.01 & AG625 & D & In & Yes & Yes \\
 &  & \nodata & \nodata  & 22.0 & 0.10$\pm$0.01 & 0.07$\pm$0.01 & ? & ? & 0.214 & 0.210 & 0.02 & 0.02 & 14A--481 & G & & & \\
 &  & \nodata & \nodata & 22.0 & \nodata & \nodata & \nodata & \nodata & 0.21 & 0.24 & 0.02 & 0.03 & Masque2017 & L & & & \\
 & 2 & 20:19:39.23 & +40:56:36.55 & 4.9 & 6.18$\pm$0.27 & 5.89$\pm$0.25 & 4.38 & 3.95 & 65.218 & 34.027 & 1.5 & 0.17 & AD219 & G & In & Yes & No \\
 &  & \nodata & \nodata & 4.9 & 10.5 & 8.7 & \nodata & \nodata & 75.853 & 6.500 & \nodata & 0.12 & AH726 & D & & & \\
 &  & \nodata & \nodata & 8.4 & 0.92 & 0.084 & \nodata & \nodata & 82.200 & 3.700 & \nodata & 0.16 & Kurtz1994 & L & & & \\
 &  & \nodata & \nodata & 8.5 & 10.6 & 9.3 & \nodata & \nodata & 64.064 & 2.120 & \nodata & 0.01 & AG625\tablenotemark{c} & D & & & \\
 &  & \nodata & \nodata & 8.5 & 9.34$\pm$0.11 & 8.84$\pm$0.10 & 4.68 & 4.59 & 67.790 & 50.116 & 0.59 & 0.09 & AK450 & G & & & \\
 &  & \nodata & \nodata & 8.5 & 11.2 & 9.4 & \nodata & \nodata & 67.675 & 15.530 & \nodata & 0.10 & AK477 & D & & & \\
 &  & \nodata & \nodata & 15.0 & 0.7 & 0.7 & \nodata & \nodata & 35.6 & 2.1 & \nodata & 0.11 & Kurtz1994 & L & & & \\
 &  & \nodata & \nodata & 23.2 & 16.3 & 11.6 & \nodata & \nodata & 61.567 & 20.240 & \nodata & 0.05 & 17A--107 & D & & & \\
 & 3 & 20:19:39.60 & +40:56:31.92 & 8.5 &0.73$\pm$0.03 & 0.65$\pm$0.03 & * & * & 0.377 & 0.422 & 0.01 & 0.01 & AG625 & G & In & No & No \\
 &  &  &  & 22.0 & 0.13$\pm$0.01 & 0.08$\pm$0.01 & * & * & 0.185 & 0.149 & 0.01 & 0.02 & 14A--481 & G & & & \\
 &  & \nodata & \nodata & 22.0 & \nodata & \nodata & \nodata & \nodata & 0.24 & 0.14 & 0.02 & 0.03 & Masque2017 & L & & & \\
\hline
753 & 1 & 20:20:39.49 & +39:38:12.38 & 4.9 & 4.71$\pm$0.16 & 4.34$\pm$0.15 & 1.61 & 0.64 & 3.218 & 2.989 & 0.154 & 0.09 & AD219 & G & Out & Yes & No \\
 &  & \nodata & \nodata & 4.9 & 5.04 & 1.79 & 1.16 & 0.64 & 3.82 & 2.86 & \nodata & 0.08 & Molinari1998 & L & & & \\
 &  & \nodata & \nodata & 5.8 & 3.5 & 1.6 & \nodata & \nodata & 1.583 & 0.636 & \nodata & 0.07 & 14A--420\_9676 & D & & & \\
 &  & \nodata & \nodata & 8.4 & 2.7 & 1.4 & \nodata & \nodata & 2.953 & 1.615 & \nodata & 0.07 & AJ239 & D & & & \\
 &  & \nodata & \nodata & 23.2 & 3.89$\pm$0.19 &3.05$\pm$0.12 & * & * & 1.162 & 1.064 & 0.05 & 0.04 & 17A--107 & G & & & \\
\hline
798 & 1 & 20:20:44.59 & +39:35:21.16 & 23.2 & 3.95$\pm$0.41 & 2.42$\pm$0.15 & ? & ? & 0.064 & 0.068 & 0.01 & 0.02 & 17A--107 & G & In & Yes & Yes \\
\hline
801 & 1 & 20:40:28.35 & +41:57:16.02 & 23.2 & 4.82$\pm$0.54 & 3.00$\pm$0.25 & * & * & 0.139 & 0.128 & 0.01 & 0.03 & 17A--107 & G & In & Yes & Yes \\
\hline
839 & 1 & 20:24:14.26 & +42:11:42.95 & 23.2 & 3.49$\pm$0.19 & 2.79$\pm$0.12 & 1.6 & 0.85 & 0.436 & 0.415 & 0.01 & 0.02 & 17A--107 & G & In & Yes & Yes \\
\hline
1112 & 1 & 20:38:36.44 & +42:37:34.82 & 4.9 & 0.50$\pm$0.02 & 0.37$\pm$0.01 & 0.17$\pm$0.03 & 0.08$\pm$0.03 & 1.7 & 1.4 & 0.1 & 0.05 & Gibb2007 & L & In & Yes & No \\
 &  & \nodata & \nodata & 4.9 & \nodata & \nodata & \nodata & \nodata & 5.3 & 3.5 & \nodata & 0.11 & Shepherd2004 & L & & & \\
 &  & \nodata & \nodata & 5.8 & 1.6 & 0.8 & \nodata & \nodata & 4.630 & 2.524 & \nodata & 0.01 & 12B--140 & D & & & \\
 &  & \nodata & \nodata & 5.8 & 1.38$\pm$0.04 & 1.05$\pm$0.06 & 0.60 & 0.20 & 3.100 & 2.696 & 0.10 & 0.04 & 14A--420\_9375 & G & & & \\
 &  & \nodata & \nodata & 8.5 & \nodata & \nodata & 0.43$\pm$0.02 & 0.17$\pm$0.02 & 3.8 &\nodata & \nodata & 0.05 & CG2010 & L & & & \\
 &  & \nodata & \nodata & 8.5 & 0.30$\pm$0.01 & 0.31$\pm$0.01 & 0.12$\pm$0.03 & 0.05$\pm$0.05 & 0.7 & 0.7 & 0.1 & 0.03 & Gibb2007 & L & & & \\
 &  & \nodata & \nodata & 14.9 & 0.74$\pm$0.12 & 0.46$\pm$0.06 & 0.57 & 0.181 & 2.779 & 1.599 & 0.11 & 0.10 & AF381 & G & & & \\
 &  & \nodata & \nodata & 15.0 & \nodata & \nodata & \nodata & \nodata & 4.0 & 1.5 & \nodata & 0.23 & Shepherd2004 & L & & & \\
 &  & \nodata & \nodata & 22.3 & \nodata & \nodata & 0.43 & 0.12 & 7.8 & 1.42 & \nodata & 0.16 & Torrelles1997 & L & & & \\
 &  & \nodata & \nodata & 44.0 & 0.2 & 0.1 & \nodata & \nodata & 4.071 & 1.583 & \nodata & 0.04 & 14A--092 & D & & & \\
 & 2 & 20:38:36.48 & +42:37:34.03 & 5.8 & 0.52$\pm$1.12 & 0.34$\pm$0.75 & 0.42 & 0.17 & 1.206 & 0.624 & 0.31 & 0.01 & 12B--140 & G & In & Yes & No \\
 &  & \nodata & \nodata & 8.5 & \nodata & \nodata & \nodata & \nodata & 0.7 & \nodata & \nodata & 0.05 & CG2010 & L & & & \\
 &  & \nodata & \nodata & 14.9 & 0.64$\pm$0.34 & 0.47$\pm$0.14 & 0.43 & 0.23 & 1.725 & 1.106 & 0.13 & 0.10 & AF381 & G & & & \\
 &  & \nodata & \nodata & 15.0 & \nodata & \nodata & \nodata & \nodata & 0.87 & \nodata & \nodata & 0.01 & CG2015 & L & & & \\
 &  & \nodata & \nodata & 15.0 & \nodata & \nodata & \nodata & \nodata & 1.5 & 1.2 & \nodata & 0.23 & Shepherd2004 & L & & & \\
 &  & \nodata & \nodata & 22.0 & \nodata & \nodata & \nodata & \nodata & 1.08 & \nodata & \nodata & 0.10 & CG2015 & L & & & \\
 &  & \nodata & \nodata & 22.3 & \nodata & \nodata & $\leqslant$0.1 & $\leqslant$0.1 & 1.6 & 1.40 & \nodata & 0.16 & Torrelles1997 & L & & & \\
 &  & \nodata & \nodata & 43.3 & \nodata & \nodata & \nodata & \nodata & 2.6 & 2.2 & \nodata & 0.31 & Shepherd2004 & L & & & \\
 &  & \nodata & \nodata & 44.0 & \nodata & \nodata & \nodata & \nodata & 1.59 & \nodata & \nodata & 0.01 & CG2015 & L & & & \\
 &  & \nodata & \nodata & 44.0 & 0.1 & 0.1 & \nodata & \nodata & 1.575 & 1.482 & \nodata & 0.04 & 14A--092 & D & & & \\
 & 3 & 20:38:36.48 & +42:37:33.42 & 4.9 & 0.30$\pm$0.10 & 0.36$\pm$0.03 & 0.03$\pm$0.01 & 0.10$\pm$0.10 & 0.9 & 0.7 & 0.1 & 0.05 & Gibb2007 & L & In & Yes & No \\
 &  & \nodata & \nodata & 4.9 & \nodata & \nodata & \nodata & \nodata & 2.7 & 2.7 & \nodata & 0.11 & Shepherd2004 & L & & & \\
 &  & \nodata & \nodata & 5.8 & 0.36$\pm$0.11 & 0.31$\pm$0.07 & 0.20 & 0.05 & 1.883 & 1.542 & 0.24 & 0.01 & 12B--140 & G & & & \\
 &  & \nodata & \nodata & 8.4 & 0.60$\pm$0.10 & 0.50$\pm$0.01 & \nodata & \nodata & 2.70 & 2.29 & \nodata & 0.21 & Hunter1994 & L & & & \\
 &  & \nodata & \nodata & 8.5 & \nodata & \nodata & 0.21$\pm$0.01 & 0.07$\pm$0.01 & 4.0 & \nodata & \nodata & 0.05 & CG2010 & L & & & \\
 &  & \nodata & \nodata & 8.5 & 0.32$\pm$0.01 & 0.20$\pm$0.01 & 0.13$\pm$0.01 & 0.06$\pm$0.01 & 2.0 & 1.6 & 0.1 & 0.03 & Gibb2007 & L & & & \\
 &  & \nodata & \nodata & 14.9 & 0.49$\pm$0.01 & 0.43$\pm$0.02 & \nodata & \nodata & 5.341 & 4.918 & 0.11 & 0.10 & AF381 & G & & & \\
 &  & \nodata & \nodata & 15.0 & \nodata & \nodata & \nodata & \nodata & 5.8 & 4.5 & \nodata & 0.23 & Shepherd2004 & L & & & \\
 &  & \nodata & \nodata & 22.3 & \nodata & \nodata & 0.09 & 0.04 & 11.9 & 7.59 & \nodata & 0.16 & Torrelles1997 & L & & & \\
 &  & \nodata & \nodata & 43.3 & \nodata & \nodata & \nodata & \nodata & 5.7 & 5.4 & \nodata & 0.31 & Shepherd2004 & L & & & \\
 &  & \nodata & \nodata & 43.5 & 0.05$\pm$0.00 & 0.04$\pm$0.00 & 0.03$\pm$0.01 & 0.02$\pm$0.01 & 5.2 & 3.1 & 0.3 & 0.30 & Gibb2007 & L & & & \\
 &  & \nodata & \nodata & 44.0 & 0.2 & 0.1 & \nodata & \nodata & 15.943 & 8.236 & \nodata & 0.04 & 14A--092 & D & & & \\
 & 4 & 20:38:36.57 & +42:37:31.47 & 5.8 & 0.39$\pm$0.04 & 0.36$\pm$0.02 & 0.24 & 0.20 & 1.811 & 1.197 & 0.04 & 0.01 & 12B--140 & G & In & No & No \\
 & 5 & 20:38:36.53 & +42:37:31.39 & 5.8 & 0.41$\pm$0.12 & 0.34$\pm$0.12 & 0.27 & 0.16 & 1.288 & 0.852 & 0.03 & 0.01 & 12B--140 & G & In & No & No \\
 & 6 & 20:38:36.54 & +42:37:29.89 & 5.8 & 0.340$\pm$0.006 & 0.332$\pm$0.006 & 0.146 & 0.139 & 0.620 & 0.508 & 0.01 & 0.01 & 12B--140 & G & In & No & No \\
 &  & \nodata & \nodata & 5.8 & 1.34$\pm$0.50 & 1.00$\pm$0.19 & * & * & 0.418 & 0.558 & \nodata & 0.04 & 14A--420\_9375 & G & & & \\
 &  & \nodata & \nodata & 5.8 & 1.29$\pm$2.02 & 0.80$\pm$0.77 & ? & ? & 0.673 & 0.966 & 0.40 & 0.27 & AH869 & G & & & \\
 &  & \nodata & \nodata & 8.5 & \nodata & \nodata & \nodata & \nodata & 0.7 & \nodata & \nodata & 0.05 & CG2010 & L & & & \\
 &  & \nodata & \nodata & 14.9 & 0.70$\pm$0.19 & 0.47$\pm$0.18 & 0.50 & 0.24 & 0.795 & 0.474 & 0.03 & 0.10 & AF381 & G & & & \\
 & 4 5\tablenotemark{e} & \nodata & \nodata & 4.9 & \nodata & \nodata & \nodata & \nodata & 0.600 & 0.500 & \nodata & 0.05 & Gibb2007 & L & In & No & No \\
 &  & \nodata & \nodata & 4.9 & \nodata & \nodata & \nodata & \nodata & 4.4 & 2.8 & \nodata & 0.11 & Shepherd2004 & L & & & \\
 &  & \nodata & \nodata & 5.8 & 1.65$\pm$0.04 & 1.51$\pm$0.04 & 0.86 & 0.54 & 3.818 & 3.000 & 0.08 & 0.08 & AF381 & G & & & \\
 &  & \nodata & \nodata & 5.8 & 1.48$\pm$0.03 & 1.13$\pm$0.03 & 0.75 & 0.56 & 2.054 & 1.547 & 0.02 & 0.04 & 14A--420\_9375 & G & & & \\
 &  & \nodata & \nodata & 5.8 & 1.65$\pm$0.16 & 1.41$\pm$0.15 & 1.03 & 0.82 & 4.632 & 2.959 & 0.21 & 0.27 & AH869 & G & & & \\
 &  & \nodata & \nodata & 8.4 & \nodata & \nodata & \nodata & \nodata & 0.85 & 0.85 & \nodata & 0.21 & Hunter1994 & L & & & \\
 &  & \nodata & \nodata & 8.5 & \nodata & \nodata & 0.75$\pm$0.05 & 0.23$\pm$0.03 & 3.3 & \nodata & \nodata & 0.05 & CG2010 & L & & & \\
 &  & \nodata & \nodata & 8.5 & \nodata & \nodata & \nodata & \nodata & 0.500 & 0.400 & \nodata & 0.03 & Gibb2007 & L & & & \\
 &  & \nodata & \nodata & 14.9 & \nodata & \nodata & \nodata & \nodata & 3.0 & \nodata & \nodata & 0.08 & CG2010 & L & & & \\
 &  & \nodata & \nodata & 15.0 & \nodata & \nodata & \nodata & \nodata & 1.7 & 1.2 & \nodata & 0.23 & Shepherd2004 & L & & & \\
\hline
1201 & 1 & 20:27:26.78 & +37:22:47.81 & 8.5 & 0.25$\pm$0.01 & 0.25$\pm$0.01 & 0.15 & 0.026 & 10.240 & 8.181 & 0.14 & 0.13 & AF362 & G & In & Yes & No \\
 &  & \nodata & \nodata & 43.3 & 0.53$\pm$0.02 & 0.42$\pm$0.02 & * & * & 33.601 & 31.643 & 0.84 & 1.14 & AR537 & G & & & \\
\hline
1225 & 1 & 20:31:58.14 & +40:18:36.05 & 24.4 & 3.72$\pm$0.22 & 2.78$\pm$0.16 & 1.41 & 1.12 & 0.188 & 0.155 & 0.00 & 0.01 & 13A--373\_9583 & G & In & Yes & Yes \\
 & 2 & 20:31:57.55 & +40:18:32.02 & 23.2 & 4.41$\pm$0.74 & 2.94$\pm$0.44 & * & * & 0.424 & 0.420 & 0.03 & 0.04 & 17A--107 & G & In & Yes & Yes \\
 &  & \nodata & \nodata & 24.4 & 3.66$\pm$0.86 & 3.49$\pm$1.46 & * & * & 0.096 & 0.064 & 0.00 & 0.01 & 13A--373\_9583 & G & & & \\
\hline
1243 & 1 & 20:39:01.99 & +42:24:59.06 & 7.0 & 1.35$\pm$0.30 & 0.94$\pm$0.18 & ? & ? & 0.688 & 0.516 & 0.04 & 0.07 & 14A--420\_9375 & G & In & Yes & Yes \\
 &  & \nodata & \nodata & 23.8 & 3.30$\pm$0.25 & 1.18$\pm$0.10 & 1.77 & 0.61 & 0.742 & 0.536 & 0.02 & 0.04 & 13A--373 & G & & & \\
\hline
1267 & 1 & 20:38:03.23 & +42:40:04.07 & 23.2 & 4.06$\pm$0.21 & 2.96$\pm$0.16 & * & * & 0.430 & 0.367 & 0.01 & 0.02 & 17A--107 & G & In & Yes & Yes \\
 & 2 & 20:38:03.90 & +42:39:31.40 & 23.2 & 3.96$\pm$0.27 & 3.11$\pm$0.28 & 2.14 & 0.35 & 0.212 & 0.177 & 0.01 & 0.02 & 17A--107 & G & Out & Yes & Yes \\
\hline
1454 & 1 & 20:32:22.10 & +40:20:16.94 & 23.2 & 4.82$\pm$0.31 & 3.32$\pm$0.20 & 2.77 & 1.27 & 0.375 & 0.277 & 0.01 & 0.02 & 17A--107 & G & In & Yes & Yes \\
 &  & \nodata & \nodata & 24.4 & 3.39$\pm$0.41 & 2.57$\pm$0.20 & 1.65 & 0.62 & 0.191 & 0.162 & 0.01 & 0.01 & 13A--373\_4954 & G & & & \\
 & 2 & 20:32:21.47 & +40:20:14.56 & 23.2 & 4.31$\pm$0.31 & 3.04$\pm$0.21 & 1.65 & 0.42 & 0.235 & 0.211 & 0.01 & 0.02 & 17A--107 & G & In & Yes & Yes \\
 &  & \nodata & \nodata & 24.4 & 3.74$\pm$0.42 & 2.42$\pm$0.32 & * & * & 0.189 & 0.155 & 0.01 & 0.01 & 13A--373\_4954 & G & & & \\
 & 3 & 20:32:22.10 & +40:20:10.23 & 23.2 & 5.09$\pm$0.63 & 3.75$\pm$0.56 & 3.12 & 2.30 & 0.402 & 0.249 & 0.02 & 0.02 & 17A--107 & G & In & Yes & Yes \\
 &  & \nodata & \nodata & 24.4 & 3.39$\pm$0.13 & 2.61$\pm$0.12 & 1.64 & 0.76 & 0.237 & 0.198 & 0.00 & 0.01 & 13A--373\_4954 & G & & & \\
 & 4 & 20:32:20.62 & +40:19:50.09 & 24.4 & 2.81$\pm$0.19 & 2.40$\pm$0.17 & ? & ? & 0.061 & 0.070 & 0.01 & 0.01 & 13A--373\_4954 & G & In & No & Yes \\
 & 5 & 20:32:21.16 & +40:20:25.68 & 24.4 & 3.43$\pm$0.22 & 1.70$\pm$0.05 & ? & ? & 0.046 & 0.058 & 0.01 & 0.01 & 13A--373\_4954 & G & In & No & Yes \\
\hline
1467 & 1 & 20:39:01.00 & +42:22:48.80 & 7.0 & 1.70$\pm$0.370 & 0.90$\pm$0.19 & 1.25 & 0.30 & 4.450 & 2.787 & 0.39 & 0.07 & 14A--420\_9375 & G & Out & Yes & No \\
 &  & \nodata & \nodata & 23.2 & 3.04$\pm$0.27 & 2.76$\pm$0.27 & * & * & 6.577 & 6.269 & 0.24 & 1.09 & 14A--241 & G & & & \\
 &  & \nodata & \nodata & 30.9 & 1.00$\pm$0.03 & 0.64$\pm$0.02 & 0.69 & 0.12 & 10.780 & 7.476 & 0.19 & 0.02 & 14B--173 & G & & & \\
 &  & \nodata & \nodata & 43.6 & 3.8 & 1.8 & \nodata & \nodata & 14.233 & 2.777 & \nodata & 0.13 & 13A--315 & D & & & \\
 & 1-1\tablenotemark{f} & 20:39:00.975 & +42:22:48.937 & 14.9 & 0.50$\pm$0.03 & 0.41$\pm$0.03 & * & * & 2.670 & 2.499 & 0.17 & 0.11 & AF381U & G & & & \\
 &  & \nodata & \nodata & 25.1 & 0.29$\pm$0.01 & 0.20$\pm$0.00 & 0.11 & 0.03 & 3.714 & 3.380 & 0.03 & 0.09 & 15A--059 & G & & & \\
 & 1-2 & 20:39:01.006 & +42:22:48.622 & 14.9 & 0.49$\pm$0.02 & 0.43$\pm$0.04 & * & * & 2.799 & 2.561 & 0.14 & 0.11 & AF381U & G & & & \\
 &  & \nodata & \nodata & 25.1 & 0.29$\pm$0.01 & 0.21$\pm$0.01 & 0.11 & 0.07 & 2.925 & 2.542 & 0.04 & 0.09 & 15A--059 & G & & & \\
& 2 & 20:39:00.612 & +42:22:43.368 & 8.5 & \nodata & \nodata & 0.22 & 0.05 & 0.226 & 0.17 & 0.01 & 0.02 & Araya2009 & L & Out & Yes & No \\
&  & & & 22.4 & \nodata & \nodata & 0.34 & $<$0.08 & 0.20 & 0.14 & 0.01 & 0.02 & Araya2009 & L & & & \\
 &  &  &  & 30.9 & 0.72$\pm$0.03 & 0.56$\pm$0.02 & ? & ? & 0.255 & 0.277 & 0.01 & 0.02 & 14B--173 & G & & & \\
\hline
2210 & 1 & 20:32:23.05 & +40:19:23.02 & 23.2 & 7.39$\pm$2.30 & 6.48$\pm$2.48 & 6.30 & 5.66 & 0.780 & 0.192 & 0.01 & 0.02 & 17A--107 & G & In & Yes & Yes \\
 &  & \nodata & \nodata & 24.4 & 4.34$\pm$0.48 & 3.26$\pm$0.31 & 3.30 & 1.86 & 0.216 & 0.113 & 0.00 & 0.01 & 13A--373\_4954 & G & & & \\
\hline
3188 & 1 & 20:38:02.21 & +42:39:48.79 & 23.2 & 3.49$\pm$0.79 & 3.20$\pm$0.66 & 1.05 & 0.80 & 0.176 & 0.162 & 0.01 & 0.02 & 17A--107 & G & In & No & Yes \\
\hline
4797 & 1 & 20:40:33.90 & +41:51:04.05 & 23.3 & 6.26$\pm$1.38 & 3.73$\pm$0.77 & 4.63 & 2.35 & 0.318 & 0.162 & 0.01 & 0.02 & 17A--107 & G & In & No & Yes \\
\hline
5417 & 1 & 20:39:00.36 & +42:24:37.15 & 7.0 & 1.35$\pm$0.10 & 0.87$\pm$0.06 & 0.72 & 0.18 & 2.122 & 1.726 & 0.05 & 0.07 & 14A--420\_9375 & G & In & Yes & Yes \\
 &  & \nodata & \nodata & 23.8 & 3.13$\pm$0.13 & 1.27$\pm$0.05 & 1.45 & 0.77 & 2.112 & 1.488 & 0.04 & 0.04 & 13A--373 & G & & & \\
\hline
\enddata
\tablecomments{The basic properties of the radio sources associated with the dust condensations. Column 1 gives the MDC names according to Cao21.
 Column 2 gives the ID of the radio sources detected in the MDC.
 Columns 3 and 4 give the coordinates of the radio sources.
 Column 5 gives the frequency of the map.
 Columns 6 and 7 give the convolved major-axis and minor-axis lengths of the sources;
 columns 8 and 9 give the sources' deconvolved major-beam sizes and minor-beam sizes.
 The fitting errors are appended to the values. A question mark (``?'') means the source has failed to be deconvolved by the Gaussian fitting task, \texttt{IMFIT};
 an asterisk (*) means it is a point source.
 Column 10 gives the integrated flux densities.
 Column 11 and 12 give the peak flux densities and the corresponding uncertainties. 
 The fitting uncertainties and beam sizes are only provided by Gaussian fitting or are obtained from the literature. Direct measurement provides total flux densities, peak flux densities, equivalent sizes of convolved major- and minor-axis only.
 Column 13 gives the RMS noises of the maps.
 Column 14 gives the corresponding VLA project code of the map.
 Column 15 gives how the parameters are measured: ``G/D/L'' means that the measurement is done by Gaussian fitting (``G''), directly measured within selected contour levels (``D''), or obtained from the literature (``L''), respectively.
 Column 16 gives the relative position between a radio source and the MDC: ``In'' or ``out'' means that the radio source is in or out of the FWHM area of an MDC, respectively.
 Column 17 gives whether a radio source is associated with a dust condensation.
 Column 18 gives whether a radio source is firstly reported by this work.}
\tablenotetext{a}{AM432 is the data used in \citet{1999RMxAA..35...97C}. Their measured flux density is 1.1$\pm$0.1 mJy.}
\tablenotetext{b}{13B--210 is the data used in \citet{2016ApJS..227...25R}. The K band flux density of 274--r1 (20343+4129 A in Rosero 2016) is about 0.917 mJy in their work. Only peak intensity of  274--r2 (20343+4129 B in Rosero 2016) is provided in Rosero 2016. We also find it very weak.}
\tablenotetext{c}{Project AG625 is the data used in \citet{2010RMxAA..46..253N}. Their measured flux densities of 327--r1 (VLA 3), 742--r2 (VLA 1), and 742--r3 (VLA 4) are 0.49$\pm$0.04, 0.57$\pm$0.05, and 0.09$\pm$0.03, respectively.}
\tablenotetext{d}{The radio source is on the edge of the maps and is severely affected by the noises after beam correction. this may cause large uncertainties in the measured flux densities.}
\tablenotetext{e}{The radio sources are unresolved in the map. Also, 1112--r4 and r5 are likely to be one radio HH-object \citep{2020MNRAS.496.3128R}.}
\tablenotetext{f}{The radio source is resolved by the observations in Q band A array. The resolved structure is ~0.001 pc and is much smaller than the focused spacial scale in this work. Here we take them as substructures instead of individual sources.}
\end{deluxetable*}
\end{longrotatetable}

%% file: Individual_MDCs.tex
\section{Radio Properties of Individual MDCs} \label{sec:rprop}

\subsection*{MDC 214, 247}\label{214_274}

MDC 214 and 247 are located in the star-forming region IRAS 20286$+$4105. The condensations of the two cores linearly distribute along the inner side of a large-scale, shell-like \ion{H}{2} region. We detect a compact radio source, 247-r1, in MDC 247. It is not associated with any dust condensation and has a 5$''$.8 offset with the closest one. \citet{2005ApJ...625..864Z, 2013ApJS..208...11L, 2015MNRAS.453..645M} have reported it as an UC \ion{H}{2} region with outflow detection; \citet{2017MNRAS.465.4753R} has also reported this source and obtained a negative spectral index of -0.5$\pm$0.05, and interpreted it as a combination of synchrotron radiation and thermal free-free emission.

\subsection*{MDC 220}\label{220}

We detected no associated compact radio source in MDC 220. A large-scale shell-like radio source is found to be surrounding the dense core. Based on the infrared data provided by Cao19, the infrared counterpart of the large-scale radio emission is slightly shifted east, indicating it is polycyclic aromatic hydrocarbons(PAHs) excited by an external radiation source(s). It is unclear whether MDC 220 and the PAHs are physically related or just visually overlapped.

\subsection*{MDC 248}\label{248}

We report a newly detected radio source, 248-r1, in MDC 248. The source is bright and is associated with the central dust condensation of MDC 248. It is slightly resolved to be extended to the northwest. Limited by the resolution, it remains unclear whether the extended part of the emission is from 248-r1 or has an independent origin.

\subsection*{MDC 274}\label{274}

MDC 274 is commonly referred to as IRAS 20343+4129 \citep{2002ApJ...566..931S}. An extended radio component is located in its southwest, which belongs to an arch-like radio source \citep{1999RMxAA..35...97C, 2002ApJ...576..313K}. The relation between the MDC and the extended source remains unknown.

\begin{figure*}[htpb]
\gridline{
	\fig{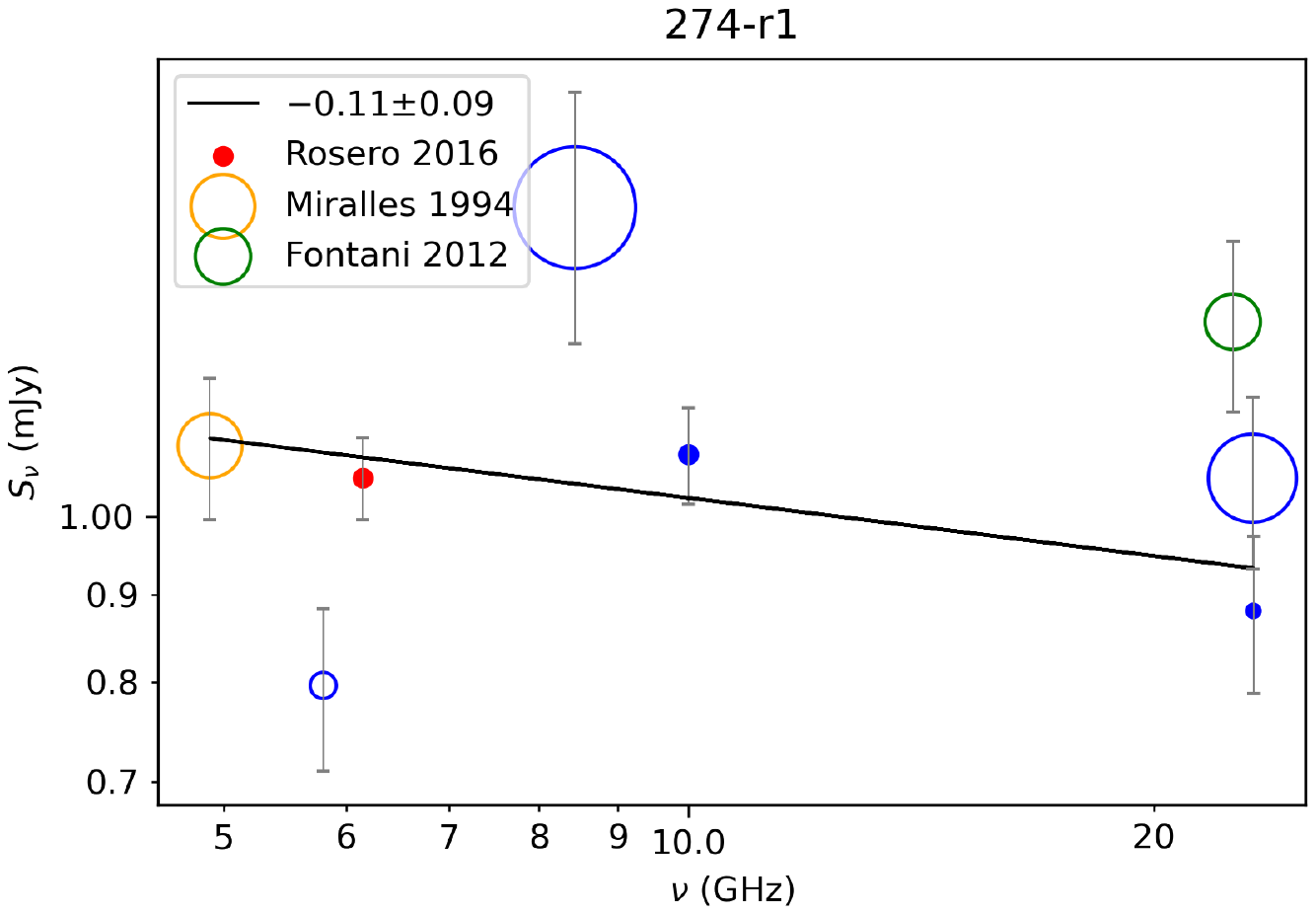}{0.5\textwidth}{}
}
\caption{\label{sed:274r1}SED of radio source 274-r1. The solid points are points adopted for fitting. The hollow ones are excluded owing to inconsistent recoverable scales. The blue markers represent the data obtained from our PI surveys and the VLA data archive; markers with the other colors represent the data obtained from the literature. The sizes of the circles are proportional to the source sizes in arc sec$^2$ measured from each continuum map. The grey segments represent the uncertainties in flux density. The solid black line represents the fitting result. The points are in correspondence with the data in Table \ref{tab:detection}. The data used for fitting are from \citet{2016ApJS..227...25R}, project 16A--301, and 13B--210.}
\end{figure*}

We detected two radio sources in MDC 274: 274-r1 and 274-r2. Radio source 274-r1 is associated with no dust condensation. It has a $\sim$3$''$ (0.02 pc) displacement with the closest condensation, which is within the typical offset between a UC \ion{H}{2} region and a millimeter core/condensation. It is slightly resolved under high resolutions (maps of projects 16A--301 and 13B--210). A flat spectral index of $-$0.11$\pm$0.09 (Figure \ref{sed:274r1}) is obtained, indicating optically thin free-free emission. This source has been reported by various works and has been identified as an UC \ion{H}{2} region with a central ionizing source equivalent to a B2 ZAMS star \citep{1994ApJS...92..173M, 2008ApJ...673..954C}. \citet{2012MNRAS.423.1691F} derived a spectral index of 0.1$\pm$0.2; \citet{2016ApJS..227...25R} obtained an index of $-$0.1$\pm$0.1. These results agree with ours within uncertainties. 

Radio source 274-r2 is weak and compact and is associated with a dust condensation. \citet{2016ApJS..227...25R} built the SED with the peak intensities and obtained a positive spectral index of 0.9$\pm$0.1. The central source of 274-r2 is considered as an embedded young stellar object with the luminosity of a B3 star \citep{2008ApJ...673..954C} or a class \uppercase\expandafter{\romannumeral1} source with the bolometric luminosity  $\sim$1000 L$_\odot$. A CO outflow is found to be associated with it \citep{2007A&A...474..911P}. 

\citet{2012MNRAS.423.1691F} reported another radio source in MDC 274 at the position of 20$^h$36$^m$08$^s$.23, +41$^\circ$40$'$02$''$.0. However, this source is neither detected in this work nor reported by \citet{2016ApJS..227...25R}. We also notice that despite higher sensitivity, \citet{ 2012MNRAS.423.1691F} did not detect 274-r1, whose peak intensity should be above 20$\sigma$ of their map. We thus refrain from including the results of \citet{ 2012MNRAS.423.1691F} in our work.

\subsection*{MDC 302, 520}\label{302_520}

The condensations of MDC 302 and 520 are distribute along the edge of the UC \ion{H}{2} region G80.363 +0.449. No associated radio source of 0.01-pc scale is detected in either MDC.

\subsection*{MDC 310}\label{310}

\begin{figure*}[htpb]
\gridline{
	\fig{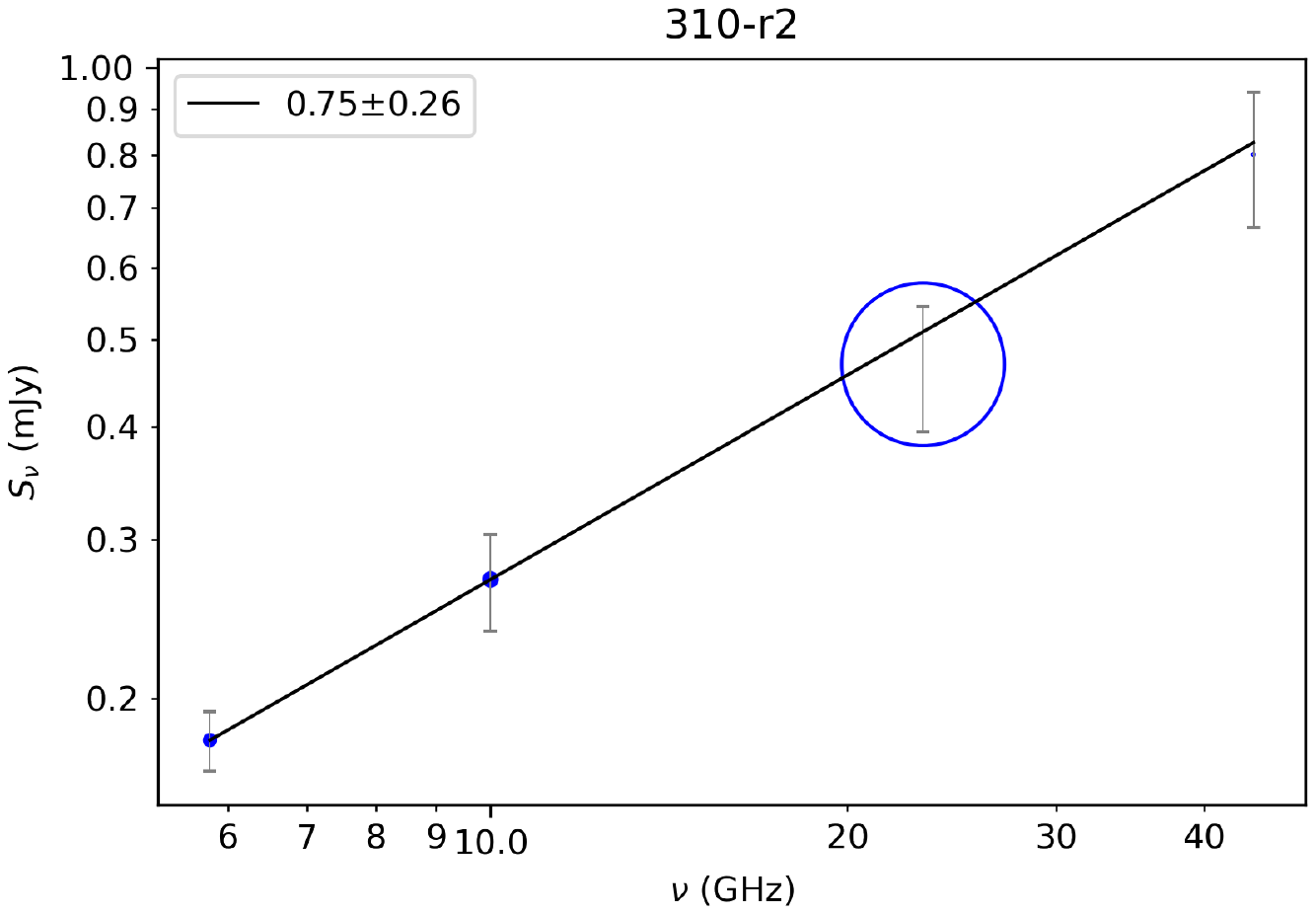}{0.5\textwidth}{}
}
\caption{\label{sed:310r2}Same convention as Figure \ref{sed:274r1} for the SED of 310-r2. The data used for fitting are from project 12B--140 and 16A--301.} 
\end{figure*}

Two radio sources, 310-r1 and 310-r2, are detected in MDC 310, both of which are associated with dust condensations. Under a very high resolution of 0$''$.04 at 44.0 GHz,  radio source 310-r1 is resolved into two resembled point sources with a projected separation of 0$''$.084, corresponding to $\sim$120 AU at a distance of 1.4 kpc (map of project 14A--092). Radio source 310-r2 remains unresolved in all our maps. Meanwhile, we notice that 310-r1 is brighter than 310-r2 in the 23.2 GHz and 44.0 GHz maps (maps of project17A--107 and 14A--092). But it is undetected in the 5.8 GHz and 10 GHz maps (project 12B--140 and 16A--301) when 310-r2 can still be well detected. This implies a very steep rising SED for 310-r1. The spectral index of 310-r2 is obtained to be 0.75$\pm$0.26 (Figure \ref{sed:310r2}), indicating thermal free-free emission from a region with moderate optical depth.

\subsection*{MDC 327, 742} \label{327_742}

MDC 327 and 742 are two MDCs very close to each other. Four radio sources are detected in this region.

We detected a compact radio source, 327-r1, in the 8.5 GHz radio continuum map (project AG625). It is associated with no dust condensation and has a 1$''$.5 $\sim$0.01 pc offset from the nearest condensation. This source is also reported by \citet{2010RMxAA..46..253N}. 

The infrared counterpart of MDC 742 is also known as IRAS 20178+4046. Three radio sources were detected: a cometary UC \ion{H}{2} region, 742-r2, and two compact sources, 742-r1 and 742-r3.

Radio source 742-r1 is associated with a dust condensation. It is so weak that is only detected in the maps with the highest sensitivities (project AG625 and 14A--481). This source has not been properly reported before this work. We notice that \citet{2017ApJ...836...96M} has also detected this source but confused it with VLA 4 reported by \citet{2010RMxAA..46..253N}.

\begin{figure*}[htpb]
\gridline{
	\fig{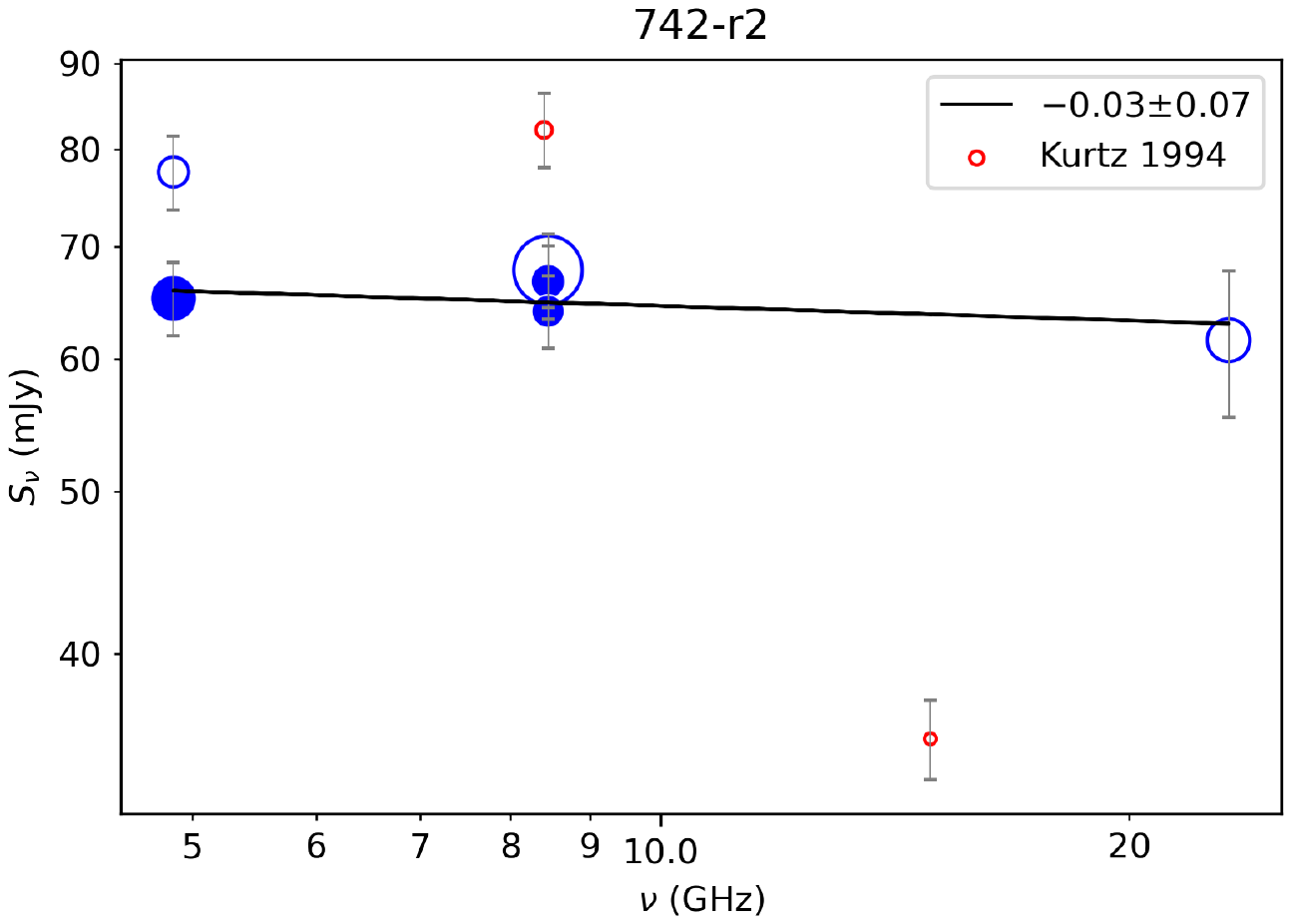}{0.5\textwidth}{}
}
\caption{\label{sed:742r2}Same convention as Figure \ref{sed:274r1} for the SED of 742-r2. The data used for fitting are from project AD219, AK477, AG625, and 17A--107.}
\end{figure*}

The UC \ion{H}{2} region, 742-r2, presents a typical cometary morphology with its peak position coincident with an condensation. Its spectral index is $-$0.03$\pm$0.07, a typical value of UC \ion{H}{2} regions with optically thin free-free emission (Figure \ref{sed:742r2}). 

Radio source 742-r3 is weak and compact. It is severely contaminated by the bright UC \ion{H}{2} region 742-r2. This source is also reported in \citet{1994ApJS...91..659K}.

\subsection*{MDC 340}\label{340}

MDC 340 is overlapping with an extended radio source, whose extended emission is filtered out by the interferometer and only leaves some compact clumps. A multi-band study by \citet{2017MNRAS.465.4753R} reveals that it is a commentary UC \ion{H}{2} region. At the condensation scale, we detected a compact radio source, 340-r1, associated with a dust condensation. This radio source is reported for the first time.

\subsection*{MDC 341}\label{341}

We detected a bright radio source in MDC 341. It is slightly elongated in the northwest-southeast direction and is associated with a dust condensation. This radio source is reported for the first time.

\subsection*{MDC 351}\label{351}

MDC 351 is located adjacent to a large-scale irregular radio source. One faint radio source, 351-r1, is detected to be associated with one of the dust condensations. This radio source is reported for the first time. 

\subsection*{MDC 370}\label{370}

MDC 370 is located adjacent to a large-scale irregular radio source. We detected no radio source associated with the MDC at the sub-0.1 pc. 

\subsection*{MDC 507, 753}\label{507, 753}

We detected three radio sources in MDC 507. Radio source 507-r1 and r2 are both compact and are associated with dust condensations. Both sources are reported for the first time. Source 507-r3 is a bright UC \ion{H}{2} region that has been reported in several works, e.g. \citet{1998A&A...336..339M, 2009A&A...501..539U}. It is not associated with any dust condensation. According to the Lyman photon rate calculated by \citet{1998A&A...336..339M}, the central ionizing source of 507-r3 is equivalent to a B3 ZAMS star. 

\begin{figure*}[htpb]
\gridline{
	\fig{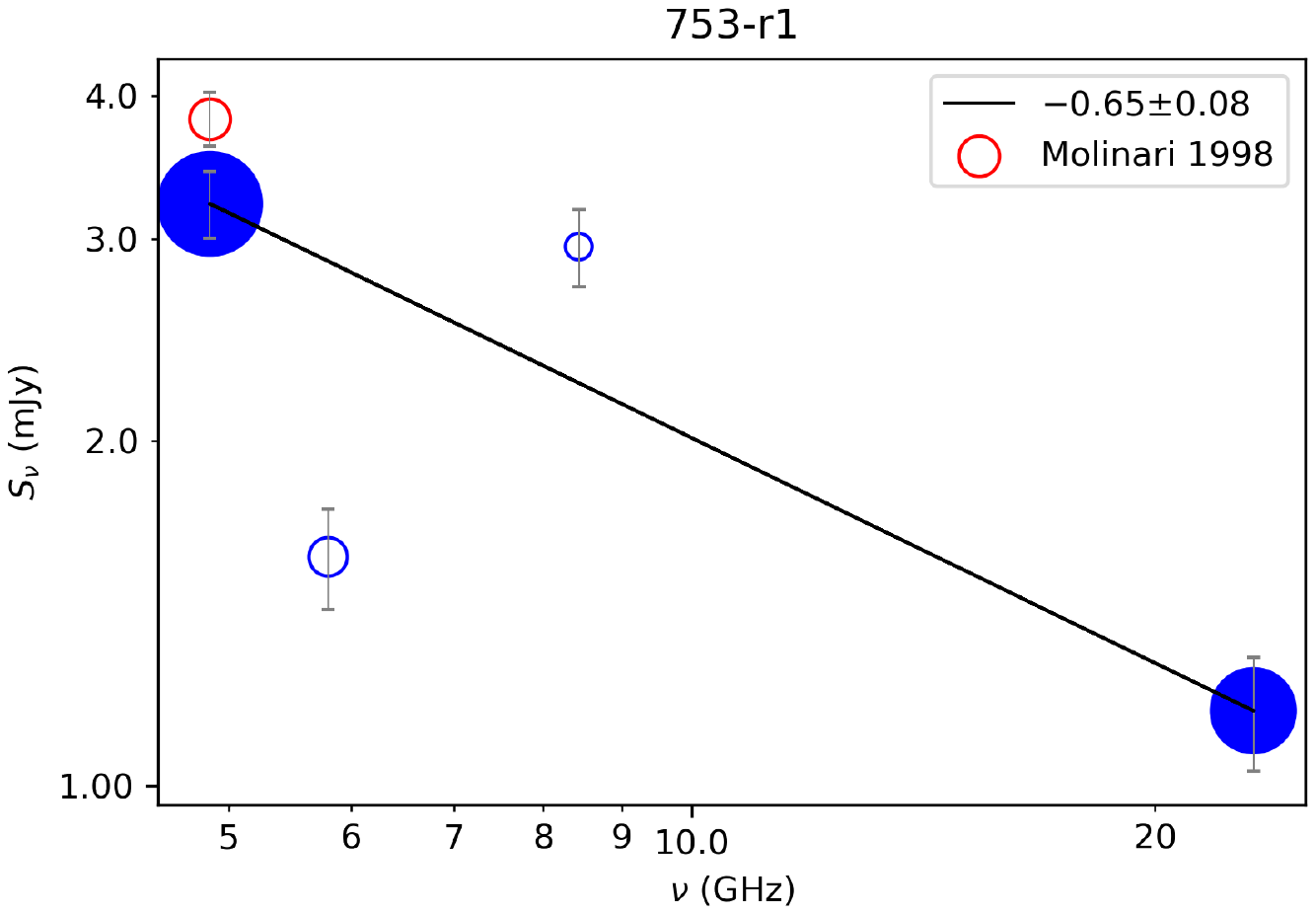}{0.5\textwidth}{}
}
\caption{\label{sed:753r1}Same convention as Figure \ref{sed:274r1} for the SED of 753-r1. The data used for fitting are from project AD219 and 17A--107.}
\end{figure*}

We detected a radio source, 753-r1, to be associated with a dust condensation in adjacent to MDC 753. It is resolved to be extended towards the northeast direction. A negative spectral index of $-$0.65$\pm$0.08 is obtained (Figure \ref{sed:753r1}).  Considering its jet-like morphology as well as the negative spectral index, we suggest that this radio source is a jet knot with non-thermal synchrotron radiation. It has also been reported by \citet{1998A&A...336..339M} and was identified as a possible precursor of UC \ion{H}{2} region. 

\subsection*{MDC 509}\label{509}

We detected a total of seven radio sources associated with MDC 509. Five radio sources, 509-r1--r5, are located in the MDC and are associated with dust condensations. They are weak and are only clearly detected in the 23.2 GHz map of project 17A-107. We also notice that these radio sources are connected by weak extended emission. Source 509-r3 and r5 are reported for the first time. Two brighter radio sources, 509-r6 and r7 are detected out of the MDC and are associated with no dust condensations. In the map of project 13B--210, which has smaller MRS, most of the extended radio emission is filtered out.

\subsection*{MDC 540}\label{540}

We detected three radio sources in adjacent to MDC 540, all of which are reported for the first time. All three sources are located on the edge of the MDC. They are weak and compact. Only source 540-r3 is associated with a dust condensation.

\subsection*{MDC 608}\label{608}

MDC 608 is located near a large-scale irregular radio source. We detected no associated radio source at the sub-0.1 pc scale.

\subsection*{MDC 640, 675}\label{640_675}

MDC 640 and 675 are located closely to each other. We detected no radio source at the sub-0.1 pc scale in MDC 640.

We detected a radio source, 675-r1, in MDC 675, which is associated with a dust condensation. The radio source is resolved to be extended to the southwest. This source is reported for the first time.

\subsection*{MDC 684}\label{684}

We detected a compact radio source, 684-r1, associated with a dust condensation in MDC 684. It is weak and barely meets our detection criteria. This source is reported for the first time.

\subsection*{MDC 698, 1179}\label{698_1179}

MDC 698 and 1179 are located closely to each other. We detected a compact radio source, 698-r1, associated with a dust condensation in MDC 698. This source is reported for the first time. We detected no radio source associated with MDC 1179 at the sub-0.1 pc scale.

\subsection*{MDC 699}\label{699}

\begin{figure*}[htpb]
\gridline{
	\fig{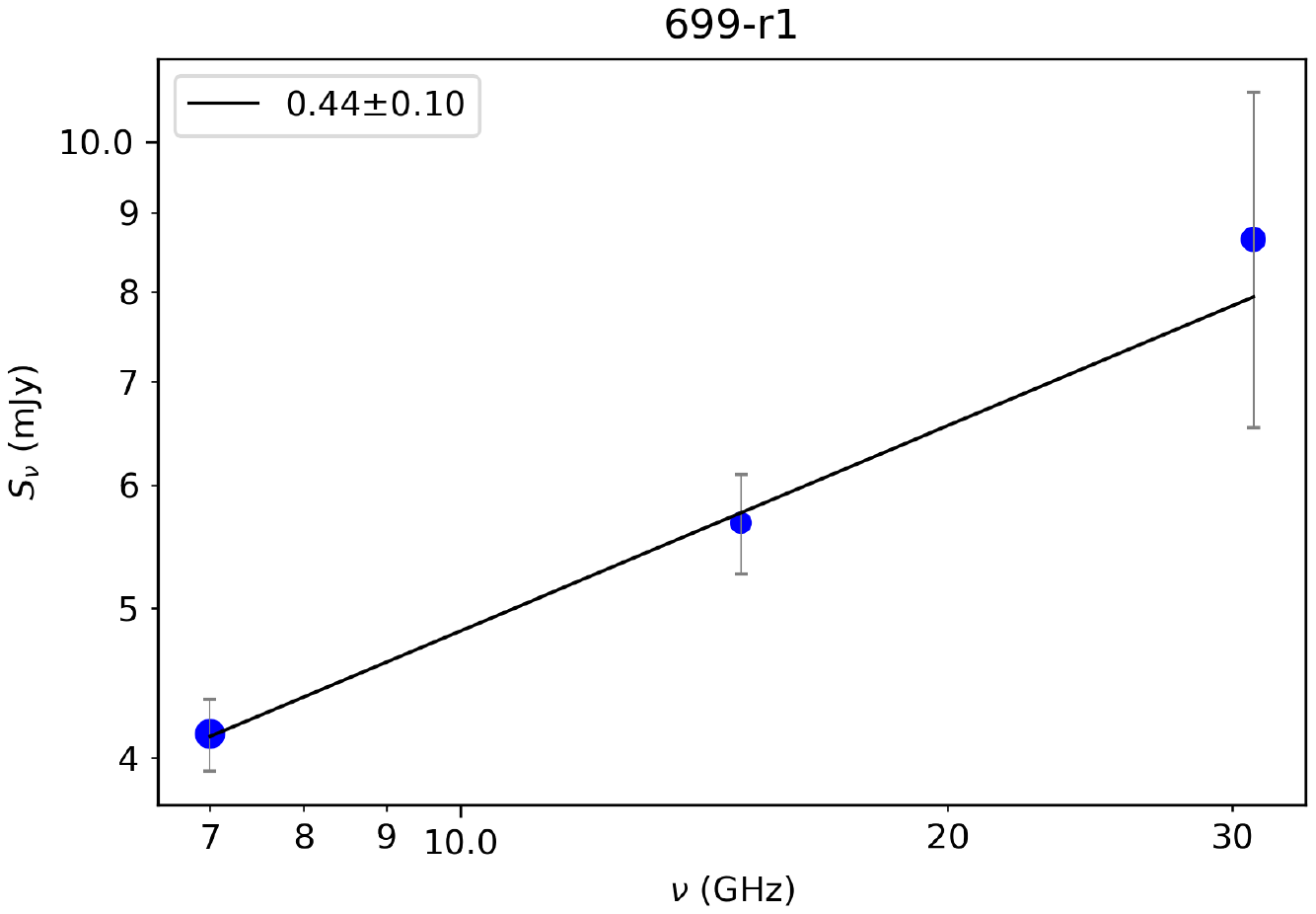}{0.5\textwidth}{}
}
\caption{\label{sed:699r1}Same convention as Figure \ref{sed:274r1} for the SED of 699-r1. The data used for fitting are from project  14A--420\_9375, AF381, and 14B--173.} 
\end{figure*}

MDC 699 is located in the DR21(OH) filament. We detected an extended radio source, 699-r1, on the edge of the MDC. The radio source is associated with no dust condensation. We obtained a spectral index of 0.44$\pm$0.10, which indicates thermal free-free emission from regions with a moderate optical depth (Figure \ref{sed:699r1}). This radio source is reported for the first time.

\subsection*{MDC 714}\label{714}

The radio emission detected to be associated with MDC 714 is elongated in the northwest-southeast direction. Although visually indistinguishable, the emission can be perfectly fitted by three Gaussian components. We thus take them as three individual sources, although we do not exclude the possibility that they are one extended radio source. All three sources are associated with dust condensations and are reported for the first time. 

\subsection*{MDC 723}\label{723}

We detected two weak and compact radio sources in MDC 723. Radio source 723-r1 is associated with a dust condensation; source 723-r2 is on the edge of the MDC and is associated with no dust condensation. Both radio sources are detected for the first time. 

\subsection*{MDC 725}\label{725}

We detected a hierarchical radio emission system in MDC 725. At a spatial scale of 1 pc, the bright and compact radio emission associated with the MDC is connected with a faint and extended radio source in the north (see the map of project 14A--420). The extended emission has two resemblance  peaks. The \textit{Spitzer} 24 $\mu$m counterpart of the extended source has been reported by Cao19, but the peak positions are not consistent with those of the radio emission.

At a scale of 0.1 pc, \textbf{two} radio source, 725-r1 and r2, are detected to be associated with the dust condensations in the MDC. In high resolution observations, radio source 725-r1 is resolved to have the typical core-halo morphology of UC \ion{H}{2} region. However, radio source 725-r2 has never been detected before and is reported for the first time. One explanation could be that its flux density has been increased in the two decades.

\begin{figure*}[htpb]
\gridline{
	\fig{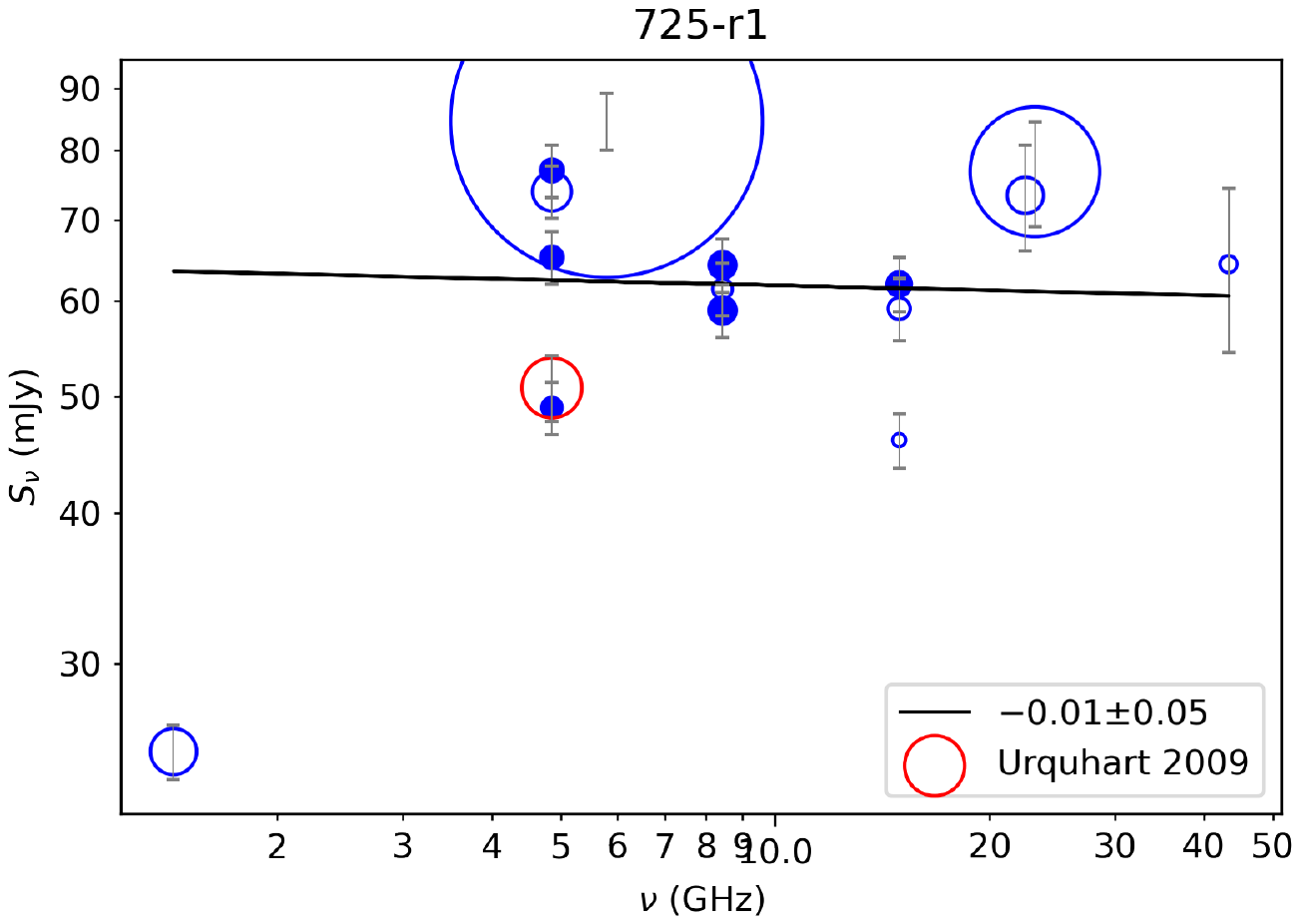}{0.5\textwidth}{}
}
\caption{\label{sed:725r1}Same convention as Figure \ref{sed:274r1} for the SED of 725-r1. The data used for fitting are from project AH369, AH549, AM446, AK355 (8.4 GHz and 14.9 GHz), and AC240 (8.4 GHz)}
\end{figure*}

The SED (Figure \ref{sed:725r1}) is built from 15 datasets that covers the L, C, X Ku, K, and Q band, with angular resolutions from $\sim$0.1$''$ to $\sim$5$''$. We selected the data that can moderately resolve the source for SED fitting. The obtained spectral index is $-$0.01$\pm$0.05, a flat spectrum indicating optically thin free-free emission. Both the morphology and the spectral index support that 725-r1 is an UC \ion{H}{2} region.

The radio properties of 725-r1 have been studied by various works. It was first detected by single-dish surveys (e.g. \citealt{1970A&A.....4..378W, 1984A&AS...58..291W}), and was further observed by interferometers \citep{1994ApJS...91..659K, 2009A&A...501..539U}. OH and Class II methanol masers were detected on or close to its edge, indicating a newly formed massive star \citep{2000ApJS..129..159A, 2003A&AT...22....1M}. The radio flux density of 725-r1 was reported to be corresponding to a B0.5 star \citet{1990AJ.....99..288O}.

\subsection*{MDC 798}\label{798}

MDC 798 is located adjacent to a bright $\sim$1 pc-scale champagne-like \ion{H}{2} region. The dust condensations are distributed closely along the outer rim of the radio source. A natural hypothesis is that the dust distribution of MDC 798 is affected by the ionized region, which has been reported by \citet{2005A&A...433..565D}. At the core scale, several radio clumps are observed. Although some of them have reached our criteria of source identification, considering the radio environment, they are more likely to be the compact components of the UC \ion{H}{2} region whose extended emission has been filtered out by the interferometer. One radio source, 798-r1, is confirmed to be a radio source associated with a dust condensation by its outflow signature (Yang et al., in preparation). This radio source is reported for the first time.

\subsection*{MDC 801}\label{801}

We detected a compact radio source, 801-r1, in MDC 801. It is associated with a dust condensation. This source is reported for the first time.

\subsection*{MDC 839}\label{839}

We detected a compact radio source, 839-r1, in MDC 839. It is associated with a dust condensation. This source is reported for the first time.

\subsection*{MDC 892}\label{892}

MDC 892 is located adjacent to a bright \ion{H}{2} region \citep{1980MNRAS.192..377C}. It is composed of a bright core and a weak tail extended to the east. The dust condensations of MDC 892 distribute right along the northeast leading front of the \ion{H}{2} region, suggesting that the core is possibly compressed by the expansion of the \ion{H}{2} region. No radio source is detected at the sub-0.1 pc scale.

\subsection*{MDC 1018, 1467}\label{1018_1467}

\begin{figure*}[htpb]
\gridline{
	\fig{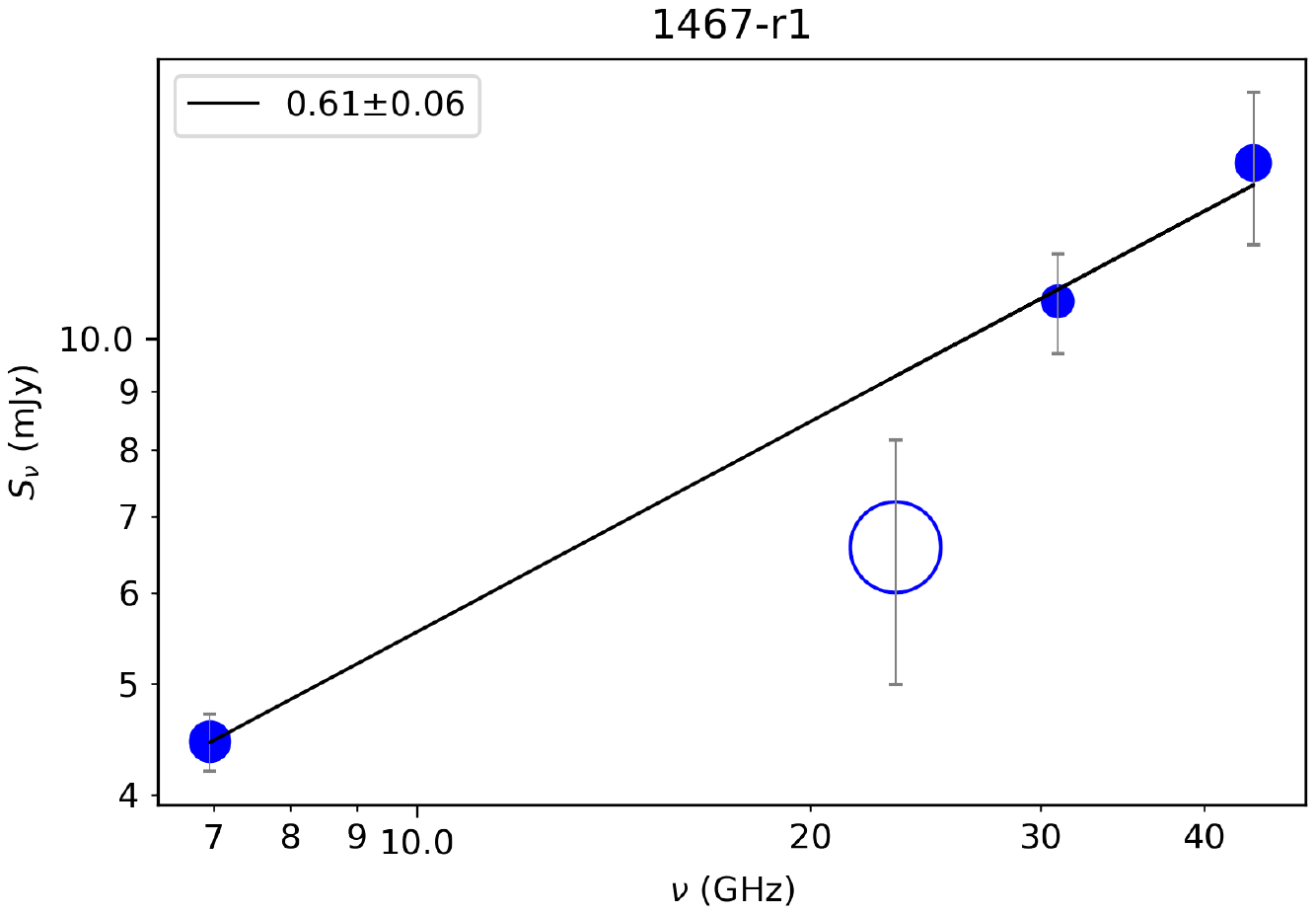}{0.5\textwidth}{}
}
\caption{\label{sed:1467r1}Same convention as Figure \ref{sed:274r1} for the SED of 1467-r1. The data used for fitting are from project 14A--420\_9375, 14B--173\_7917, and 13A--315.} 
\end{figure*}

MDC 1018 and 1467 are located in the DR21(OH) filament and are close to each other. No radio source is detected in MDC 1018. Two radio sources are detected adjacent to MDC 1467, both of which are associated with the dust condensations. Radio source 1467-r1 is resolved to have a compact core and a faint irregular envelope under high resolution with moderate MRS map of project 14B--173. The dense core is further resolved to be composed of two Gaussian-like sources under even higher resolution (map of project 15A--059). \citet{2009ApJ...698.1321A} have resolved two additional weak compact sources from the envelope. The spectral index of 1467-r1 is 0.61$\pm$0.06 (Figure \ref{sed:1467r1}), indicating free-free emission from regions with moderate optical depth.Radio source 1467-r2 is weak and compact. It is only detected in the high sensitivity map (project 14B--173).

\subsection*{MDC 1112}\label{1112}

MDC 1112 (or W75N(B)) is one of the most well-studied star-forming regions in the Cygnus X complex. It is first observed at centimeter wavelengths by \citet{1981ApJ...244...76H} and is resolved to have three bright radio continuum components by the follow-up studies (e.g. \citealt{1994A&A...284..215H, 1997ApJ...489..744T}): 1112-r1 (also know as VLA 1 or W75N(Ba)), 1112-r2 (also known as VLA 2), and 1112-r3 (also know as VLA 3 or  W75N(Bb)). All of them are associated with the brightest dust condensation of MDC 1112. Another three weaker radio sources, 1112-r4, 1112-r5 (together known as W75N(Bc)), and 1112-r6 (also known as VLA 4), are detected by later high-sensitivity observations (e.g. \citealt{2001ApJ...546..345S, 2010AJ....139.2433C}). None of them are associated with dust condensations.

The nature of the radio sources has been long under debate. \citet{1997ApJ...489..744T} considered that 1112-r1 was a thermal biconical jet and 1112-r2 and r3 were both UC \ion{H}{2} regions. This conclusion was questioned by \citet{2001ApJ...546..345S} and \citet{2010AJ....139.2433C}, who claimed that 1112-r1 was insufficient to power the large-scale outflow \citep{1994A&A...284..215H} and the main powering source should be 1112-r3. The most recent works by \citet{2010AJ....139.2433C, 2015Sci...348..114C} and \citet{2020MNRAS.496.3128R} concluded that 1112-r1 was an UC \ion{H}{2} region, 1112-r2 a thermal, ionized wind, and 1112-r3 a thermal jet. They also concluded that 1112-r4, r5, and r6 were all Herbig-Haro (HH) objects with r4 and r5 together as a bright radio HH object and r6 as an obscured one.

The spectral indices of the radio sources in MDC 1112 have been provided by several works. However, these results show great inconsistencies that some may even suggest different physical mechanisms. We suggest that the inconsistencies are mainly caused by two reasons. Firstly, the radio sources are highly time-variable \citep{2010AJ....139.2433C, 2015Sci...348..114C}. The SEDs should be fitted with data obtained from the same epoch, e.g. less than five years (inferred from Figure 2 in \citealt{2010AJ....139.2433C}). Among our references, only \citet{1997ApJ...489..744T} and \citet{2015Sci...348..114C} meet this requirement. Secondly, since the sources are resolvable with moderate resolutions, it is essential to fit the SEDs with data sensitive to similar spatial scales (see Section \ref{ssec: idx}). We are not able to check the $uv$ coverages of the literature data, whereas it is possible that the data are sensitive to different spatial scales implied by the highly inconsistent beam sizes.
\clearpage

\startlongtable
\begin{deluxetable}{c|c|c}
\tablecaption{Spectral indices of the radio sources in 1112 \label{tab:1112}
}
\tablehead{
\colhead{Source} & \colhead{Spec. Idx.} & \colhead{Reference}
}
\startdata
1112-r1 & 0.7$\pm$0.1 & \citet{1997ApJ...489..744T} \\
            & 0.2$\pm$0.3 &  \citet{2004ApJ...601..952S} \\
            & $-$1.60         & \citet{2007MNRAS.380..246G} \\
            & $-$0.4$\pm$0.1 & \citet{2010AJ....139.2433C} \\
\hline
1112-r2 & $\geqslant$ 1.0 & \citet{1997ApJ...489..744T} \\
            & 0.4$\pm$0.1     &  \citet{2004ApJ...601..952S} \\
            & 2.2$\pm$0.3     & \citet{2010AJ....139.2433C} \\
            & \textbf{0.61} & \citet{2015Sci...348..114C} \\
\hline
1112-r3 & 1.5$\pm$0.1 & \citet{1997ApJ...489..744T} \\
            & 0.5$\pm$0.3 &  \citet{2004ApJ...601..952S} \\
            & 0.79$\pm$0.15 & \citet{2007MNRAS.380..246G} \\
            & 0.6$\pm$0.1    & \citet{2010AJ....139.2433C} \\
\hline
1112-r4,5 & $-$0.3$\pm$0.6 &  \citet{2004ApJ...601..952S} \\
               & $-$0.34              &   \citet{2007MNRAS.380..246G} \\
               & $-$0.2$\pm$0.2 & \citet{2010AJ....139.2433C} \\
\hline
1112-r6   & 0.4$\pm$0.5 & \citet{2010AJ....139.2433C} \\
\enddata
\end{deluxetable}

\subsection*{MDC 1201}\label{1201}

\begin{figure*}[htpb]
\gridline{
	\fig{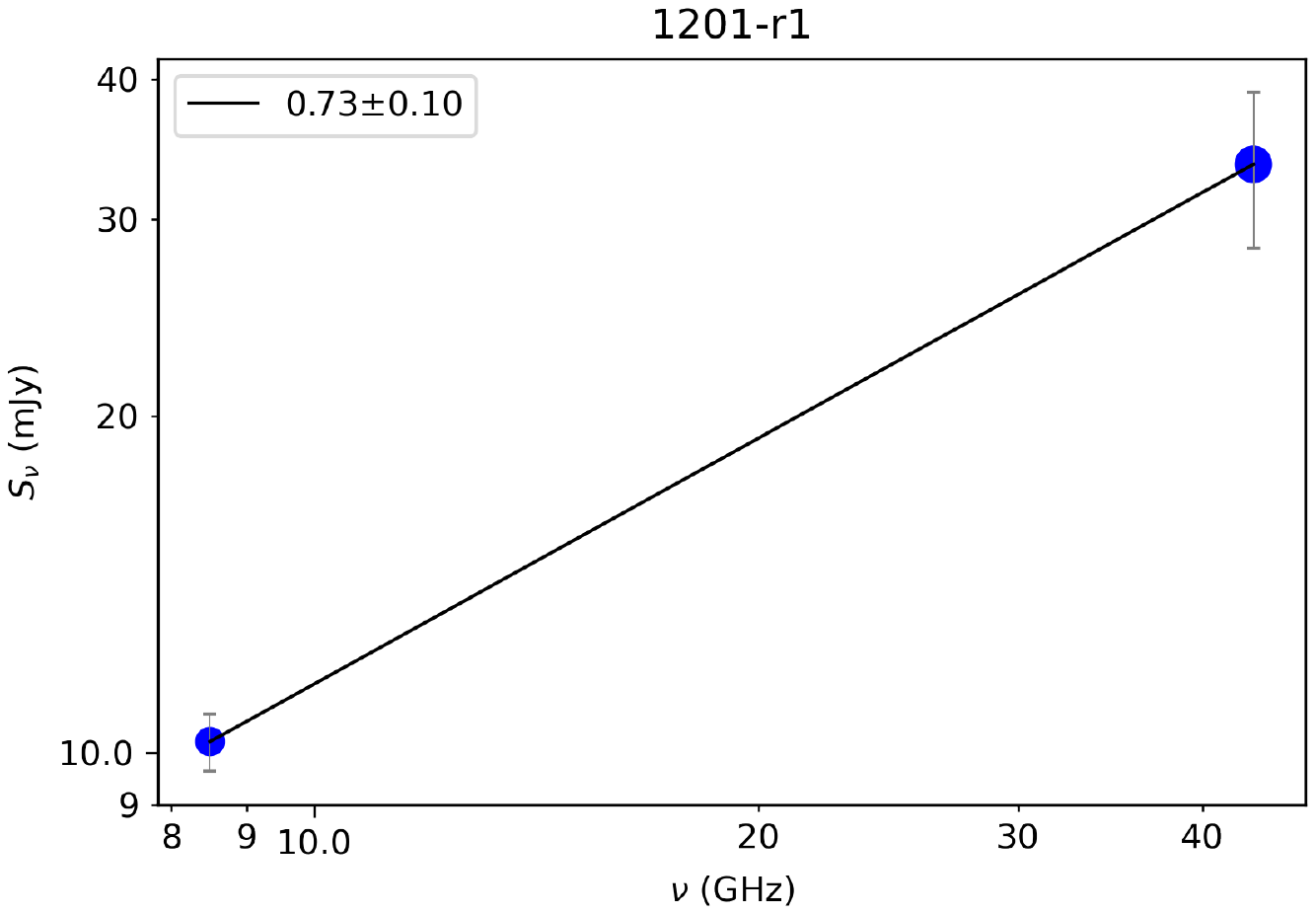}{0.5\textwidth}{}
}
\caption{\label{sed:1201r1}Same convention as Figure \ref{sed:274r1} for the SED of 1201-r1. The data used for fitting are from project AF362 and AR537.} 
\end{figure*}

MDC 1201 is located in the S106 region. S106 is one of the best-studied bipolar \ion{H}{2} regions at a distance of 1.3 kpc \citep{2018A&A...617A..45S}. As part of the Cygnus X rift, it is believed to be physically connected with the Cygnus X complex \citep{1982ApJ...255...95S, 2007A&A...474..873S}. The most prominent radio emission in this region is a bright and compact source, 1201-r1, whose infrared counterpart is S106 IR. It has been interpreted as an extremely powerful source that powers the large-scale bipolar \ion{H}{2} outflow \citep{1983ApJ...272..154B, 1999ApJ...525..821F, 2007MNRAS.380..246G}. The radio source remains unresolved at all resolutions. We construct the SED of this source and derive a spectral index of 0.73$\pm$0.11 (Figure \ref{sed:1201r1}). It is comparable to those of \citet{1983ApJ...272..154B} (0.73) and \citet{2007MNRAS.380..246G} (0.65$\pm$0.11).

\subsection*{MDC 1225}\label{1225}

We detected two radio sources in MDC 1225, each of which are associated with a dust condensation and are reported for the first time. In the 24.4 GHz map observed in 2013 (project 13A--372), both radio sources are detected and are connected by extended emission, whereas 1225-r1 is not detected in the 23.2 GHz map observed in 2017 (project 17A--107) owing to a higher RMS noise. Both radio maps are at K band with similar resolutions. We notice that radio source 1225-r2 has a dramatic increase in flux density from 0.1 mJy to 0.4 mJy in merely four years (2013--2017), which means an average increase rate of $\sim$40\% per year. It also shows a change in morphology from compact to elongated. Since 1225-r2 should be at an early evolution stage, which has low radio luminosity, no emission at 8 $\mu$m, and weak emission at 24 $\mu$m, we consider it as an ionized jet. The time variability of thermal radio jets has been observationally confirmed by various studies (see \citet{2018A&ARv..26....3A} and the references therein), most of which are no more than 20\% but some extreme cases can have flux density doubled in two years \citep{2012A&A...537A.123R}. Our data are not yet enough to reveal the mechanism of the time variation. According to the previous studies, high time variability ($>$20\%) can be caused by the periastron passage in a close binary system \citep{2018A&ARv..26....3A} or more commonly, the ejection of bright radio knots \citep{2010ApJ...712.1403P, 2012A&A...537A.123R}.
 
\subsection*{MDC 1243, 1599, 5417}\label{1243_1599_5417}

The three MDCs are linearly distributed in the DR21(OH) filament. We detected no radio source in MDC 1599 and one radio source in each of MDC 1243 and 5417. Both radio sources are compact and associated with the dust condensations. They are reported for the first time.

\subsection*{MDC 1267, 3188}\label{1267_3188}
 
MDC 1267 and 3188 are located closely to each other. MDC 1267 is associated with two radio sources: source 1267-r1 is compact and is associated with a dust condensation; source 1267-r2 is located out of the MDC. MDC 3188 has one radio source, 3188-r1. All the radio sources are reported for the first time.

\subsection*{MDC 1454, 2210}\label{1454_2210}

MDC 1454 and 2210 are located in DR15, one of the most prominent star-forming regions in the southern Cygnus X complex. We detected five compact radio sources in MDC 1454, in which 1454-r1, r2, and r3 are associated with dust condensations. All the radio sources are reported for the first time.

We detected a weak and extended radio source, 2210-r1, in MDC 2210. It is associated with a dust condensation. This source is reported for the first time.

\subsection*{MDC 1460}\label{1460}

We detected no radio emission at the sub-0.1 pc scale associated with MDC 1460. At a larger scale, the dust condensations are distributed on the outer rim of its central UC \ion{H}{2} region, G080.634+00.684. A hypothesis is that the expanding UC \ion{H}{2} region compresses the ambient material.

\subsection*{MDC 1528}\label{1528}

MDC 1528, also known as DR21, is a very bright and well-studied high-mass star-forming region. We detected no radio emission at the sub-0.1 pc scale. At larger scale, a bright extended radio source was detected on the north of the MDC. The large-scale radio source shows a very complex structure. Its major components are two patches of cometary emission: a small one in the north and a larger one in the south.  Two weak and long ``tails''  extend to the east. The nature and origin of the radio emission have been studied by various works. \citet{2003ApJ...596..344C} suggested that they are two cometary UC \ion{H}{2} regions produced by wind-blowing stars moving through the molecular cloud; \citet{2013ApJ...765L..29Z} raised the hypothesis that they are the products of an explosive event related to the disintegration of a massive stellar system.

\subsection*{MDC 4797}\label{4797}

MDC 4797 is surrounded by a bright large-scale shell-like \ion{H}{2} region, whose infrared counterpart has been reported by Cao19. A compact source, 4797-r1, is marginally detected. This source is reported for the first time.